\pdfoutput=1
\RequirePackage{fix-cm}
\documentclass[twocolumn]{svjour3}          % twocolumn
\smartqed  % flush right qed marks, e.g.\ at end of proof
\usepackage{amsmath}
\usepackage{graphicx}
\usepackage[T1]{fontenc}
\usepackage[utf8,latin1]{inputenc}
\usepackage{epsfig}
\usepackage{txfonts}
\usepackage{amssymb}
\usepackage{siunitx}
\usepackage{natbib}
\usepackage{hyperref}
\usepackage{url}
\usepackage{xcolor}

\usepackage{bm}
\usepackage{sgl-commands}
\usepackage{aas_macros}
\journalname{}

\title{Wave Mechanics, Interference, and Decoherence in Strong Gravitational Lensing}
\subtitle{}
\date{}

\titlerunning{}

\author{Calvin Leung \and Dylan Jow \and Prasenjit Saha \and Liang Dai \and Masamune Oguri \and L\'eon V.~E.~Koopmans}
\institute{Calvin Leung \at MIT Kavli Institute for Astrophysics and Space Research, Massachusetts Institute of Technology, 77 Massachusetts Ave, Cambridge, MA 02139, USA}
\begin{document}

\maketitle
% \author[0000-0002-4209-7408]{Calvin Leung}
%   \affiliation{MIT Kavli Institute for Astrophysics and Space Research, Massachusetts Institute of Technology, 77 Massachusetts Ave, Cambridge, MA 02139, USA}
%   \affiliation{Department of Physics, Massachusetts Institute of Technology, 77 Massachusetts Ave, Cambridge, MA 02139, USA}
% \author[0000-0003-0136-2153]{Prasenjit Saha}
%   \affiliation{Physik-Institut, University of Zurich, 8057 Zurich, Switzerland}
% \author[0000-0003-3236-8769]{Dylan L.~Jow}
%   \affiliation{Canadian Institute for Theoretical Astrophysics, 60 St.~George Street, Toronto, ON M5S 3H8, Canada}
%   \affiliation{Department of Physics, University of Toronto, 60 St.~George Street, Toronto, ON M5S 1A7, Canada}
% \author[0000-0003-2091-8946]{Liang Dai}
%   \affiliation{Department of Physics, University of California, 366 Physics North MC 7300, Berkeley, CA 94720, US}
% \author[0000-0003-1840-0312]{Léon V.~E.~Koopmans}
%   \affiliation{Kapteyn Astronomical Institute, University of Groningen, P.O. Box 800, 9700AV Groningen, the Netherlands}
% \author[0000-0003-3484-399X]{Masamune Oguri}
%   \affiliation{Center for Frontier Science, Chiba University, 1-33 Yayoi-cho, Inage-ku, Chiba 263-8522, Japan}
%   \affiliation{Department of Physics, Graduate School of Science, Chiba University, 1-33 Yayoi-Cho, Inage-Ku, Chiba 263-8522, Japan}
% Unique acks:
\newcommand{\allacks}{
C.L. was supported by the U.S. Department of Defense (DoD) through the National Defense Science \& Engineering Graduate Fellowship (NDSEG) Program.
L.D. acknowledges research grant support from the Alfred P. Sloan Foundation (Award Number FG-2021-16495).
M.O. was supported by JSPS KAKENHI Grant Number JP22K18720.
}

\begin{abstract}
Wave-mechanical effects in gravitational lensing have long been predicted, and with the discovery of populations of compact transients such as gravitational wave events and fast radio bursts, may soon be observed. We present an observer's review of the relevant theory underlying wave-mechanical effects in gravitational lensing. Starting from the curved-spacetime scalar wave equation, we derive the Fresnel-Kirchoff diffraction integral, and analyze it in the eikonal and wave optics regimes. We answer the question of what makes interference effects observable in some systems but not in others, and how interference effects allow for complementary information to be extracted from lensing systems as compared to traditional measurements. We end by discussing how diffraction effects affect optical depth forecasts and lensing near caustics, and how compact, low-frequency transients like gravitational waves and fast radio bursts provide promising paths to open up the frontier of interferometric gravitational lensing.

\end{abstract}

\tableofcontents
%Colleagues who could potentially review prior to submission: Dylan Jow, Ue-Li Pen, Olaf Wucknitz, Job Feldbrugge
%External: Nakamura, Deguchi, Watson, Joachim Kopp

\section{Introduction}
Gravitational lensing in the geometric optics approximation has enjoyed immense success as a unique and precise probe of cosmology and astrophysics. However, electromagnetic and gravitational radiation both obey wave equations, have a wave-like nature, and interfere and diffract as they propagate to Earth from the cosmos. Just how the first theoretical predictions of gravitational lensing~\citep{chwolson1924uber} were made decades before the first observation of an Einstein ring~\citep{hewitt1988unusual}, wave effects in gravitational lensing were predicted long before there was any hope of detecting them.

Now, there is hope on multiple fronts: Lensed supernovae~\citep{kelly2015multiple} have already been detected. In principle, wave-mechanical effects should be observable in the lensing of gravitational wave (GW) signals from binary black
holes~\citep{abbott2016observation}, binary neutron stars~\citep{abbott2017observation}, and even black hole-neutron star mergers~\citep{abbott2021observation}, as well as a large
population of fast radio bursts (FRBs) at low frequencies~\citep{chimefrb2021first}. Growing observational efforts in these areas puts astronomers well on their way to detecting interference effects in gravitational lensing. Measuring these interference effects enables unique tests of gravity: for example, testing the equivalence principle tests using precise measurements of time lags between electromagnetic and gravitational-wave images~\citep{takahashi2017arrival}. It is also of interest from the perspective of testing the foundations of quantum mechanics~\citep{miller1997delayed,jacques2007experimental,doyle2009quantum,leung2018astronomical}. 

Wave-mechanical effects in gravitational lensing also offer a significant enhancement of gravitational lensing as an astrophysical tool. In traditional lensing, only two numbers (the flux magnification and time delay) can be measured, leading to difficult degeneracies in lensing-based inference: the mass-sheet degeneracy is perhaps the best-known example~\citep{falco1985model,saha2000lensing}. In the wave-optics limit, much more information is present, since the flux magnification, and potentially the phase of the wave, may both be measured as a function of wavelength. In certain circumstances, time delays may also be extracted interferometrically with dramatically-increased precision. As an example, lensing of FRBs can allow for interferometric measurements of lensing time delays. These time delays can be measured with a precision of $\sim 10^{-9}-10^{-6}$ seconds: several orders of magnitude more precise than any lensing delay ($\sim$ 1 millisecond) inferred from measured FRB light curves alone \citep{wucknitz2021cosmology,jow2020wave}. 

Wave-mechanical effects may also shed light on the emission physics of compact objects (such as GWs and FRBs), by using gravitational lenses as astrophysical-scale interferometers. This concept has been demonstrated in the study of pulsar scattering in the ionized interstellar medium to resolve the nanoarcsecond-scale emission region of the Crab pulsar \citep{Lin2022crab}. The observation of interference effects in gravitational lensing is a new frontier in using lensing to learn about our universe.

The theory of interference effects in gravitational lensing was developed since as far back as 1973~\citep{ohanian1973focusing,ohanian1974on,peters1974index}. Further developments have mainly focused on GW lensing~\citep{deguchi1986wave,deguchi1986diffraction,nakamura1998gravitational,nakamura1999wave,baraldo1999gravitationally,takahashi2003wave,macquart2004scattering} and because of the relative simplicity of the gravitational waveforms produced by compact binary coalescences, and the low (kilohertz) emission frequencies at which interference effects are most pronounced. 
However, wave-mechanical effects in lensing are of growing importance at frequencies far higher than gravitational waves. An early example was in using ``femtolensing'' to probe extremely low mass scales by analyzing the spectra of gamma ray bursts (GRBs)~\citep{gould1992femtolensing,stanek1993features,nemiroff1993searching} More recently, various other applications using lensing to search for exotic phenomena have had to contend with wave-mechanical effects, which limit the reach of the search. Some examples include optical microlensing searches for primordial black holes (PBHs; see e.g.~\citet{sugiyama2020on},
searches for PBHs using lensed FRBs~\citep{katz2020looking,kader2022high,leung2022constraining}, searches for small-scale density perturbations in minihalos or e.g.\ axion dark matter~\citep{venumadhav2017microlensing,dai2020gravitational}, and even searching for lensing by cosmic strings~\citep{xiao2022detecting}.

In this review, we aim to present the fundamental concepts underlying wave interference in gravitational lensing. We derive the Fresnel-Kirchhoff integral from first principles (Sec.~\ref{secy:first_principles}), and analyze it in two opposite regimes. We first analyze the stationary-phase (a.k.a. Eikonal) regime in Sec.~\ref{secy:coherent_geo}. Then, we present some analytical solutions from the diffractive (a.k.a. wave optics) regime (Sec.~\ref{secy:diffractive}) and comment on numerical solution methods for this regime, in which the saddle-point approximation fails. We discuss how the finite angular sizes of sources pose a challenge to detecting interference effects (Sec.~\ref{secy:angular_broadening}). 

In anticipation of detections of wave-mechanical effects, we discuss several observationally-relevant features of wave mechanics in lensing. First, we discuss how the chromaticity of lensing in the wave optics regime can be observed and used to significantly increase the information extracted out of systems which exhibit wave optics effects (Sec.~\ref{secy:breaking}). We then discuss the effect of diffraction near caustics (Sec.~\ref{secy:finite},~\ref{secy:fold}) and talk about how wave optics alters calculations of the lensing optical depth (Sec.~\ref{secy:optical_depth}) in searches for exotic objects. We finally comment on how various compact transients offer opportunities for detections in the future (Sec.~\ref{secy:discussion}).

\section{The curved-spacetime scalar wave equation}\label{secy:first_principles}
Much of gravitational lensing is captured by the scalar wave equation in a curved spacetime:
\begin{align}
    \qquad \qquad \qquad \qquad 0 &= g^{ab} \nabla_a \nabla_b \phi.
    \label{eqy:laplacian}
\end{align}
Scalar wave theory captures interference and diffraction phenomena in gravitational lensing~\footnote{but ignores e.g. polarization-dependent effects.}, and is sufficient for analyzing most interference phenomena in gravitational lensing. The starting point of scalar wave theory is the Fresnel-Kirchoff integral, from which most lensing phenomena can be derived. It requires a little bit of general relativity, but we will assume the reader has little background in that area.

The formalism of gravitational lensing is similar to the Newtonian dynamics of point masses orbiting in a potential. In the same way that the external potential is fixed for point masses, we consider the dynamics of a scalar field representing our wave, $\phi$, on a static background spacetime, whose curvature is specified by the metric tensor $g^{ab}$. In a weak gravitational field~\footnote{Confusingly, ``strong lensing", which refers to the regime in lensing in which multiple images arise, occurs mostly in the weak-field limit; we will discuss the strong-field limit briefly in Sec.~\ref{secy:beyond}} with only scalar degrees of freedom, the metric tensor can be written in terms of the gravitational potential as $g_{00} = - (1 + 2 U)$ and $g_{ii} = 1 - 2 U$ for $i = 1,2,3$, and where $U$ is the Newtonian gravitational potential, and where we use units in which $c = 1$ for now. 

In Eq.~\ref{eqy:laplacian}, $\nabla_a$ refers to the covariant derivative. The covariant derivative needs to be expanded into partial differential equations using several identities from general relativity. We introduce the notation $g = \det(g^{ab}) = g^{00} g^{11} g^{22} g^{33}$ for our diagonal metric. Since scalars are Lorentz-invariant, the covariant derivative of a scalar is the same as its partial derivative:
\begin{align}
    \qquad \qquad \qquad \qquad \nabla_b \phi &= \partial_b \phi.
\end{align}
The analogous rule for (the covariant divergence of) a vector is\footnote{This in turn arises from an identity for a contracted Christoffel symbol: $\Gamma^{c}_{c a} = \dfrac{\partial_a (\sqrt{-g})}{\sqrt{-g}}$; see e.g.~\citet{lightman1975problem} (Exercise 7.7j) for more details.}
\begin{align}
    \qquad \qquad \qquad \qquad \nabla_a V^a &= \dfrac{1}{\sqrt{-g}}\partial_a(\sqrt{-g} V^a).
    \label{eqy:divergence}
\end{align}
Eq.~\ref{eqy:divergence} implies
\begin{align}
    g^{ab}\nabla_a \nabla_b \phi &= \nabla_a(g^{ab} \nabla_b \phi) = \nabla_a( g^{ab} \partial_b \phi) \\
    &= \dfrac{1}{\sqrt{-g}} \partial_a(\sqrt{-g} g^{ab} \partial_b \phi) = 0.\label{eqy:wave_eq}
    \intertext{Eq.~\ref{eqy:wave_eq} is the wave equation often used as a starting point in studying gravitational lensing in wave optics, e.g.~\citet{nakamura1999wave,macquart2004scattering}. We now expand the four non-zero terms in $U$, since we are working the weak-field limit, $|U| \ll 1$; this justifies neglecting terms of $\mathcal{O}(U^2)$. Since the potential is static, the 00 term evaluates to}
    \dfrac{1}{\sqrt{-g}}\partial_0(\sqrt{-g} g^{00} \partial_0 \phi) &= -\dfrac{1}{1+2U} \partial^2_{t} \phi \approx (-1 + 2U)\partial^2_t \phi.
\intertext{The $a = b = 1$ term is}
\dfrac{1}{\sqrt{-g}}\partial_1(\sqrt{-g} g^{11} \partial_1 \phi) &= \phi'' + (-4 \phi' U' + 2 \phi'') U.\label{eqy:middleterm}
\end{align}
where the primes denote $\partial_1$ (the $a = b = 2$ and $a = b = 3$ terms are similar). From dimensional arguments, $U' \sim |U| / R_s$ (the curvature scale of the spacetime) and the scale of $\phi' \sim 1/\lambda$, and $\phi'' \sim 1 / \lambda^2$. We are justified in neglecting the $\phi' U'$ term of Eq.~\ref{eqy:middleterm}, which is smaller than the other two by a factor of $\mathcal{O}(\lambda / R_s)$\footnote{This is called the \textit{Eikonal} approximation, from which we derive the equations of diffractive lensing. This should not be confused with the Eikonal / stationary-phase limit of the Fresnel-Kirchhoff diffraction integral.}. The full equation is
\begin{align}
    0 &= (-1 + 2U) \partial_0^2 \phi - (1 + 2U) \sum_{i=1,2,3} \partial_i^2\phi + \mathcal{O}(U^2)\\
    \intertext{which can compactly be written as}
    0 &= \nabla^2 \phi -(1 - 4U) \partial^2_t \phi \label{eqy:propagation_timedep}
    \intertext{where the $\nabla$ (without the indices) refers to the flat-space Laplacian. We now have a standard PDE for which various solution approaches exist. For a point-mass (or a point charge~\citep{davydov2013quantum}), one solution taken by~\citet{peters1974index,deguchi1986diffraction} is to substitute an ansatz of the form $\phi(\vec{r},t) = \tilde{\phi}(\vec{r}) \exp(i\omega t)$. This gives us a Schr\"odinger-like equation for
    a free particle in a Coulomb potential $V = 4 \omega U$ and whose ``mass'' is $\omega/2$:} 
-i\partial_t \phi  &= -\dfrac{1}{\omega} \nabla^2 \phi + 4 \omega U \phi. \label{eqy:propagation_schrodinger}\\
    \intertext{We take an alternate approach using the Feynman path integral to derive the Fresnel-Kirchhoff diffraction integral by summing over all possible paths through the lens plane~\citep{nakamura1999wave,yamamoto2003path}. Further simplifying Eq.~\ref{eqy:propagation_schrodinger} gives another commonly-used starting point:} 
4\omega^2 U \tilde\phi &= (\nabla^2 + \omega^2) \tilde\phi. \label{eqy:propagation_timeindep}\\
    \intertext{To solve this, we use the ansatz $\tilde \phi(\vec{r}) =  F(\vec{r})\exp(ikr)/r$ with Eq.~\ref{eqy:propagation_schrodinger}, giving}
4 \omega^2 U F &= -k^2 F + 2 i k \partial_r F + \partial^2_r F + \dfrac{1}{r^2} \nabla_{\theta}^2 F + \omega^2 F.~\label{eqy:propagation_spherical}
\intertext{In this equation, we have decomposed the spherical Laplacian into its radial and angular parts:}
\nabla^2 &= \dfrac{1}{r^2}\dfrac{\partial}{\partial r}\left( r^2 \dfrac{\partial}{\partial r}\right) + \nabla_{\theta}^2 \\
\intertext{Note that in the un-lensed case ($U = 0$), $F = 1$ solves Eq.~\ref{eqy:propagation_spherical} (assuming $\omega = k$). The ansatz also captures the free wave intensity decaying as $r^{-2}$. Therefore, to an overall normalization constant which we define later, we expect $F$ to capture the interference pattern generated by lensing. $F(\vec{r})$ is called the ``amplification factor'' of the wave \textit{amplitude} (not the flux). The characteristic transverse scale over $F$ varies is estimated by considering a spherical wave of wavelength $\lambda$ emanating from a point towards a plane at some distance $D \gg \lambda$. In the tangential (transverse) direction, $F$ varies over a characteristic scale $R_F = \sqrt{D \lambda}$ (the Fresnel scale). Therefore $\partial_r^2 F$ is smaller than $2 i k \partial_r F$ by a factor of $\mathcal{O}(\sqrt{\lambda / D})$ and can be neglected. Finally, using $\omega = k$ yields Eq.~\ref{eqy:amplification}, which can be thought of as the ``flattened'' version of Eq.~\ref{eqy:propagation_schrodinger}, where the propagation (radial) direction is collapsed. In the flat-sky approximation, we can work in transverse coordinates $\boldsymbol q = r \boldsymbol\theta$, with $\boldsymbol{\dot q} = r\boldsymbol{\dot\theta}$, and $\nabla^2_q = \dfrac{1}{r^2}\nabla^2_\theta$.}
i\partial_r F &= -\dfrac{1}{2 \omega}\nabla^2_{q} F + 2 \omega U F \label{eqy:amplification}
    \intertext{Since Eq.~\ref{eqy:amplification} also resembles a Schr\"odinger equation, we may solve it using the Feynman path integral. The classical Hamiltonian corresponding to Eq.~\ref{eqy:amplification} governs transverse deflections of rays evolving over the propagation path; the radius $r$ from the waves' origin plays the role of time in the classical analogy. Hence, the evolution of the transverse trajectories of rays is specified by $\boldsymbol{q}$ and $\boldsymbol{\dot q}$. The conjugate momentum is $\boldsymbol{p} = m r \dot{\boldsymbol{\theta}}$, where the analogous ``mass''\footnote{not to be confused with the lens mass!} $m = \omega$ and the classical potential is $V = 2 \omega U$. The Lagrangian corresponding to this Hamiltonian can be obtained by a Legendre transform:}
L({\boldsymbol q},{\boldsymbol{\dot q}},r) &= {\boldsymbol{\dot q}} \cdot (m \dot{\boldsymbol{q}}) - H(\boldsymbol{q},\boldsymbol{p}) = \omega \left[\dfrac{{\boldsymbol{\dot q}}}{2} - 2 U \right]
\intertext{where $H = \dfrac{\boldsymbol{p}^2}{2m} + V(r,\boldsymbol{q})$. Solutions to Eq.~\ref{eqy:amplification} can be written as an integral over all possible ray trajectories $\boldsymbol{q}(r)$:}
F(\omega,\vec{r}) &= \int \mathcal{D}[\boldsymbol{q}(r)] \exp\left[i \int L(\boldsymbol{q},\boldsymbol{\dot q},r)~dr\right] 
\end{align}
    The kinetic and potential terms in the Lagrangian correspond to the geometric time delay ($\tau_\mathrm{geo}$) and the Shapiro time delay ($\tau_\mathrm{grav}$) in optics\citep{blandford1986fermats}. Putting the factors of $c$ back in, the geometric term is 
\begin{align}
    \hat\tau_\mathrm{geo} &= \int \dfrac{r^2 \boldsymbol\theta(r)^2}{2}~dr \approx \dfrac{D_{L} D_S}{2D_{LS} c}|\boldsymbol{\theta}_l -  \boldsymbol{\beta}|^2. \label{eqy:thin_geo}
    \end{align}
where $\boldsymbol{\theta}_l$ and $\boldsymbol{\beta}$ are angles from the observer's perspective, and where we have approximated the paths as straight rays bent ``instantly'' once a ray hits the lens plane (see Fig. 1 in~\citep{feldbrugge2020gravitational}). This is the same as assuming that far away from the lens plane, the index of refraction of the medium is homogeneous enough such that rays travel in approximately straight lines.\\
The gravitational term is
\begin{align}
    \hat\tau_\mathrm{grav} &= -\dfrac{2}{c^3} \int U(\boldsymbol{q}, r)~dr.
    \intertext{This equation integrates the gravitational potential along the line of sight in the source frame. In the lens frame, we approximate this integral by collapsing the gravitational potential over the $z$ axis (the thin-lens approximation). Defining the physical impact parameter $\boldsymbol{b}$ ($= r_l\boldsymbol{\theta}$ from the source's perspective, and $\boldsymbol{b}=D_L \boldsymbol{\theta}_l$ from the observer's perspective) we define}
    \hat\psi(\boldsymbol{b}) &= \dfrac{2}{c^3}\int U(\boldsymbol{b} + z \boldsymbol{\hat z})~dz \approx - \hat\tau_\mathrm{grav}.\label{eqy:thin_grav}
\end{align}
The approximations in Eq.~\ref{eqy:thin_geo} and~\ref{eqy:thin_grav} reduce the infinite-dimensional Feynman path integral to the Fresnel-Kirchhoff diffraction integral:%
\begin{equation}
F(\omega,\boldsymbol\beta) \propto \int d^2\boldsymbol{\theta_l} \exp \left\{ i \omega \left[ \dfrac{D_{L} D_{S}}{2D_{LS} c} | \boldsymbol{\beta} - \boldsymbol{\theta}_l|^2 - \hat\psi(D_{L}\boldsymbol{\theta}_l) \right] \right\}~\label{eqy:fresnel}
\end{equation}
where $\omega$ is the observing frequency and $\boldsymbol{\beta}$ is the source's angular position with respect to the lens. In a cosmological context, the distances thus far should be replaced by angular diameter distances, and $\omega \to \omega(1+z_l)$.

At this point it is convenient to define a characteristic scale to measure deflection angles. While any scale may be chosen, each lens model comes with a different ``natural'' choice of the angular scale. For instance, the angular scale is often chosen such that the dimensionless angular radius of the Einstein ring is 1. For example, for a singular isothermal sphere characterized by its velocity dispersion $\sigma$, the deflection angle is a constant and defines the angular scale $\theta_E$ by which to normalize Eq.~\ref{eqy:fresnel}:
\begin{equation}
\theta_E = 4\pi\dfrac{\sigma^2}{c^2} \dfrac{D_{LS}}{D_{S}}
\end{equation}

While the singular isothermal sphere is common in geometric optics, its solution in wave optics is quite complicated. Instead, in wave optics, the point mass model is much simpler and more commonly analyzed. We write Eq.~\ref{eqy:fresnel} in terms of dimensionless variables using a notional point-mass lens model.
\begin{align}
\theta_E &= \sqrt{\dfrac{4GM}{c^2}\dfrac{D_{LS}}{D_{L} D_{S}}} \\
    \boldsymbol{\theta}_l &\to \mathbf{x} \theta_E\\
    \boldsymbol{\beta} &\to \mathbf{y} \theta_E\\
    \hat\psi(\boldsymbol{\theta}_l) &\to \psi(\boldsymbol{x})\theta_E^2\dfrac{D_L D_S}{D_{LS}}\\
    \omega &\to \Omega \dfrac{c D_{LS}}{D_L D_S\theta_E^2}\label{eqy:omega_definition}
\end{align}
Note that
\begin{equation} 
 \Omega = 4 G M \omega / c^3 = 4\pi R_s / \lambda\label{eqy:omega_ratio}
 \end{equation}
has a physical interpretation as the answer to the question: \textit{how large is a wavelength relative to the Schwarzschild radius, $R_s = 2 GM / c^2$?} The dimensionless frequency, $\Omega$, therefore plays a crucial role in determining whether geometric optics is a valid approximation. Eq.~\ref{eqy:fresnel} becomes
\begin{align}
F(\Omega,\boldsymbol y) &= \dfrac{\Omega}{2\pi i} \int d^2\boldsymbol{x} \exp ( i \Omega \tau(\boldsymbol{x},\boldsymbol{y}) )\label{eqy:fresnel_xy} \\
\intertext{where we have also introduced a normalization constant into $F(\Omega, \boldsymbol y)$ such that when $\psi \to 0$, $F\to 1$, and where the dimensionless delay is}
\tau(\boldsymbol{x},\boldsymbol{y}) &= \dfrac{|\boldsymbol{x} - \boldsymbol{y}|^2}{2} - \psi(\boldsymbol{x}).\label{eqy:td_potential} 
\end{align}
Eqs.~\ref{eqy:fresnel_xy} and~\ref{eqy:td_potential} are valid in the limit of weak gravity and small angular deflections. However, this non-dimensionalization scheme is not restricted to a point-mass lens. The constant $M$ can be anything, but it is understood to be some characteristic mass scale of the lens. Similarly $\theta_E$ need not be the Einstein radius, but is understood to be some characteristic scale of deflections.

\section{Different Regimes in Wave Optical Gravitational Lensing}
Eq.~\ref{eqy:fresnel_xy} looks simple, but describes extremely diverse phenomena, including all of standard (geometric optics-based) gravitational lensing phenomena. This is because the complex-valued amplification $F$ (not to be confused for the positive-valued flux magnification from geometric optics) carries both amplitude and phase information, the latter of which can result in constructive and destructive interference. For an amplification factor of $F$, the flux of the source gets magnified by $|F|^2$, in analogy with the Born rule from quantum mechanics. To extend the analogy, the overall phase of $F$ is arbitrary, as the choice of the origin of the source plane $\boldsymbol y$ was arbitrary. However, we will see later that relative phase between the different contributions to $F$ (the Morse phase) can alter the observables.

Just like in traditional geometric optics, we care about the image positions, defined as stationary points of $\tau(\boldsymbol{x},\boldsymbol{y})$ obtained by solving
\begin{equation} 
\dfrac{\partial \tau}{\partial{\boldsymbol{x}}} = 0.
\end{equation} 
However, in wave optics, we sum over all possible paths through the lens plane: the resulting amplification is proportional to the size of the coherent ``patches,'' or Fresnel zones, surrounding the stationary points in the lens plane. The Fresnel zone is defined here as the patch surrounding each stationary point in the lens plane for which the phase has not changed by 2$\pi$. 

\begin{figure*}
    \centering
    \includegraphics{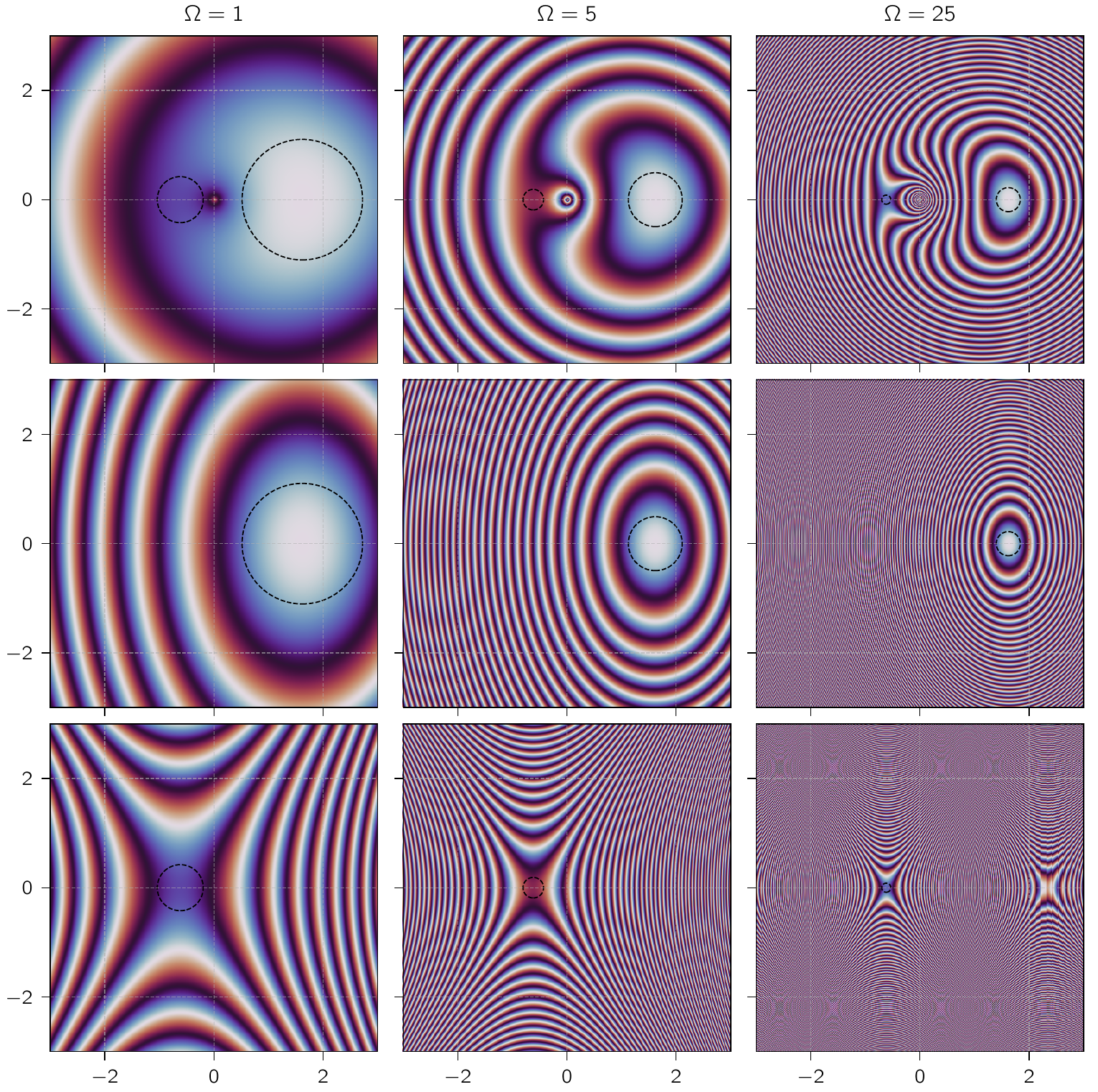}
    \caption{\textbf{Top row:} The phase $\Omega T(\boldsymbol x)$ corresponding to the lensing potential $T(\boldsymbol x,\boldsymbol y) = \dfrac{1}{2}|\boldsymbol x - \boldsymbol y|^2 - \ln(|x|)$, for three values of $\Omega = 1$ (heavily diffracted), $\Omega = 5$, $\Omega = 25$. We fix the source at $\boldsymbol y = (1,0)$ and evaluate the phase as a function of position in the lens plane, visualized with a $2\pi$-periodic colormap. To guide the eye, we have drawn two circles centered on the loci of the image positions $\boldsymbol{x}_{\pm}$ evaluated in the geometric optics limit. The radii of the pairs of circles are proportional to the flux magnification ratios $\mu_\pm$, and are scaled as $1/\Omega$. \textbf{Top left:} In the heavily-diffracted regime ($\Omega = 1$), the time delay between images is less than $2\pi$ radians. The images are superimposed in the time domain within one wavelength, and are rendered indistinguishable due to diffraction. \textbf{Top center:} As the frequency increases ($\Omega = 5$), the images are separated by more than one wavelength, and there are several oscillations between the two stationary points of the lens. The images are therefore no longer blended by diffraction; they become distinguishable in both the time domain and the angular domain. The interference leads to  constructive and destructive interference of the phase of the integrand when integrated over the whole lens plane, and shows a simple sinusoidal functional form when plotted against $\Omega$. \textbf{Top right:} In the high-frequency limit (here shown as $\Omega = 25$ for visualization purposes), the two images in the time domain are very well-resolved. The total magnification may be treated as a sum of discrete images -- the discrete stationary points of the lens potential -- using geometric optics. \textbf{Middle/bottom rows:} The saddle-point approximation to the ``plus''/``minus'' image individually, for the same three values of $\Omega$. It can be seen that the saddle-point approximation works better and better for higher values of $\Omega$, as the relevant parts of the integrand become increasingly confined around the stationary points. Note that a second saddle point appears on the right side of the bottom right panel; this is a image rendering artifact.
    }\label{figy:fresnel}
\end{figure*}

\begin{figure*}
    \centering
    \includegraphics{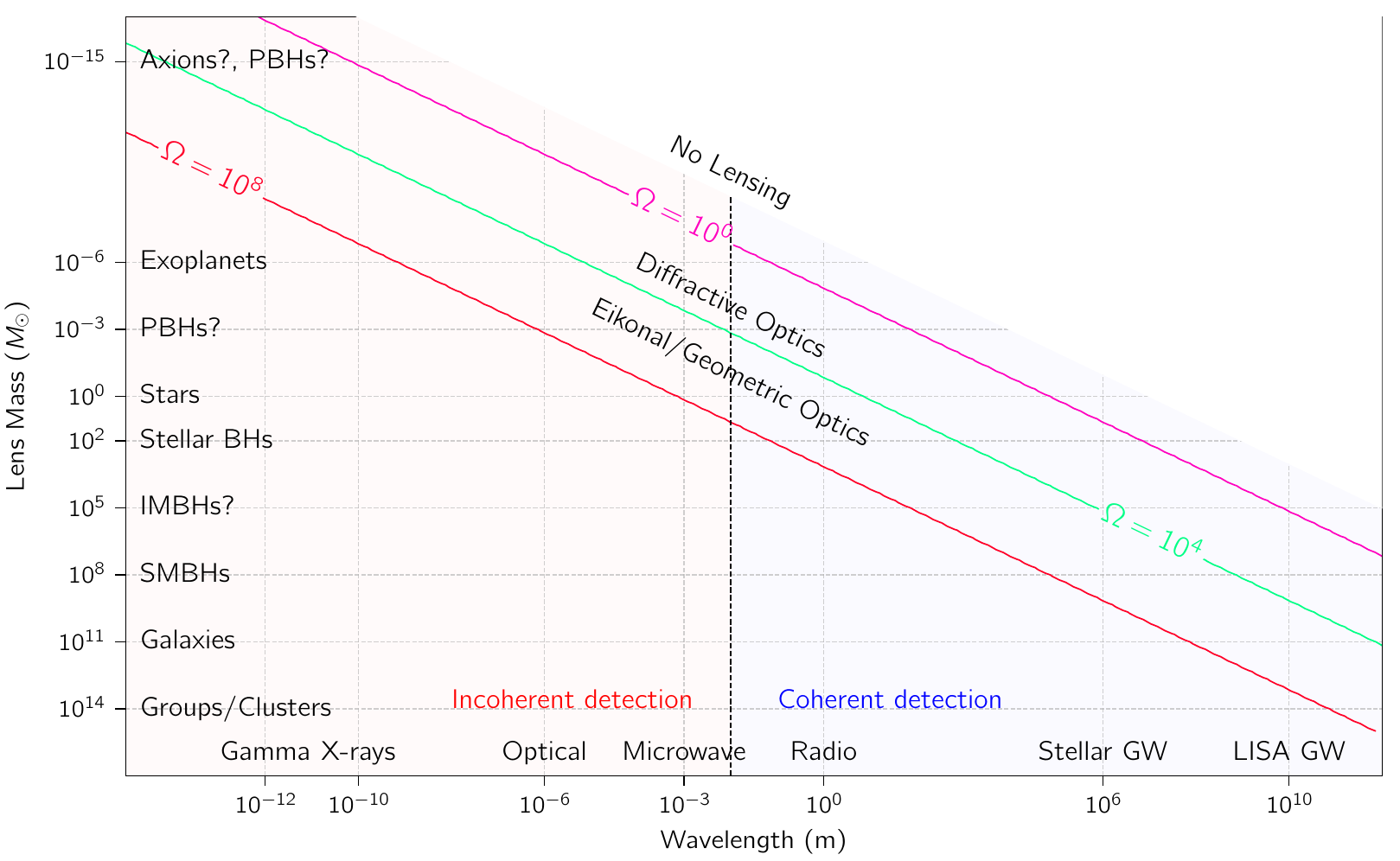}
    \caption{The transition from geometric to diffractive optics, as quantified by $\Omega = 4\pi R_s / \lambda$, as function of lens mass and observing wavelength. \textbf{Right half:} At longer wavelengths, advances in instrumentation (e.g.\ wideband voltage recording in radio telescopes and gravitational wave detectors) have enabled measurements of the amplitude (coherent detection) instead of the flux (incoherent detection). \textbf{Left half:} At high $\Omega$, the stationary-phase approximation holds, and geometric/Eikonal optics applies. In most scenarios, this is an excellent approximation. However, the stationary-phase approximation breaks down for long wavelengths or shallow local minima in the time-delay potential, ($\Omega \ll 1$). In this regime lensing is referred to as diffractive, and the unique frequency-dependence of diffractive lensing may be observable. In the $\Omega \ll 1$ regime, the lenses can be thought of as being ``too small'' to impart any phase on the passing wave.}\label{figy:regimes}
\end{figure*}

In the top row of Fig.~\ref{figy:fresnel} we illustrate the phase delay potential $\Omega \tau(\boldsymbol{x},\boldsymbol{y})$ modulo $2\pi$ over the lens plane for several values of $\Omega$. We hold a source fixed at $\boldsymbol{y}_0 = (1,0)$ (one Einstein radius away from the origin of the source plane) for various choices of $\Omega$, with a point-mass lens $(\psi(\boldsymbol x) = \log(|\boldsymbol{x}|)$. To compare the coherent patches with traditional geometric optics, we place dotted circles centered at the two stationary points of $\tau$, at $\boldsymbol{x} = (x_{\pm},0)$ for 
\begin{equation}
    x_\pm = \dfrac{y \pm \sqrt{y^2 + 4}}{2}.
\end{equation}

The radii of the circles are chosen according to their geometric-optics magnifications, and are scaled as $1/\Omega$ to ensure that at high $\Omega$, due to the prefactor in Eq.~\ref{eqy:fresnel_xy}, the flux magnification $|F|^2$ stays constant.

The phase over the lens plane varies significantly as a function of $\Omega$, suggesting that the amplification, and therefore the lens magnification, can become frequency-dependent, unlike in traditional gravitational lensing, where magnification ratios are completely achromatic. 

This frequency dependence arises because Eq.~\ref{eqy:omega_ratio} introduces a characteristic length scale (the wave period), and therefore chromaticity, into gravitational lensing. Lensing phenomenology can be classified depending on whether the integral in Eq.~\ref{eqy:fresnel_xy} can be treated using the saddle-point approximation near its stationary points.

The wisdom is that $\Omega$ is the most relevant quantity; however we shall see later (in Eq.~\ref{eqy:taylor_exp}) that it is really the product $\Omega A$ which sets the size of a coherent patch in the lens plane around the individual stationary points of $\tau(\boldsymbol{x},\boldsymbol{y})$~\citep{jow2022regimes,berry2021scalings}, where $A$ is the determinant of the magnification matrix:
\begin{equation}
A_{ab} = \dfrac{\partial^2 \tau}{\partial \boldsymbol{x}_a \partial\boldsymbol{x}_b} = 1 - \dfrac{\partial^2 \psi}{\partial \boldsymbol{x}_a \partial\boldsymbol{x}_b}.
\end{equation}
When the phase of the integrand varies rapidly (large values of $\Omega A$), the individual stationary points are separable into discrete images on the sky (and into discrete events in the time domain, for transient events). The first Fresnel zone of each image is observationally separable from the others, and they interfere with each other to form an oscillatory fringe pattern. This is called the \textit{eikonal} limit. When the observing frequency is so high (extremely large values of $\Omega A$) that the oscillations are washed out by summing over the finite extent of the source in the \textit{source} plane, ordinary geometric optics applies. This is the regime in which most gravitational lensing observations have been conducted. 

In this work, we make the distinction between the Eikonal limit (i.e. the stationary-phase approximation of Eq.~\ref{eqy:fresnel_xy}) and geometric optics (which we take to be the regime in which the Eikonal images are added together incoherently at the observer). Some authors do not make this distinction and simply refer to the Eikonal limit as geometric optics as well. \citet{grillo2018wave} distinguish the incoherent and coherent regimes as ``zeroth" and ``first-order" geometric optics. We also note that the Eikonal limit of optics that we refer to in this section is distinct from the Eikonal approximation used in Sec.~\ref{secy:first_principles} in order to go from Eq.~\ref{eqy:wave_eq} to Eq.\ref{eqy:propagation_timedep}.

For small values of $\Omega A$, the first Fresnel zones for each image blend together. In this regime the images are inseparably blurred together on the sky, and are separated by less than a period in the time domain. There are still multiple stationary points of the time-delay potential, so the lensing is still ``strong lensing'', but it is impossible to associate a particular photon to a particular image. For this reason, it is helpful to think of this regime as a perturbation of the unlensed ($F = 1$) case. We will see that the lensing is indeed suppressed in this limit.

\begin{figure}
    \centering
    \includegraphics[width = 0.47\textwidth]{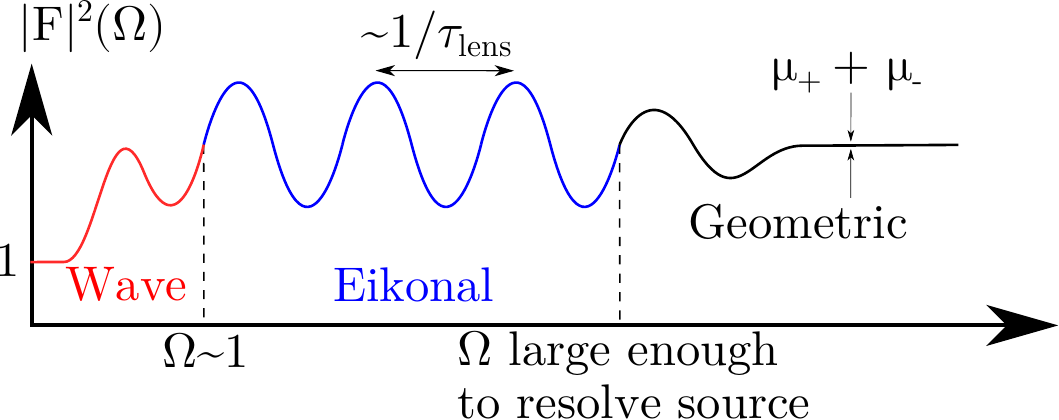}
    \caption{A schematic depicting different regimes in gravitational lensing as a function of $\Omega$. \textbf{Left red:} When $\Omega \sim 1$, we are in the diffractive or wave optics regime where the lensing is present but suppressed ($|F|^2 \to 1$) for $\Omega << 1$. \textbf{Center blue:} In the eikonal regime, the interference fringe between the two images, here represented as two amplitudes $F_+$ and $F_-$, is clearly visible. The total magnification has a perfect sinusoidal dependence (see Eq.~\ref{eqy:f2_geometric}) with a characteristic spectral oscillation scale of $1/\tau_{lens}$. \textbf{Right black:} Finally, at high frequencies, we enter the geometric optics (traditional lensing) regime when the lens's resolving power is sufficiently great to resolve the finite extent of the source. In this limit, the interference effects become washed out, and the lensing magnification becomes achromatic. In this limit the total flux magnification is the sum of the image flux magnifications: $|F_+|^2 + |F_-|^2$.}
    \label{figy:spectra_regimes}
\end{figure}

To organize these limits, we consider possible values of $\Omega$ and $\kappa$. With modern observational capabilities ranging from ground-based (and soon, space-based) gravitational wave detectors to gamma-ray telescopes, $\Omega$ can vary by 30 orders of magnitude for different lenses and wavelengths, and $A$ can vary from order unity to $\sim 10^3$ in rare extreme-magnification scenarios. Roughly speaking, we can focus on $\Omega$, keeping in mind that $A$ may play an important role under circumstances where the stars align. Fig.~\ref{figy:regimes} shows $\Omega$ as a function of the mass scale of the lens and the wavelength of the source, while Fig.~\ref{figy:spectra_regimes} schematically depicts the spectrum of a lensed source as a function of $\Omega$.

Within Fig.~\ref{figy:regimes}, we have also delineated a second boundary, which denotes whether the radiation is detected incoherently (without phase information, e.g.\ with a flux detector) or coherently (with phase information, e.g.\ with an antenna). At long wavelengths (radio and gravitational waves), it is possible with current technology to detect the phase of the radiation field using electromagnetic/gravitational antennas. These situations where we have access to the phase of the electromagnetic/gravitational field are extremely powerful and enable qualitively different observations to be carried out. For example, recording the field amplitude instead of the
intensity (voltages instead of light-curves) allows antennas to reach Nyquist-limited time resolution for resolving shorter lensing delays, regardless of the flux variability timescale of the source. This is similar to radio interferometry, where coherent voltage recording allows precise time-delay measurements between different receivers even when the sources do not exhibit time variability in their light curves. Improving time resolution through coherent detectors and recording instrumentation therefore opens up new phase space towards smaller time delays (corresponding to smaller mass scales) in gravitational lensing.

\subsection{Beyond Scalar Wave Optics}\label{secy:beyond}
These aforementioned regimes in scalar wave optics cover much of the relevant phenomenology of interference effects in gravitational lensing. From an observational standpoint, the eikonal and wave optics regimes are still uncharted territory. Virtually all observations of gravitational lensing thus far fall squarely in the geometric optics regime.

However, in an era of gravitational-wave observations from the ground and from space, there are compelling reasons to push further theoretically. Notably, we have worked with scalar waves in the weak gravity limit, ignoring cases where $U \sim 1$, for which Eq.~\ref{eqy:fresnel} is not valid. Numerically simulating wave propagation in strongly-curved spacetimes becomes necessary and is an active area of research in general relativity~\citep{zenginoglu2012caustic,yang2014scalar}. These numerical solutions of scalar wave propagation reveal rich phenomenology, particularly near caustics. To give an idea of the phenomenology,~\citet{zenginoglu2012caustic} computes the Green's function of Eq.~\ref{eqy:laplacian} near the Schwarzchild radius of the lens, and observe multipath propagation for integer numbers of cycles around the photon ring. The observed ``caustic echo'' has a period in good agreement with the light-crossing time of the photon ring~\citep{stewart1989solutions}.

Other aspects of lensing are not captured by the scalar theory, such as polarization-dependent effects. Two examples from general relativity are the gravitational Faraday rotation induced by rotating gravitational potentials~\citep{ishihara1988gravitational,Nouri-Zonoz1999,Asada2000,Li2022gfr} and the gravitational spin-Hall effect, in which different polarizations of light traverse different trajectories through a gravitational potential ~\citep{gosselin2007spin,oancea2020gravitational}. 

Note also that while the scalar wave formalism for both gravitational and plasma lensing are effectively identical (the only difference being that gravitational potentials are frequency independent, but plasma lensing potentials have a $\sim f^{-2}$ dependence), magnetic fields in ionized plasma cause different polarizations to experience different refractive indices in a phenomenon known as birefringence \citep{Li2019bwbiref}. Thus, while the scalar wave theory of gravitational lensing and plasma lensing share significant overlap, notable differences arise in the vector propagation.

A final limitation is that we have so far limited our discussion to the effects of scalar perturbations to the metric on the propagation of waves through space-time. In general, metric perturbations may themselves have vector and tensor components (e.g. gravitational waves). Gravitational waves will induce additional time delays in the propagation of light from background sources which may be observable (indeed, this is the basis for pulsar timing array experiments). Gravitational waves themselves may therefore lens coherent sources of radiation \citep{boyle2012pulsar,rahvar2018gravitational} inducing characteristic interference patterns; this is an under-explored and active area of research. Polarization effects, such as the gravitational Faraday rotation, may also be more pronounced in lensing by tensor and vector perturbations, and may have observable impacts on CMB lensing~\citep{Dai2014cmb}. However, in light of the complexity and magnitude of these effects, for the remainder of this review paper, we will focus on scalar wave optics in the aforementioned regimes: eikonal scalar optics, diffractive scalar optics, and the transition from coherent to incoherent (traditional) gravitational lensing.

\section{Eikonal Optics}\label{secy:coherent_geo}
Lensing in eikonal optics is similar to treating the source as a plane wave hitting an opaque screen with a set of well-separaged pinholes of negligible size. Each pinhole's size and phase relative to the other pinholes is set by the traditionally-calculated flux magifications and time delays. To see this mathematically, we simplify Eq.~\ref{eqy:fresnel_xy} using the saddle-point approximation. Conceptually, this collapses the integral over the lens plane to a coherent superposition of a small number of paths defined by the stationary points of the lens potential. Taking the limit as $\Omega \to \infty$ results in rapid oscillations of the phase of the integrand as a function of $\boldsymbol{x}$. In this limit, only the stationary points of $\Omega \tau(x,y)$ as a function of $x$, which we index by $j$, contribute constructively to the integrand. These positions $\boldsymbol{x}_j$ correspond to the image positions as calculated with geometric optics. Near the $j^{th}$ stationary point, the time delay potential can be approximated by a Taylor expansion:
\begin{align}
    \tau_j(\boldsymbol{x},\boldsymbol{y}) \approx \tau(\boldsymbol{x}_j) &+ \dfrac{1}{2} A^j_{ab} (\boldsymbol{x} - \boldsymbol{x}_j)_a (\boldsymbol{x} - \boldsymbol{x}_j)_b + \ldots \label{eqy:taylor_exp}
    \intertext{where $A^{j}_{ab} = \partial_a \partial_b \tau(\boldsymbol{x}_j)$ is the inverse magnification matrix evaluated at the $j$th image. In Fig.~\ref{figy:fresnel}, we plot in the top row the phase $\Omega \tau(\boldsymbol{x},\boldsymbol{y})$ modulo $2\pi$, and in the bottom two rows the saddle-point approximation of the phase potential about each of the two stationary points. We approximate the Fresnel-Kirchhoff integral as the coherent superposition of both of
    these expansions. This is called the stationary-phase (or saddle-point) approximation, which relies on the fact that contributions far away from the stationary points cancel each other out.}
    \int d^2\boldsymbol{x} \exp(i\Omega \tau(\boldsymbol x)) &\approx \sum_j \int d^2\boldsymbol{x} \exp(i\Omega \tau_j(\boldsymbol x)) \\
    = \sum_j \exp(i\Omega \tau(\boldsymbol{x}_j)) &\int d^2\boldsymbol{x} \exp(i\Omega A^j_{ab} (\boldsymbol{x} - \boldsymbol{x}_j)_a  (\boldsymbol{x} - \boldsymbol{x}_j)_b / 2)
\intertext{
We evaluate the above expression analytically, being a sum of complex Gaussian integrals. Since $A^j_{ab}$ is a symmetric matrix, it is diagonalizable via a rotation of $\boldsymbol x$; we call its eigenvalues $\lambda^j_{1}$ and $\lambda^j_{2}$. The Gaussian integral associated with each image contributes a factor of $\dfrac{2\pi i}{\Omega \sqrt{\lambda_1^j}\sqrt{\lambda^j_2}}$. The eigenvalues $\lambda^j_1,\lambda^j_2$ may be both positive if the image is a minimum of the time-delay potential, both negative if the image is a maximum of the time-delay potential, or they may have different signs if the image is at a saddle point of the potential. If we define $F_j$ by
}
    F_j^{-1} &= \sqrt{\lambda_1^j}\sqrt{\lambda^j_2}
    \intertext{the amplification becomes}
    F(\Omega) &= \sum_j F_j \exp(i\Omega \tau(\boldsymbol x_j) - i\pi n_j).\label{eqy:f_geometric}
\end{align}
where $n_j =0,1/2,1$ keeps track of the extra factor of $-i$ arising from the sign of the eigenvalues. This phase shift arises from the image parity and is called the Morse phase; in the eikonal limit it is discretely-valued\footnote{However, this is not necessarily true in the diffractive regime when the saddle-point approximation breaks down -- see Appendix A of~\citep{jow2022regimes}}. It has been known for decades in~\citep{blandford1986fermats,nakamura1999wave} but only recently have the observational ramifications become observationally relevant, in particular in the parameter space of gravitational-wave lensing~\citep{dai2017on,dai2020search}.
The overall Morse phase is not predicted by waveform models. In the specific setting of gravitational waves generated by the nearly-circular inspirals of compact binaries, an overall Morse phase shift of $180^\circ$ is degenerate with a $180^\circ$ orbital phase shift of the binary as long as we restrict our analysis to the dominant quadrupole mode of the radiation. Orbital eccentricity or the inclusion of higher-order modes in model waveforms can also break this degeneracy~\citep{janquart2021on,ezquiaga2021phase}.

The relative Morse phase between two images of the same waveform can be used to reduce backgrounds in searches for lensed GWs~\citep{dai2020search}. In the time domain, a relative Morse phase of $180^\circ$ induces a sign flip of the electromagnetic/gravitational waveform (Type II images; $n_j = 1$) relative to a Type I image. In the case of saddle points (Type III images; $n_j = 1/2$), the Morse phase applies a Hilbert transform to the unlensed waveform. This latter distortion is non-trivial and leads to biases in parameter estimation unless this is properly taken into account in search templates~\citep{vijaykumar2022detection}. 

Interference effects are imprinted in not only the phase, but also the amplitude information from a lensing event. Interference induces ripple-like modulations of the intrinsic spectrum of a source. In this scenario, knowing the intrinsic spectrum allows us to measure $|F|^2$ as a function of $\Omega$. Expanding Eq.~\ref{eqy:f_geometric}, we compute

\begin{equation}
    |F|^2 = \sum_j |F_j|^2 + 2 \sum_{j<k} F_j F_k \cos(\Omega (\tau(\boldsymbol{x}_j) - \tau(\boldsymbol{x}_k)) - \pi (n_j - n_k)) \label{eqy:f2_geometric}
\end{equation}
where we recognize $|F_j|^2 = 1/\det{A^j_{ab}}$ as the flux magnification of each image present in geometric optics. Eikonal optics induces interference terms which induce a ripple-like modulation, similar to pulsar scintillation, in the spectrum of the observed source over a bandwidth $\Delta \Omega = 2\pi / (\tau(\boldsymbol{x}_j) - \tau(\boldsymbol{x}_k))$, as shown in Fig.~\ref{figy:spectra_regimes}. These modulations are potentially quite detectable due to the favorable scaling of Eikonal optics
over geometric optics~\citep{jow2020wave,takahashi2003wave}. To see this, consider a detector which detects the brightest image, which has magnification $|F_j|^2$, and whose noise floor is expressed as some fraction $\varepsilon^2$ of this: $\varepsilon^2 |F_j|^2$. In geometric optics, the detectability of a dimmer image $k$ if it is distinguishable from the brighter one (e.g. has a large angular separation on the sky) depends whether the ratio $|F_k|^2 / |F_j|^2 \sim \varepsilon^2$. In Eikonal optics, however, an image $k$ can be detected and spectrally separated from image $j$ using just the combined spectrum of both images. The detection is made by searching for the interference term in the combined spectrum, which induces sinusoidal modulations in the spectrum whose size is $|F_j F_k| / |F_j|^2 \sim \varepsilon$. 

The Eikonal limit, or stationary-phase approximation, is a particularly useful approximation because the behaviour of the stationary-phase points, a.k.a. the images, is described by the rich mathematics of catastrophe theory. Catastrophe theory characterizes the topology of caustics, which are the boundaries in parameter space at which the number of stationary-phase points changes. The essence of what makes catastrophe theory so powerful is the result that the topology of caustics can be classified by a small number of ``elementary catastrophes" \citep{thom1967}. The result is that lenses will behave qualitatively similarly based on which elementary catastrophes characterize the lenses, even though they may differ significantly in functional form \citep{Nye1999}. While, technically, the Eikonal limit breaks down at caustics (at caustics, two or more of the inverse magnification matrices are degenerate, leading to formally infinite magnifications), the area of the region around the caustic for which the Eikonal approximation fails goes to zero as $\Omega \to \infty$. Moreover, the full wave optics diffraction patterns of some of the elementary catastrophes are known \citep{berryupstill1980}. For example, as we will see in Sec.~\ref{secy:fold}, the diffraction pattern near a fold catastrophe is given exactly by the Airy function (see also \citep{nakamura1999wave}). Therefore, the known diffraction patterns near caustics coupled with the Eikonal limit away from the caustics can provide a full description of a generic lens. This approach is taken in \citet{grillo2018wave}.

The effects mentioned so far -- Morse phase shifts, spectral modulation of the amplification factor, and favorable scaling of lensing cross-sections -- arise as generic interference phenomena present in Eikonal optics. Going into the fully diffractive regime adds additional features which we will discuss next.

\section{Diffractive optics ($\Omega \gtrsim 1$)}\label{secy:diffractive}
In the case where $\Omega$ is not much greater than $1$, Eq.~\ref{eqy:fresnel_xy} cannot be simplified using the saddle-point approximation. For an axisymmetric lens, the potential $\psi(\boldsymbol{x}) \to \psi(x)$, and the amplification can be written as
\begin{align}
    F(\Omega,y) = &-i\Omega\exp(i\Omega y^2 / 2)~\times \\ 
    &\int_0^{\infty}~dx x J_0(\Omega x y) \exp\left[i\Omega ( x^2/2 - \psi(x) ) \right] \label{eqy:f_diff}
\end{align}
where $J_0$ is the zeroth-order Bessel function. In certain cases this may be computed analytically. For a point mass where $\psi(x) = \log(x)$ this simplifies to
\begin{align} 
    F(\Omega,y) &= \exp\left[ \dfrac{\pi \Omega}{4} + i\dfrac{\Omega}{2}\left(\ln\left(\dfrac{\Omega}{2}\right) \right) \right]\times \\
    &\Gamma\left(1 - \dfrac{i\Omega}{2}\right) ~_1F_1\left(\dfrac{i\Omega}{2},1;\dfrac{i\Omega y^2}{2} \right). \label{eqy:f_point_mass}
\end{align}
In this equation (see also~Eq.~21 of~\citep{deguchi1986wave}, Eq. 2.16~\citep{katz2018femtolensing} for slightly different forms), $\Gamma$ is the gamma function and $_1F_1$ is the hypergeometric function\footnote{\href{https://functions.wolfram.com/HypergeometricFunctions/Hypergeometric1F1/}{Wolfram MathWorld: Hypergeometric $_1F_1$}}. 

% \PS{I think we can replace $\frac12(1-\psi''(x_E))$ in the following
%   with $1-\kappa$.  Changing briefly to $x,y,r$ coordinates, we have
%   $$ \psi_x = \frac xr\psi_r \qquad \psi_y = \frac yr\psi_r $$
%   Then
%   $$
%   \begin{aligned}
%    \psi_{xx} &= \frac1r\psi_r - \frac{x^2}{r^3}\psi_r
%               + \frac{x^2}{r^2}\psi_{rr} \\
%    \psi_{yy} &= \frac1r\psi_r - \frac{y^2}{r^3}\psi_r
%               + \frac{y^2}{r^2}\psi_{rr}
%   \end{aligned}
%   $$
%   This gives
%   $$ \psi_{xx} + \psi_{yy} = \frac1r\psi_r + \psi_{rr} $$
%   Now at the Einstein radius, $\frac1r\psi_r=1$, hence
%   $$ 1 - \psi''(x_E) = 2(1-\kappa) $$
% }

The point mass is the simplest analytical solution for the amplification factor in wave optics in gravitational lensing. However, other analytical solutions exist which we briefly mention: the solution for a Kerr (rotating) black hole is very similar to that of the point mass; it includes a shift in the point-mass lensing potential $\psi(\boldsymbol{x}) = \log(x) - \dfrac{\boldsymbol x \cdot \boldsymbol \alpha}{x^2}$, where the shift $\vec \alpha = \hat n \times \vec{J} / (M R_E)$ depends on the physical Einstein radius $R_E$ and the lens's angular momentum vector per unit mass $\vec{J}/M$ relative to the line-of-sight unit vector $\hat{n}$. The interference pattern is the same as that of Eq.~\ref{eqy:f_point_mass}, being circularly-symmetric, albeit shifted by an angle $\alpha$~\citep{baraldo1999gravitationally}. This is similar to the Sagnac effect for a rotating optical interferometer. Analytic solutions for the singular isothermal sphere~\citep{matsunaga2006finite}, an infinite cosmic string~\citep{suyama2006exact}, and a one-dimensional sinusoidal phase screen~\citep{beach1997diffraction} are also known.

Solutions to the diffraction integral beyond the analytical ones presented here are often computed numerically. However, this can be difficult because the phase in integrals such as Eq.~\ref{eqy:fresnel_xy} oscillates faster for larger values of the argument $\boldsymbol x$. Some of the first numerical efforts based on contour integration were developed for the calculation of femtolensing spectra~\citep{ulmer1995femtolensing}; this method has also been applied
successfully to gravitational
waves~\citep{mishra2021gravitational}. Approximate methods using Eikonal optics~\citep{grillo2018wave} are effective for large lensing phase shifts ($\Omega  T \gg 1$) and near caustics (see also discussion of Sec. 1 of~\citep{jow2021imaginary}).~\citet{grillo2018wave} also point out that the Fresnel integral can be written as a convolution and evaluated via FFT. This performs well near caustics, but requires the lens plane to be densely sampled with a finite and regular two-dimensional grid fine enough to capture the details of the oscillatory integrand. 

More recently, Picard-Lefschetz theory has been used to assist in the analysis and evaluation of Fresnel integrals~\citep{feldbrugge2019oscillatory,feldbrugge2020gravitational,jow2021imaginary,shi2021plasma,jow2022regimes}. These methods essentially involve deforming the two lens plane integrals into a set of carefully-chosen contours in the complex plane along which contour integrals are non-oscillatory and
absolutely convergent. One feature of Picard-Lefschetz integration is that while it converges slowly for $\Omega \sim 1$, performance improves as the integrands become \textit{more} oscillatory ($\Omega \gg 1$), so it is complementary to many other techniques. A limitation is that the lens phase function $\phi(\boldsymbol{x})$ (e.g. $\phi(\boldsymbol{x}) = \Omega \tau(\boldsymbol{x},\boldsymbol{y})$) must be able to be evaluated when the domain is complexified from $\mathbb{R}^2\to \mathbb{C}^2$. This is straightforward for the broad class of rational functions (quotients of polynomials), which have no branch cuts (also a desirable property) and a small number of isolated poles. The point mass lens has $\psi(x) = \log(|x|)$ which \textit{does} have a branch cut; in a follow-up work to~\citep{feldbrugge2019oscillatory}, a technique to evaluate the isolated and the binary point lens potentials by reparametrizing the radial integral was reported~\citep{feldbrugge2020gravitational}. However, well-known lens models like the singular isothermal
sphere ($\psi(x) = |x|$) have not yet been evaluated using this method. Despite this limitation, the utility of Picard-Lefschetz theory is not limited to its numerical application, but also includes its conceptual power. By extending the analysis of optics into the complex plane, Picard-Lefschetz theory gives a well-defined prescription for separating contributions to the total observed flux from individual geometric images, even at low frequencies. Picard-Lefschetz theory can be seen as a way to extend the ``discrete images'' description from geometric optics lensing deep into the wave regime. In this way, Picard-Lefschetz theory can conceptually bridge the gap between these two regimes\citep{jow2022regimes}.

\begin{figure*}
    \centering
    \includegraphics[width = 0.24\textwidth]{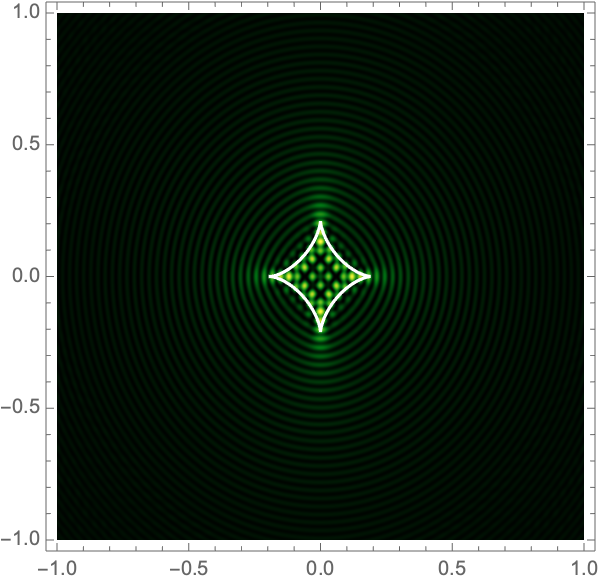}
    \includegraphics[width = 0.24\textwidth]{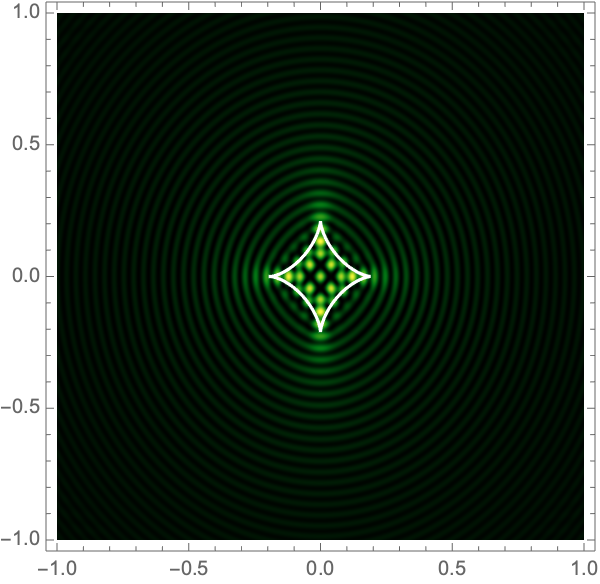}
    \includegraphics[width = 0.24\textwidth]{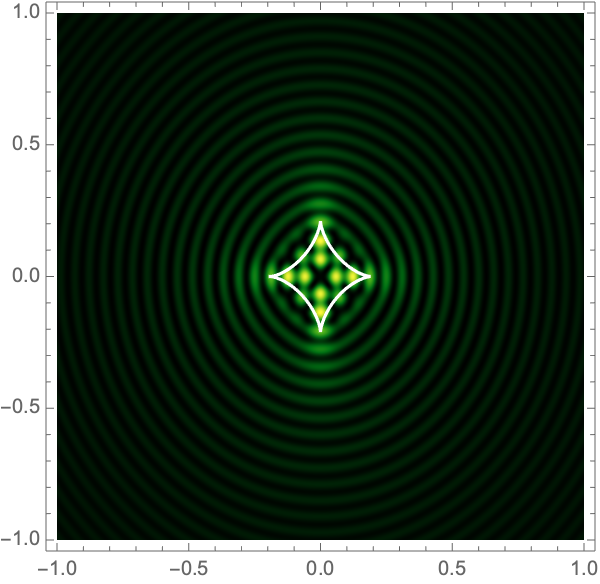}
    \includegraphics[width = 0.24\textwidth]{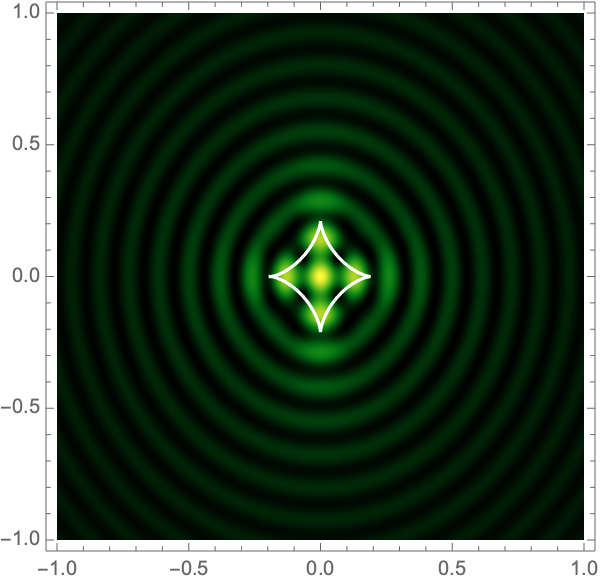}
    
    \includegraphics[width = 0.24\textwidth]{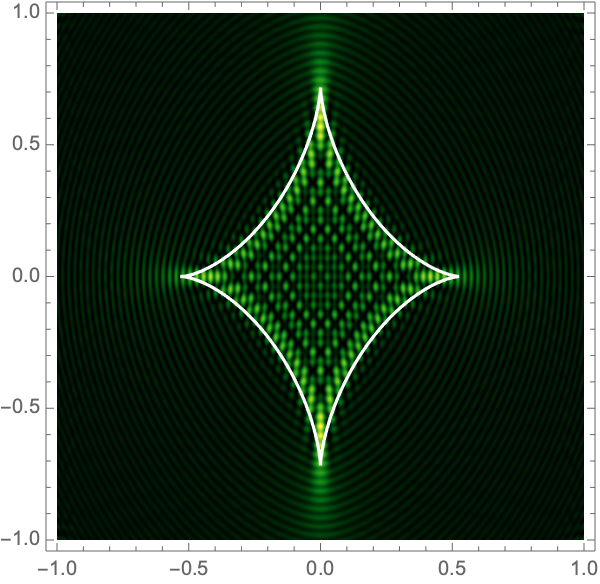}
    \includegraphics[width = 0.24\textwidth]{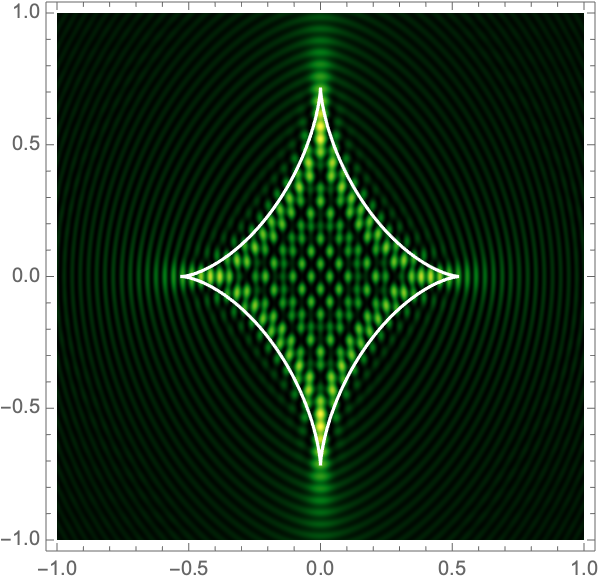}
    \includegraphics[width = 0.24\textwidth]{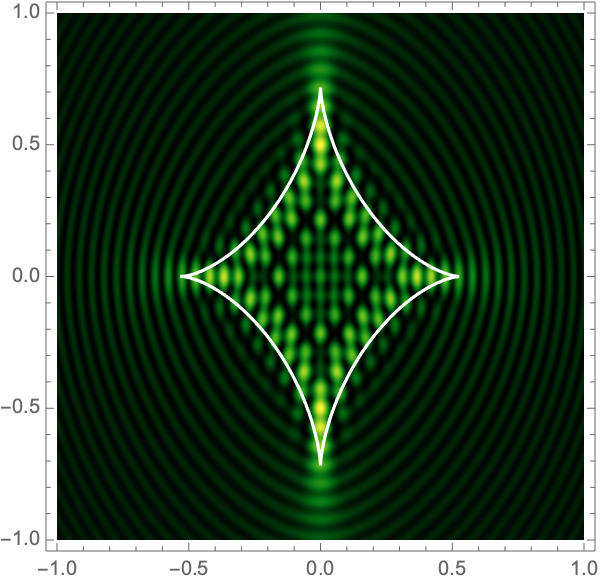}
    \includegraphics[width = 0.24\textwidth]{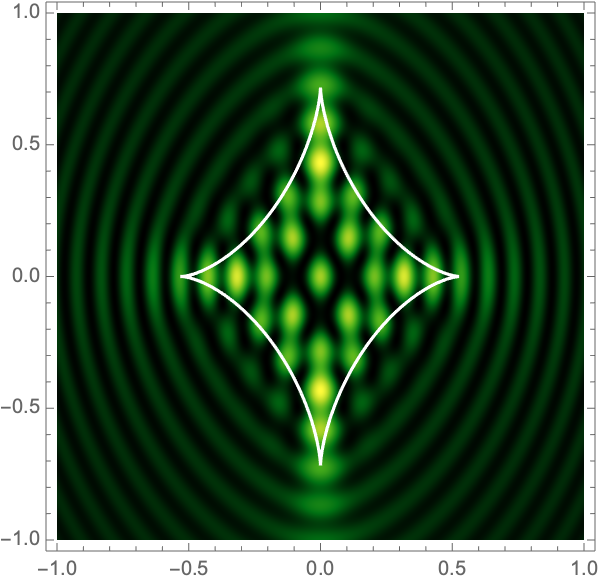}
    
    \includegraphics[width = 0.24\textwidth]{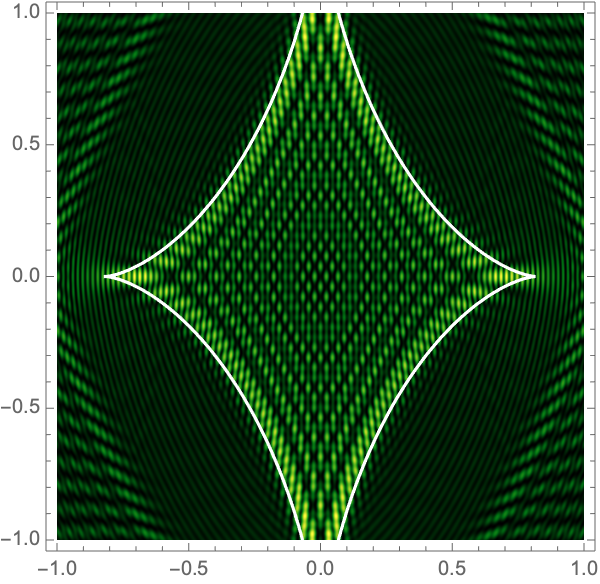}
    \includegraphics[width = 0.24\textwidth]{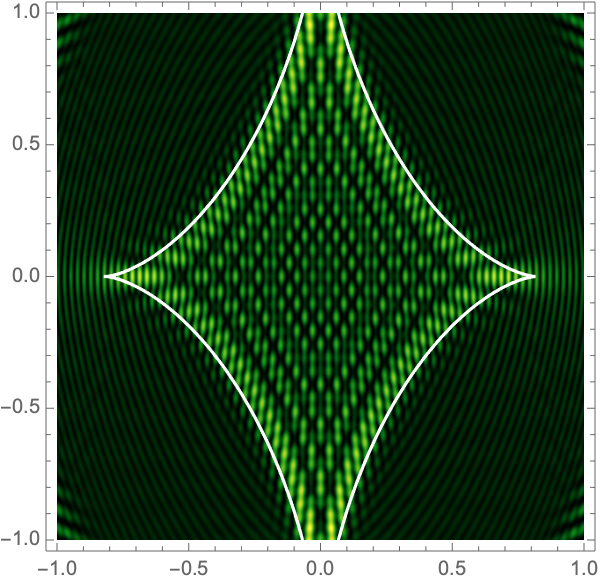}
    \includegraphics[width = 0.24\textwidth]{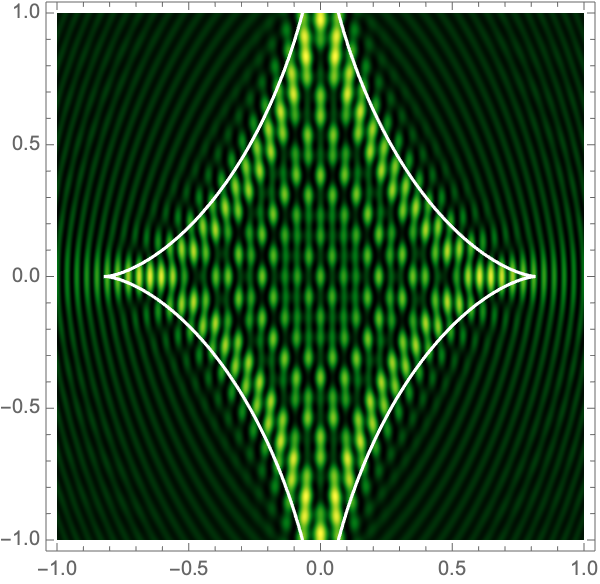}
    \includegraphics[width = 0.24\textwidth]{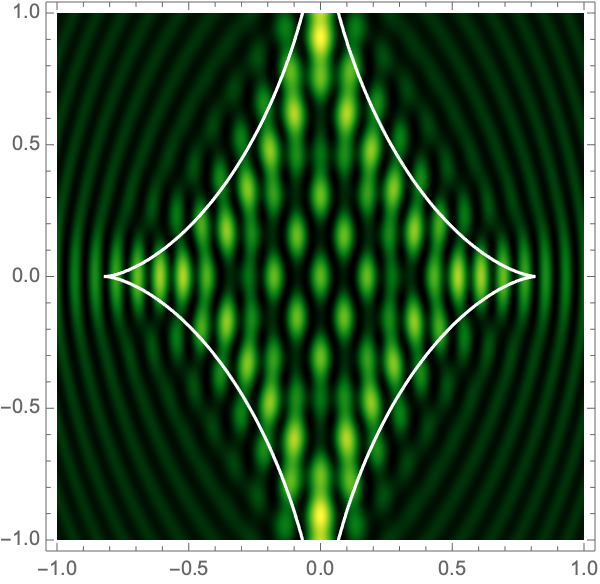}
    \caption{A map of $|F|^2(\Omega, \boldsymbol{\mu})$ for a point mass+external shear model (Eq.~\ref{eqy:pmxs}), where the map coordinates $\boldsymbol\mu$ are the scaled source plane coordinates $\boldsymbol{y}$. \textbf{Columns:} From left to right, each column corresponds to $\Omega = 100,75,50,25$. \textbf{Rows:} From top to bottom, each row corresponds to $\gamma = 0.1,0.3,0.5$. The caustic, at which the geometric optics magnification diverges, is overlaid as a white diamond on each map. Reproduced with permission from source code used for~\citep{feldbrugge2020gravitational}.}\label{figy:feldbrugge_maps}
\end{figure*}

In Fig.~\ref{figy:feldbrugge_maps} we show several magnification maps, evaluated using the Picard-Lefschetz method for
several values of $\Omega$. The map coordinates $\boldsymbol{\mu} = (\mu_x,\mu_y)$ are related to the source plane and observer plane coordinates ($\boldsymbol{y}$ and $\boldsymbol{y}_\textrm{obs}$ respectively) by $\boldsymbol\mu = \boldsymbol{y} \dfrac{D_{L}}{D_{S}} + \boldsymbol{y}_\textrm{obs} \dfrac{D_{LS}}{D_{S}}$. This parametrization reflects the symmetry of the source and the observer in optics: in gravitational lensing we may think of fixing $\boldsymbol{y}_\mathrm{obs} = 0$, and treat
$\boldsymbol{\mu}$ as a source plane coordinate, while in a laboratory diffraction experiment we may instead fix $\boldsymbol{y} = 0$ and observe the diffraction pattern on the observer's screen, which spans many values of $\boldsymbol{y}_\mathrm{obs}$. The Fresnel integral evaluated is that of the Chang-Refsdal lens~\citep{chang1979flux}, for which
\begin{equation}
    \tau(\boldsymbol x,\boldsymbol y) = \dfrac{1}{2}|\boldsymbol x - \boldsymbol \mu|^2 - \log(|\boldsymbol{x}|) - \dfrac{\gamma}{2}(x_1^2 - x_2^2)\label{eqy:pmxs}
\end{equation}

whose single free parameter $\gamma > 0$ describes the external shear.

\subsection{A useful analytical approximation}
The point mass model is extremely useful, but in practice, hypergeometric functions with complex arguments can be cumbersome to numerically evaluate~\citep{pearson2014numerical}. There is also a useful ``semiclassical'' approximation to Eq.~\ref{eqy:f_point_mass} for situations when $\Omega \gg 1$ but when the sinusoidal form of Eq.~\ref{eqy:f_geometric} is too simple. Solving for the stationary points of
\begin{align}
    \tau(x) = &\dfrac{x^2}{2} - \psi(x) \\
    \intertext{gives the geometric-optics image positions $x_E$ such that $x_E = \psi'(x_E)$. To second order in $\Delta x = x - x_E$,}
    \tau(x) &= \tau(x_E) + \dfrac{1}{2}(1 - \psi''(x_E)) \Delta x ^2 = \tau(x_E) + (1 - \kappa) \Delta x^2.\\
    \intertext{The region of the lens plane where the rays interfere constructively is defined by}
    2\pi \gtrsim & (1 -\psi''(x_E)) \Delta x^2 \Omega,
    \intertext{and this domain has a width of $\Delta x \sim (\Omega |1 - \psi''(x_E)|)^{-1/2}$. Within this domain, the Bessel function argument of Eq.~\ref{eqy:f_diff} changes by $\Omega \Delta x y$, and outside this domain, it oscillates quickly. If $y \lesssim \Omega^{-1/2}$, then $1 \gtrsim \Omega \Delta x y$, and the Bessel function does not change over the important range of the integration. Pulling $x J_0(\Omega x y)$ out of the integral sign leaves a Gaussian integral from $0$ to $\infty$:}
    |F_\textrm{wave}|^2 &\approx  \pi \Omega x_E^2 |1 - \psi''(x_E)|^{-1} J_0^2(\Omega x y)\label{eqy:f_wave_pm} 
    \end{align} 

$$\leq \pi \Omega x_E^2 |1 - \psi''(x_E)|^{-1} = 2\pi \Omega x_E^2 |1 - \kappa(x_E)|^{-1}. \footnote{Note that this approximation to the maximum magnification attainable in wave optics leaves out a factor of $1-\exp(-\pi \Omega)$ in the denominator which accounts for the suppression of waves as $\Omega \to 0$.}$$

where we have used $2(1-\kappa(x_E)) = 1 - \psi''(x_E)$.

\section{Angular broadening/finite-size effects}\label{secy:angular_broadening}
So far, we have assumed a point-like source of monochromatic radiation in the source plane. In reality, we observe radiation from a finite patch of $\boldsymbol y$-values in the source plane. In general, integrating over a finite patch of the source plane washes out the chromatic interference patterns seen in both the eikonal and wave optics regimes, e.g. Eqs.~\ref{eqy:f2_geometric} and~\ref{eqy:f_point_mass}. This is the main obstacle to observation of interference effects in gravitational lensing.

Before we dive into a deeper discussion of decoherence, it is worth defining what type of coherence we care about. Under one definition, ``phase-coherence'' is strictly a property of the emitted radiation, which can be considered as a time series. If one can predict the field amplitude as a function of time (or frequency, if one prefers the Fourier domain) with a model, that source would be considered phase-coherent.

Gravitational waves are an excellent example of a phase-coherent source by this definition. Given the time of the merger, it is possible to write down a model for the strain amplitude as a function of time. The reason this is possible is that the emitter is a simple binary system, and that the emission region is much smaller than a GW wavelength. It is therefore possible to coordinate the entire emission region at once.

Most electromagnetic waves emitted by astrophysical sources are not phase-coherent. Quantitatively, this is because the transverse size of virtually any astrophysical emission region is usually much larger than a wavelength of electromagnetic radiation. This means that different patches of the emission region cannot be causally connected, i.e. synchronized to better than one wavelength. While individual emitters may be phase-coherent (e.g. a single electron orbiting a field line emitting synchrotron radiation is phase coherent), the fact that there exists an ensemble of emitters, all radiating with some random phase with respect to each other, means that the phase of the total time series is not predictable.

Source emission regions do not need their transverse sizes to smaller than one wavelength (i.e. they do not need to be phase coherent) in order for wave-mechanical features to be observed. The actual requirement on the source size is much looser; it is a constraint on the \textit{angular} coherence of the source and involves the observer's geometry in addition to the emission physics. In the terminology of interferometry and interstellar optics, this is similar to a source appearing ``unresolved'' when viewed with a particular receiver setup or through an inhomogeneous medium with a certain geometry.

Whether a source is angularly-coherent, or unresolved, depends the apparent angular extent of the source. In some cases this corresponds to the true size of the emission region. However, in the case of radio waves, interstellar scattering can enlarge the apparent size of the source via angular broadening. In addition, in areas of the source plane with high magnifications (caustics), small displacements in the source plane correspond to large changes in the image positions on the lens plane (and therefore the lensing time delays). These effects all work to increase the apparent angular size of the source. Astrophysical masers, however, may have physical sizes which are \textit{larger} than their apparent sizes~\citep{goldreich1972astrophysical}, which have been observed to be inversely correlated with their isotropic luminosities~\citep{johnston1997apparent}.

We begin with a generic analysis of angular decoherence in the eikonal and the wave optics regimes, and study specific cases later (point caustics in Sec.~\ref{secy:finite} and briefly comment on the case of fold caustics in Sec.~\ref{secy:fold}). Our quantitative criteria for angular decoherence is whether the chromatic oscillations (e.g. those depicted in Fig.~\ref{figy:spectra_regimes}) are washed out by integrating over a finite patch of the source plane.
We require the total phase across the extent of the source to change by $\leq 1$ radian, \textit{when viewed at a distance through the gravitational lens}. We find nontrivial differences by considering this criteria in the Eikonal and the diffractive regimes.
\begin{figure*}
    \centering
    \includegraphics[width = \textwidth]{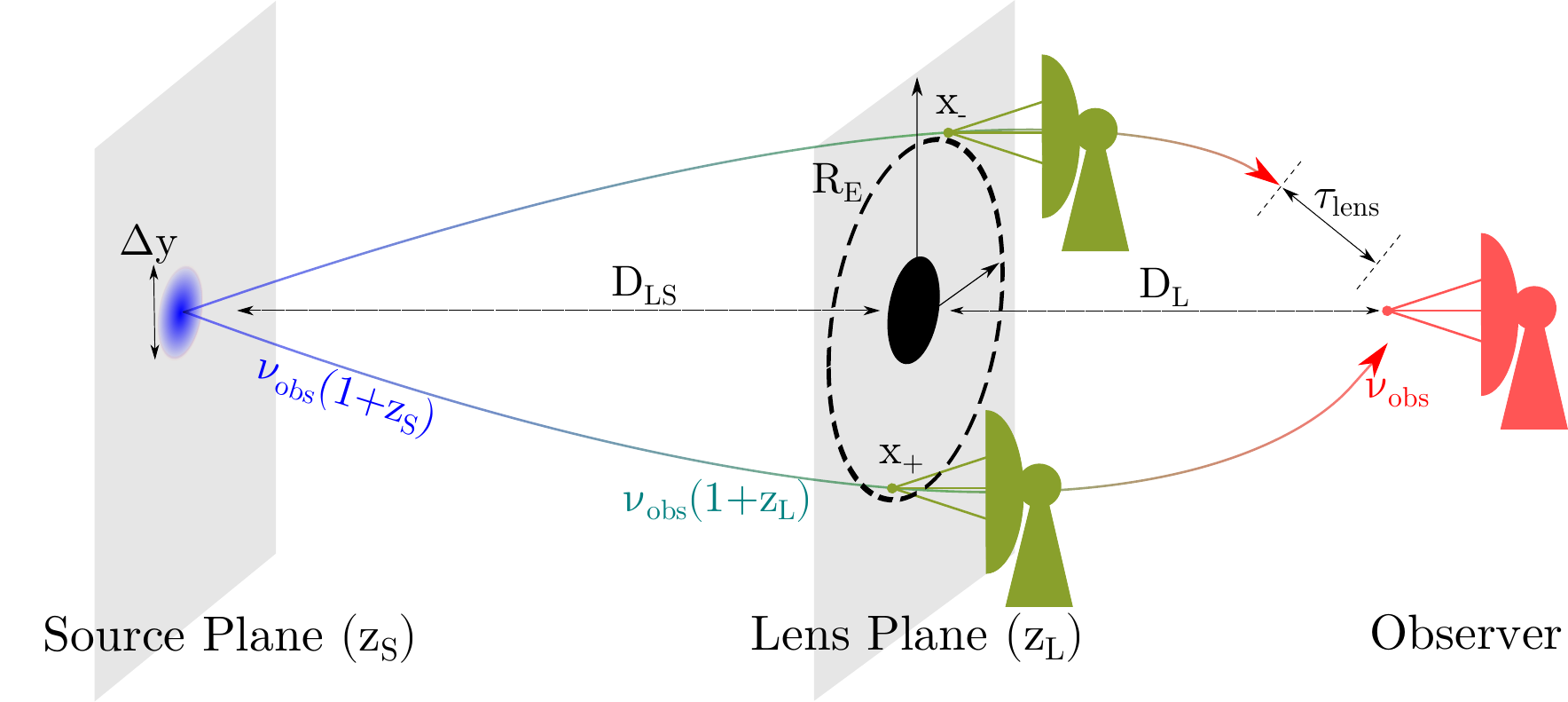}
    \caption{In the eikonal limit, a lensed source remains coherent if the spectral oscillations do not significantly change over the source's finite angular size, here shown as a blue patch. When the images are well-separated, ``rays'' from the source traverse the lens plane through the image positions $\boldsymbol x_{\pm}$ calculated via geometric optics. For the point-mass lens, one image passes outside the Einstein ring at a transverse physical scale $R_E$, and the other passes inside the Einstein ring on the opposite side of the lens. The interference fringes from lensing are present if the source looks point-like as observed by an interferometer with a baseline  of Eq.~\ref{eqy:rayleigh} (i.e. a pair of antennas placed in the lens plane at the apparent locations of the gravitational lens images in the lens plane).}
\label{figy:vlbi_resolution}
\end{figure*}

In the Eikonal limit that the stationary points are well-separated into discrete images $j$ and $k$, there exists a well-defined time delay $T_j - T_k = T_{jk}$ which can be calculated for a point-like source at any location in the source plane. We require that when integrating over a patch of radius $\Delta \boldsymbol y$, changes in $T_{jk}$ must be small. In other words, different patches of the source must not experience different time delays; otherwise, we are averaging over many fringe spacings in the spectrum as depicted in Fig.~\ref{figy:spectra_regimes}. Denoting $\overline{\boldsymbol{y}}$ as the displacement from the center of the source, and $T_j$ and $T_k$ denote the time delays of two distinct images generated by the gravitational
lens. The phase shift is
\begin{align} 
    1 \geq \Delta \varphi &= \Omega \dfrac{\partial T_{jk}}{\partial y_a} \cdot \overline{\boldsymbol{y}}_a \label{eqy:phi_finite}\\
    &= \Omega (\boldsymbol{x}_j - \boldsymbol{x}_k)_a \overline{\boldsymbol{y}}_a.
    \intertext{Using Eq.~\ref{eqy:omega_definition} gives the width of the Fresnel zone:}
    \Delta y &\lesssim \dfrac{1}{\theta_E} \dfrac{\lambda}{2\pi (1+z_L) D_\textrm{img}}.\label{eqy:deltay_geo}
    \intertext{$D_\mathrm{img}$ resembles the Rayleigh resolution limit for a radio interferometer observing at a wavelength of $\lambda / (1+z_L)$ with baseline corresponding to the physical scale of the image separation}
    D_\textrm{img} &= \dfrac{D_L D_S}{D_{LS}} |\boldsymbol{\theta}_j - \boldsymbol{\theta}_k|,\label{eqy:rayleigh}.
\end{align}

This can be interpreted as the effective separation of two telescopes located in the lens plane~\citep{schneider1983mutual}, at the geometric-optics positions of the images, observing the source at a frequency of $\Omega(1+z_L)$, as shown in Fig.~\ref{figy:vlbi_resolution}. The exact form of the fall-off is studied in~\citep{matsunaga2006finite}, and applied to the
particular case of femtolensing in~\citep{stanek1993features}. 
In the diffractive limit, where the images in the time domain are separated by less than $\approx 1/\Omega$, the two stationary points blend together coherently, i.e. they belong to the same Fresnel zone (see e.g. the image positions denoted in the top left panel of Fig.~\ref{figy:fresnel}). Therefore, the criterion we derived in Eq.~\ref{eqy:deltay_geo} treating the paths as well-separated does not apply. 
In this regime, in the absence of a template for the unlensed signal, it does not make sense to think of the two images as well-separated, distinguishable paths through the lens plane, because all parts of the lens plane within the first few Fresnel zones contribute significantly to $F(\Omega, \boldsymbol{y})$.

\begin{figure*}
    \centering
    \includegraphics[width = \textwidth]{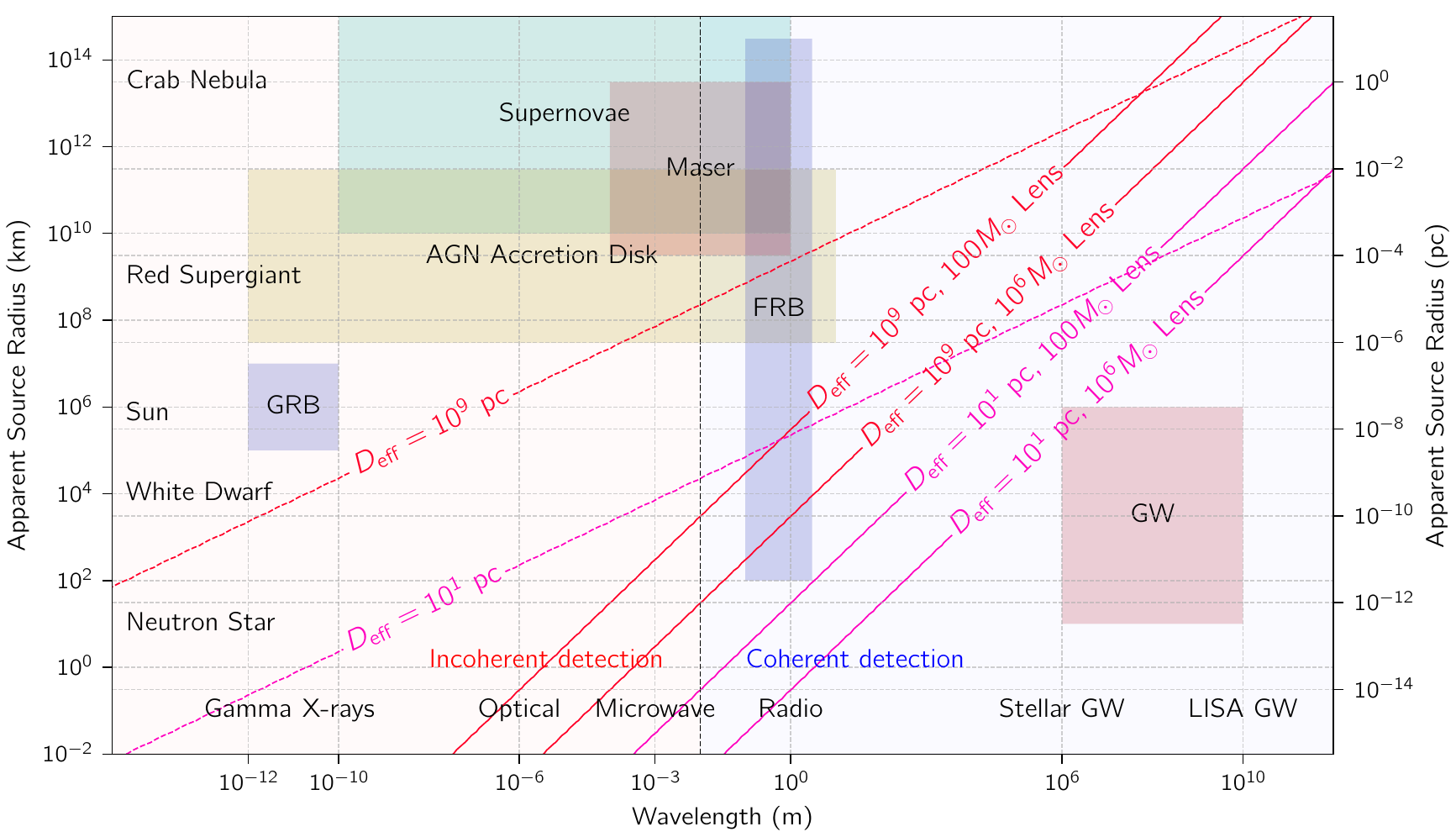}
    \caption{Regimes of coherence in gravitational lensing. We plot the apparent radius of the source against the observing wavelength (ignoring redshift effects). \textbf{Solid lines:} If the lens mass is in the eikonal limit, then the source must be below the solid lines corresponding to Eq.~\ref{eqy:deltay_geo}. \textbf{Dashed lines:} In the diffractive/wave optics regime, the source size must be below the dashed lines corresponding to Eq.~\ref{eqy:deltay_diff}. We have roughly labeled the apparent sizes of various sources and wavelengths at which they have been observed in the various boxes~\citep{oguri2019strong,leung2022constraining,katz2018femtolensing}.}
    \label{figy:fresnel_src}
\end{figure*}

If slightly different patches of the source have very different image configurations, this could lead to very different time delays. One might expect this effect to be small, but there are scenarios where this might become significant. Near the Einstein ring caustic in a point-mass lens, for example, small displacements in the source plane can lead to large differences in the image positions. 

Near the $j$th stationary point of $T$, where $dT/d\boldsymbol{x} = 0$, we analyze the curvature of $T$ with respect to $\boldsymbol{x}$:
\begin{equation}
  \Delta \varphi = \Omega \Delta T = \frac12
  \Omega\frac{\partial^2 T}{\partial x_a\partial x_b}
  \overline{\boldsymbol{x}}_a \overline{\boldsymbol{x}}_b
    = {\textstyle\frac12} \Omega (A^j_{ab})^{-1}\overline{\boldsymbol{x}}_a \overline{\boldsymbol{x}}_b
\end{equation}
where $A^{j}_{ab}$ is the magnification matrix evaluated at the stationary point $j$.  Hence the allowable displacement on the image plane that causes a one-radian phase change is defined by
\begin{equation} \label{eq:phaseflip}
   1 \geq \Delta \varphi = \Omega A^{-1}_{ab}\overline{\boldsymbol{x}}_a \overline{\boldsymbol{x}}_b
\end{equation}
Now we equate $\overline{\boldsymbol{x}}_a$ to the shift in the geometric-optics image for the source whose extent is $\overline{\boldsymbol{y}}_c$. From the definition of the
magnification matrix, we have
\begin{equation}
    \overline{\boldsymbol{x}}_a = A^j_{ac} \overline{\boldsymbol{y}}_c
\end{equation}
Eq.~\ref{eq:phaseflip} then becomes
\begin{equation} \label{eqy:fresnel-mag}
    1 \geq \Delta \varphi = \dfrac12\Omega A_{cd} \overline{\boldsymbol{y}}_c \overline{\boldsymbol{y}}_d.
\end{equation}
Substituting Eq.~\ref{eqy:omega_definition} gives our estimate of the Fresnel zone's shortest axis, if $\lambda_\mathrm{max}$ is the largest eigenvalue of $A_{cd}$: 
\begin{equation}
    \Delta y \lesssim \dfrac{1}{\theta_E |\lambda_\mathrm{max}|}\sqrt{\dfrac{\lambda}{2\pi (1+z_L)D_\mathrm{eff}}}\label{eqy:deltay_diff}
\end{equation}
where $D_\mathrm{eff} = \dfrac{D_L D_S}{D_{LS}}$. Eq.~\ref{eqy:deltay_diff} tells us that if the source size is on the order of the Fresnel angular scale ($\sqrt{\lambda / D_\mathrm{eff}}$), the pattern in Fig.~\ref{figy:fresnel_src} will get washed out because the image positions shift significantly over the source's finite extent, yielding different delays relative to the lensing delay evaluated towards the center of the source. Magnification plays a role in that higher flux magnifications shrinks the Fresnel scale in the source plane. 
We illustrate Eqs.~\ref{eqy:deltay_geo} and~\ref{eqy:deltay_diff} in Fig.~\ref{figy:fresnel_src}, using $R_{src} \sim \Delta y \theta_E D_\mathrm{eff}$ and ignoring factors of $1+z_L$. We see that Eq.~\ref{eqy:deltay_geo} is the more stringent condition that applies in most astrophysical situations, though for short wavelengths and extreme magnification scenarios (e.g.\ proposals to use the sun as a gravitational lens~\citep{maccone2010focal,turyshev2017wave,engeli2022optical}), the latter may become relevant.

\section{Exploiting Chromaticity in Lensing}\label{secy:breaking}
As illustrated in Fig.~\ref{figy:spectra_regimes}, one powerful feature of interference effects in lensing is the introduction of an additional length scale (the wavelength) into the problem. In the eikonal and wave optics regimes, the magnification factor picks up an oscillatory dependence on the wavelength (through $\Omega$) and geometry (through $\boldsymbol y$). By observing the source over a wide enough range of wavelengths to measure the spectral oscillations, more astrophysical information can be recovered than in the geometric-optics case alone. One particularly interesting setup is a multi-messenger observation of a lensed gravitational-wave + electromagnetic source. This would allow for exquisite measurements of wave optical effects, since due to their low
$\Omega$, gravitational waves are potent probes of diffractive effects, while electromagnetic waves are typically in the eikonal or the geometric optics regimes. Observing simultaneously-emitted low-frequency (gravitational) images and high-frequency (electromagnetic) images can expose wave optics effects such as modifications to the phase velocity of diffractively-lensed images~\citep{takahashi2017arrival}, though analyses of the causal ``wavefront''~\citep{suyama2020on,ezquiaga2020apparent} reveals that no violation of causality occurs, though an apparent violation of causality may arise from incorrect separation of the observed image superposition into individual images~\citep{tanaka2023kramers}.

Another example is using the chromaticity of the interference pattern to break the mass-sheet degeneracy in the eikonal and the wave optics regimes.~\citet{cremonese2021breaking} shows that the point mass can be disentangled from a mass sheet in both regimes using the spectral dependence of the amplification factor (the oscillatory terms in Eq.~\ref{eqy:f_geo_pm} and Eq.~\ref{eqy:f_wave_pm}, which encode information about the mass scale $M$). In the diffractive limit, even though the spectral dependence becomes more complex than in the simple eikonal limit, the lensed images become more difficult to separate from one another since their temporal separation is less than one wavelength.

In a similar vein, the chromatic oscillations can be used to directly measure time delays in lensing in both the eikonal and diffractive limits. This idea is applicable to various physical systems, e.g. in radio pulsar/fast radio bursts lensed by sub-solar mass compact objects and planets~\citep{jow2020wave} and sub-Hertz gravitational waves generated by chirping binaries lensed by $\sim 10^8 M_{\odot}$ objects~\citep{takahashi2003wave}. For example, consider a microlensing scenario where a transiting foreground lens with proper motion $\mu$ achieves a minimum impact parameter $y_0$ at a time $t_0$. In diffractive optics, the magnification varies as a function of $(t,\Omega,t_E,y_0,t_0)$, where $t_E \approx \dfrac{1}{\mu} \sqrt{R_E / D_\mathrm{eff}} \propto \sqrt{M/D_\mathrm{eff}} / \mu$ is the Einstein crossing time of the lens. The lens mass enters through both $\Omega$ and $t_E$. In geometric optics, the magnification varies as a function of ($t, t_E, y_0, t_0$) only (since there is no frequency dependence). Since the mass dependence enters only through $t_E$ it is perfectly degenerate with the lens distance $D$. Without an independent distance measurement, it is difficult to measure the lensing mass in this situation. 

These examples show that there is clearly additional information encoded by the frequency dependence, it is helpful to conceptualize, using a thought experiment, where the information really comes from. Consider a delta-function electric-field pulse emitted by a source which gets lensed and subsequently recorded by a detector which records the raw electric field as a function of time. In this situation it would be possible to simultaneously measure the magnification ratio from the light curves, as well as the Shapiro delay ($\sim 20 \mu\mathrm{s} \times (M/M_{\odot})$), which would give a handle on the mass. By Nyquist's theorem, the time
resolution of the Shapiro delay measurement is limited by the inverse bandwidth ($1/\Delta \nu$) of the detector. Larger bandwidths allow measuring $F(\Omega,y)$ for a wide range of $\Omega$ at a single value of $y$; this multi-frequency coverage allows the mass to be constrained. As was shown earlier in Eq.~\ref{eqy:f_geometric}, the oscillations in $F(\Omega,y)$ vary on scales proportional to the inverse Shapiro delay. In a Fourier-transform sense, measuring light curves over a wide range of wavelengths provides a new way to directly constrain the Shapiro time delay, and therefore the mass scale. This is completely distinct from the method of measuring the characteristic timescale of a lensing transit, which indirectly probes the mass via long-term monitoring and measurement of the Einstein crossing timescale and the relative velocities of the source and observer.

These Shapiro delay measurements are not limited to hypothetical delta-function pulses, or even astrophysical transients. Shapiro delay measurements can be made on non-variable sources, if they are compact enough, since the measurement uses fluctuations in the electric field rather than the total power. However, the stringent size constraints on preserving the interference fringes excludes most, if not all, currently known persistent sources from this type of measurement. Transient sources from compact objects are precisely those which are known to be sufficiently point-like to realize the interferometric measurement of Shapiro delays using the chromatic modulations of the dynamic spectrum.

In summary, using chromaticity to extract astrophysical information from lens systems offers an interesting trade-off -- the stringent requirements for source compactness demand transients, which often do not repeat. For these one-off transients, we have given up our ability to monitor the Einstein crossing timescale for a one-time, interferometric measurement of the Shapiro delay timescale free of uncertainties from estimating the relative velocities of the source and lens. Using a lens model, the Shapiro delay and the flux ratio completely determine the (redshifted) lens mass and impact parameter (see also~\citep{itoh2009method}). In the case of persistent emitters (e.g.\ optical microlensing) or recurring transient events such as pulsar pulses, measurement of a third quantity (the Einstein crossing timescale) can be used to extract additional information (such as the relative motion of the lens and the source).
\begin{figure*}
    \centering
    \includegraphics[width = \textwidth]{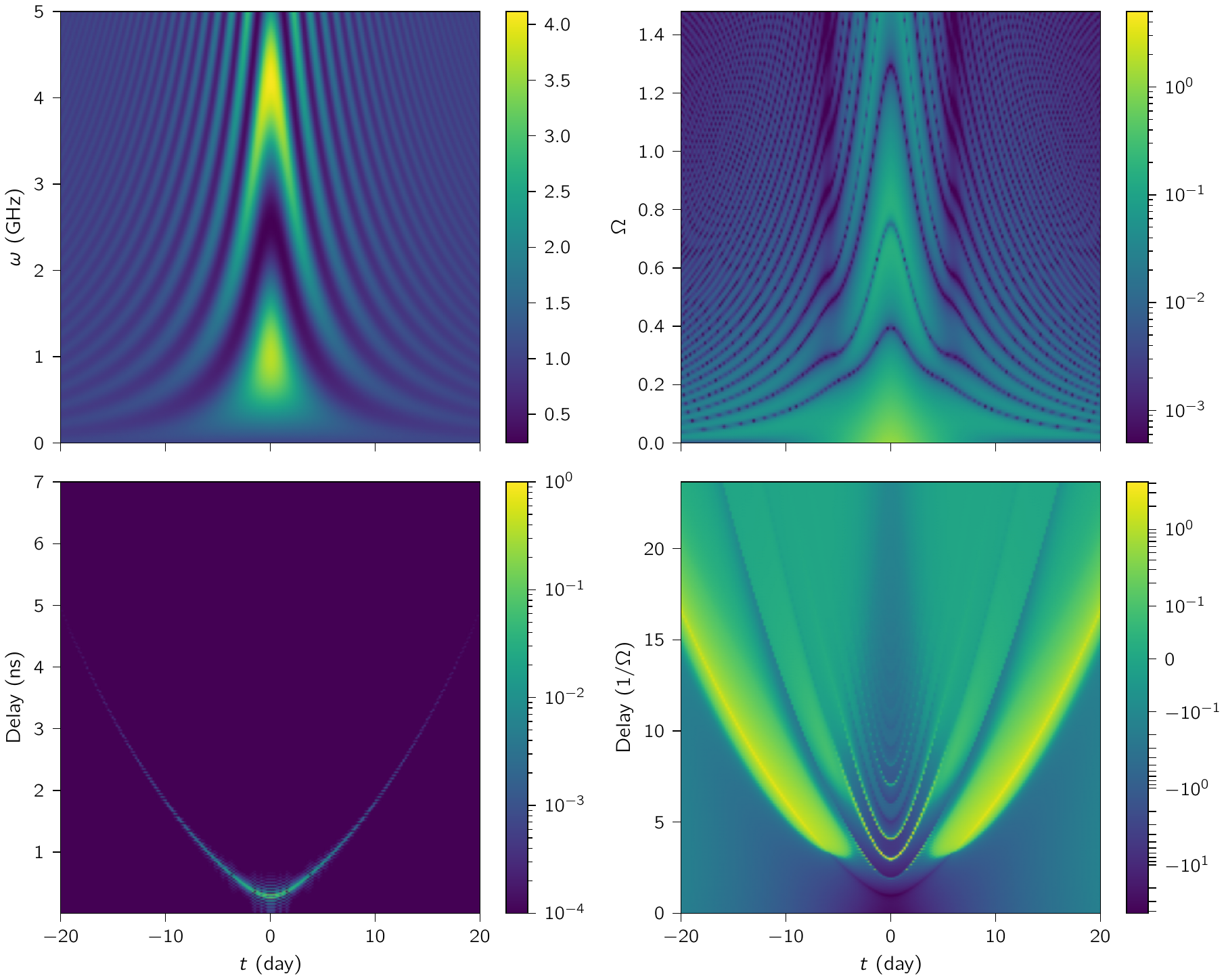}
    \caption{\textbf{Top left:} Frequency-dependent amplification $|F_{\textrm{wave}}(\Omega, y)|^2$ spectrum of a compact, broadband radio sources as it transits behind a point-mass lens of $5 M_{\otimes}$, $D = 1$ kpc with a minimum impact parameter of $y_0 = 0.5$. The amplification oscillates like $\cos^2(\Omega (\tau_+ - \tau_-))$ as a function of frequency for large impact parameters where the Eikonal optics (see Eq.~\ref{eqy:f2_geometric}) approximation does not break down.  \textbf{Top right:} Fractional discrepancy between $|F_{\textrm{wave}}(\Omega, y)|^2$ and $|F_{\textrm{geo}}(\Omega, y)|^2$ for the point-mass lens. Diffractive effects distort the regular fringe pattern predicted from geometric optics. \textbf{Bottom left:} The Fourier transform of the top left panel along the frequency axis, also referred to as the ``secondary spectrum''. The arc tracks the instantaneous lensing delay $(\tau_+(t) - \tau_-(t))$ as it changes throughout the transit. \textbf{Bottom right:} The Fourier transform of the top right panel along the frequency axis. Adapted with permission from~\citep{jow2020wave}.
    }\label{figy:jow2020fig3}
\end{figure*}

Another way of understanding this tradeoff is by analogy to radio scintillation (see e.g.\ \citep{narayan1992physics,goodman1987effects}). The interference pattern in Fig.~\ref{figy:spectra_regimes} is similar to diffractive scintillation in pulsar astronomy, where inhomogeneous plasma in front of a compact background source causes an intrinsically featureless spectrum to exhibit stochastic ripples over some characteristic bandwidth. Like in radio scintillation, the only sources which ``twinkle'', or scintillate, are not necessarily those which show intrinsic variability, but those which appear compact and unresolved by the cosmic telescope created by the plasma inhomogeneities: a star need not exhibit intrinsic variations to twinkle in our atmosphere.

To push the analogy further, another manifestation of scintillation is \textit{refractive} scintillation, where the clumpy spectrum caused by diffractive scintillation changes slowly over longer timescales as the inhomogeneities drift with respect to the observer with some characteristic crossing time (hours to days for scintillation of Galactic pulsars). Monitoring the dynamic spectrum of a source over long timescales probes the motion of plasma screens via
scintillation (see e.g.\ \citep{walker2004interpretation}), similar to how one might measure Einstein crossing timescales in lensing. Scintillation shows up as parabolic, arc-like features in the time-lag correlation function, estimated from the Fourier transform of the source's spectrum, when the time-lag correlation function is concatenated over many days such that it has both a delay axis (measuring timescales on the scale of the inverse bandwidth) and a time axis (measuring timescales of days or weeks). 

A drifting gravitational lens, or a collection of many gravitational lenses drifting together, can similarly produce a parabolic arc in this space (see lower panel of Fig.~\ref{figy:jow2020fig3}) as it crosses the lens. The effect of a collection of moving lenses causing gravitational scintillation has been considered;~\citep{macquart2004scattering} derives analytical expressions for the phase perturbations sourced by clumpy matter distributions;~\citep{congedo2006gravitational} considers the particular case of gravitational scintillation by stellar clusters in the Galactic center.~\citep{takahashi2005scattering} considers the validity of the Born approximation and the impact of chromatic (diffractive) effects in gravitational scintillation. Diffractive effects as a probe of a clumpy matter distributions on small scales are considered in~\citep{takahashi2006amplitude,oguri2020probing,inamori2021universal,oguri2022amplitude}.

\section{Finite magnifications near the Einstein ring}\label{secy:finite}
Having generically analyzed eikonal and wave optics and the transition to incoherent (traditional) gravitational lensing, we can now turn to studying more specific phenomena which arise in wave optics.
One failure of geometric optics is that for certain regions of the source plane, magnification ratios become infinite. These regions are typically isolated points or one-dimensional curves in the source plane where geometric-optics rays ``cross'' and as such are regions where interference effects are most crucial. These are caustics~\citep{ohanian1983caustics}: the most famous example of a caustic is the Einstein ring for a point-mass lens. Rays emanating from the point caustic at $\boldsymbol{y}_0 = 0$ pierce the lens
plane at a ring of points called the ``critical curve'' (here, the critical curve is defined by $|\boldsymbol{x}_0| = 1$) on their way to the observer.

\begin{figure}
    \centering
    \includegraphics{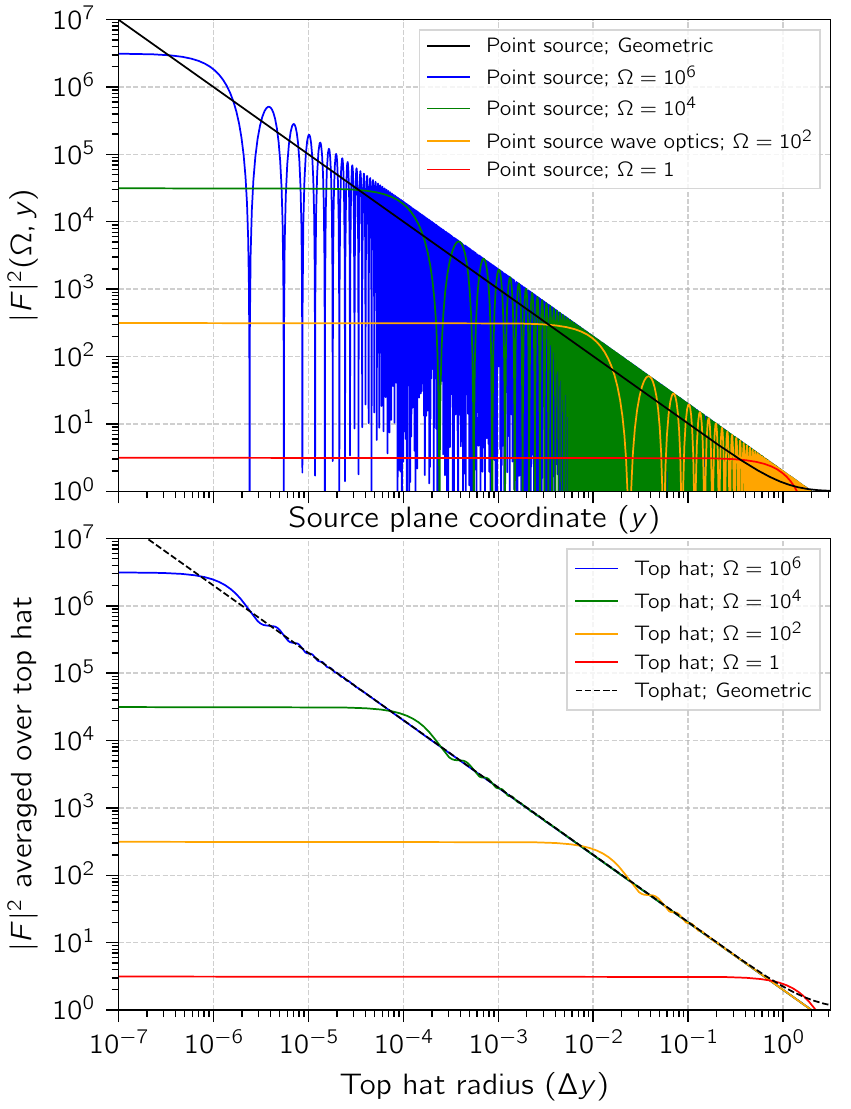}
    \caption{The behavior of the magnification $|F|^2$ taking into account finite source size and wave optics effects near the Einstein ring ($y \to 0$). \textbf{Top:} The magnification map as a function of $\boldsymbol{y}$ near the origin saturates at $\boldsymbol{y} \sim 1/\Omega$ at a magnification of $\pi\Omega$. \textbf{Bottom:} the magnification map for an on-axis source, now as a function of source radius $\Delta y$.}\label{figy:finite}
\end{figure}

The infinite magnification of geometric optics lensing is removed by two distinct effects: finite source size (only an infinitely small patch is magnified infinitely), and diffraction~\citep{herlt1976wave,benson1979high}. In other words, to truly attain infinite magnification, one would need an infinitely-small point source which radiates at an infinitely-short wavelength. In Fig.~\ref{figy:finite} we calculate magnification maps for both a point source close to the caustic, and a finite-sized source centered on the caustic, in the case of infinite and finite frequency (geometric optics and wave optics respectively).

The magnification profiles in geometric optics and diffractive optics can be computed as follows. In geometric optics the total magnification of a point source at some impact parameter $y$ is
\begin{align}
    |F_{\textrm{geo}}|^2 &= \dfrac{y^2+2}{y\sqrt{y^2+4}}~\label{eqy:f_geo_pm}
    \intertext{To obtain the finite-size result, we average Eq.~\ref{eqy:f_geo_pm} over a finite-sized top-hat of radius $\Delta y > 0$ in the source plane centered at $y=0$. This gives a finite total brightness. The amplification factor for a circular, on-axis source of radius $\Delta y$ is}
    \langle |F_\textrm{geo}|^2 \rangle_{\Delta y} &= \sqrt{1+\dfrac{4}{\Delta y^2}}~\label{eqy:f_geo_tophat}
    \intertext{In wave optics, we repeat the top-hat averaging over the point-mass magnification map derived from wave optics (Eq.~\ref{eqy:f_wave_pm}). This yields}
    \langle |F_\textrm{wave}(\Omega)|^2 \rangle_{\Delta y} &= \dfrac{\pi \Omega x_E^2 [J_0(\Omega x_E \Delta y)^2 + J_1(\Omega x_E \Delta y)^2] }{|1 - \psi''(x_E)|}~\label{eqy:f_wave_tophat}
\end{align}
In Fig.~\ref{figy:finite}, we plot Eqs.~\ref{eqy:f_geo_pm} and~\ref{eqy:f_wave_pm} in the top panel, and Eqs.~\ref{eqy:f_wave_tophat} and~\ref{eqy:f_geo_tophat} in the bottom panel. While both diffraction and finite source size effects modify the behavior of $|F|^2$ near $y = 0$, they have very different qualitatitive features -- even a tiny departure from a point source washes out the large oscillations characteristic of diffractive optics. It is also evident that when the source size
is on the caustic, the finite source size solution deviates considerably from the point source solution~\citep{witt1994can,sugiyama2020on}. The discrepancy between using a point-source formula and the extended source matter most when the source dimension $\Delta y \lesssim \Omega^{-1}$; this is most relevant for gravitational waves, which have low values of $\Omega$ and have source sizes $\lesssim\SI{100}{\kilo\meter}$~\citep{oguri2019strong}.

\section{Finite magnifications near fold caustics}\label{secy:fold}
The well-known Einstein ring is an example of a caustic: loci where the geometric optics magnification diverges. Caustics may be isolated points or one-dimensional curves in the source plane. Similar to the Einstein ring, the infinite magnifications near ``fold caustics'' get regularized by finite source size as well as diffraction. Here, following~\citep{nakamura1999wave}, we consider the Fresnel integral (Eq.~\ref{eqy:fresnel_xy}) in the scenario where a source at $\boldsymbol{y}_0$ lies
directly on the caustic. We expand the time-delay potential about a point $\boldsymbol{x}_0$ on the critical curve. We can choose coordinate systems in both the source and lens planes with unit vectors ($\hat{\boldsymbol{y}}_1$, $\hat{\boldsymbol{y}}_2$), ($\hat{\boldsymbol{x}}_1$, $\hat{\boldsymbol{x}}_2$). What does the time delay potential look like near the critical curve in the neighborhood of $\boldsymbol{x}_0$? We can begin with a third-order Taylor expansion about $\boldsymbol{x}_0$ and reason through each term. Here, $\overline{\boldsymbol{x}} = \boldsymbol{x} - \boldsymbol{x}_0$, and $\overline{\boldsymbol{y}} = \boldsymbol{y} - \boldsymbol{y}_0$, and the indices $a,b,c$ run over the two dimensions of the lens plane.

\begin{align}
    \tau(\boldsymbol{x},\boldsymbol{y}) - \tau(\boldsymbol{x_0},\boldsymbol{y}) &= \dfrac{\partial \tau}{\partial x_a}  \boldsymbol{\overline{x}_a} + \dfrac{1}{2}\dfrac{\partial^2 \tau}{\partial x_a \partial x_b}  \boldsymbol{\overline{x}_a} \boldsymbol{\overline{x}_b}\\ &+\dfrac{1}{6}\dfrac{\partial^3 \tau}{\partial x_a \partial x_b \partial x_c}  \boldsymbol{\overline{x}_a} \boldsymbol{\overline{x}_b} \boldsymbol{\overline{x}_c} \\
    \intertext{The gradient of the potential is straightforward to evaluate at $\boldsymbol{x}_0$ and is $-\overline{\boldsymbol{y}}$. The magnification at $\boldsymbol{x}_0$ ($\det(A^0_{ab})$) diverges if $\boldsymbol{y} = \boldsymbol{y}_0$; one of the eigenvalues of $A^0$ (say, $\lambda_1$) is zero. The potential is therefore free of curvature along the $\hat{\boldsymbol{x}}_1$ direction (thought it has a slope of $-\overline y_1$). Along the $\hat{\boldsymbol{x}}_2$ direction, the curvature
    is $1 - \psi_{22}$, where the subscripts in $\psi_{22}$ refer to the two derivatives taken (e.g.\ $\psi_{22} =\partial^2_{x_2} \psi$).} 
    \tau(\boldsymbol{x},\boldsymbol{y}) - \tau(\boldsymbol{x_0},\boldsymbol{y}) &\approx - \overline{\boldsymbol{x}}_0 \cdot \overline{\boldsymbol{y}}_0 + \dfrac{1 - \psi_{22}(\boldsymbol{x_0})}{2} x_2^2 \\ 
    \intertext{As we turn $\boldsymbol{y} \to \boldsymbol{y}_0$, the first derivative term vanishes. This forms an image at $\boldsymbol{x}_0$ whose magnification is roughly the size of the Fresnel zone (the patch in the lens plane over which the phase changes by $2\pi$; see Fig.~\ref{figy:fresnel} for visualization). Traversing the $\boldsymbol{\hat{x}}_2$ direction (fixing $x_1 = 0$), the second derivative of the potential sets the width of the Fresnel zone as usual. Along the $\boldsymbol{\hat{x}}_1$ axis (fixing $x_2 = 0$),
    since the potential has no second derivative, the \textit{third} derivative sets the extent of the Fresnel zone along the $\boldsymbol{\hat{x}} x_1$ direction:}
    \partial^2 \tau/\partial x_1^3 &= -\psi_{111}(\boldsymbol{x}_0).
    \intertext{If we define $p = (1 - \psi_{22}(\boldsymbol{x}_0))^{-1}$ and $q = -2/\psi_{111}(\boldsymbol{x}_0) > 0$, the integrand is a quadratic along the $x_2$ direction and cubic along the $x_1$ direction:}
    F = \dfrac{\Omega}{2\pi i} &\int d\overline{x}_1\exp\left[i\Omega \left( -\overline{x}_1 \overline{y}_1 + \dfrac{\overline{x}_1^3}{3q}\right)\right] \times \\
     &\int d\overline{x}_2\exp\left[i\Omega\left(-\overline{x}_2 \overline{y}_2 + \dfrac{\overline{x}^2_2}{2p} \right)\right] \\
     \intertext{Taking $\mu_* = |p|q^{2/3} \Omega^{1/3}$, $Y = q^{-1/3} \Omega^{-2/3}$, and letting Ai denote the Airy function\footnote{see \href{https://mathworld.wolfram.com/AiryFunctions.html}{Wolfram MathWorld}}, the amplification factor is}
     |F|^2 = 2\pi \mu_* \textrm{Ai}(-\overline{y}_1 / Y)
\end{align}
The characteristic fringe spacing near the fold caustic is $\sim Y \propto \Omega^{-2/3}$, which has a weaker frequency dependence than the point caustic $(\sim 1/\Omega)$.~\citet{jaroszynski1995diffraction} discusses diffractive effects in caustic crossings at optical wavelengths, and points out that it is possible to use the caustic crossing timescale to measure the finite linear size of a source.

\section{Modified lensing probabilities in wave optics}\label{secy:optical_depth}
Modifying the amplification factor's dependence on the impact parameter near the Einstein ring modifies any optical depth calculations for which the observations are flux-limited. For forecasts and constraints, it is important to characterize the conditions under which diffraction modifies the optical depth relative to the standard case of geometric optics. One effect, discussed previously in Sec.~\ref{secy:coherent_geo}, is that interference terms, if they are not washed out, are easier to detect than autocorrelation terms. With a search method that utilizes spectral information, this leads to an increase in the optical depth~\citep{katz2020looking,jow2020wave,kader2022high} which becomes significant when the source is much brighter than the limiting flux threshold. 

\begin{figure}
    \centering
    \includegraphics{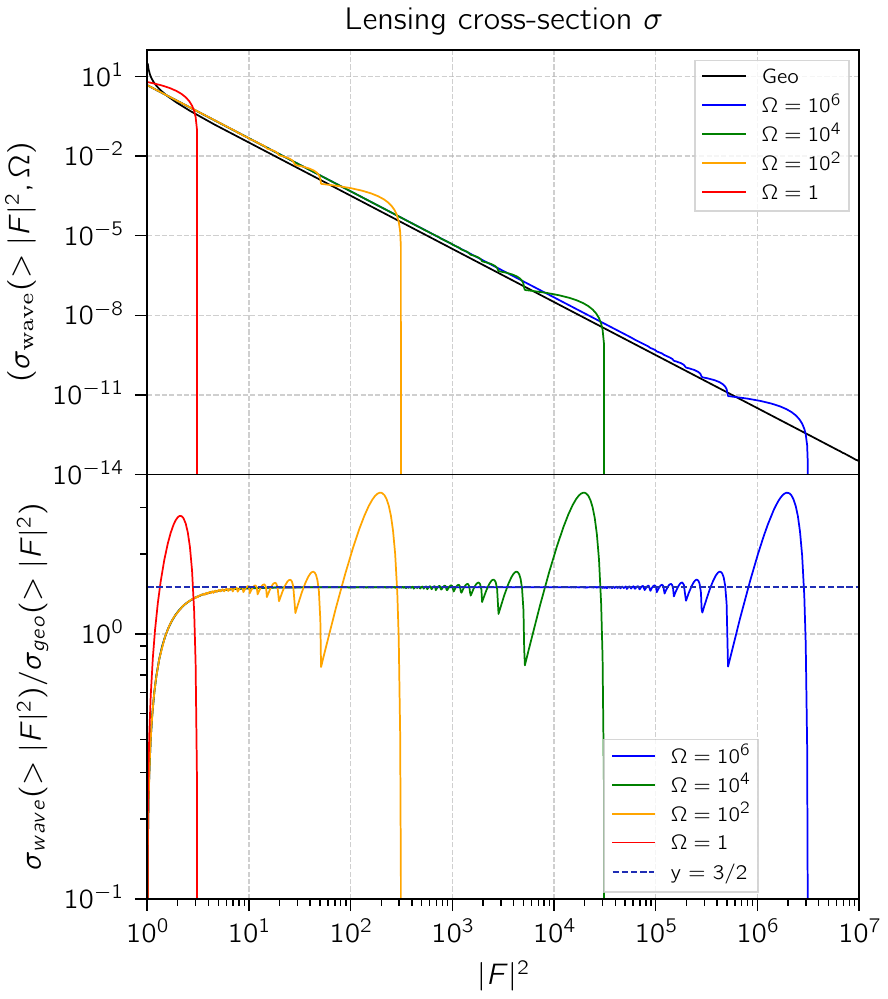}
    \caption{The lensing cross-section $\sigma$ for a minimum detectable amplification factor $|F|^2$. Top: we plot several values of $\Omega$ as well as the cross-section calculated in geometric optics. Bottom: the ratio of $\sigma_\textrm{wave} / \sigma_\textrm{geo}$ is well-approximated by a boxcar between $|F|^2 = [1,2\pi\Omega]$ with a height of $y = 3/2$. The conclusion is that interference effects enhances lensing cross-sections for high magnifications by a factor of $\approx 3/2$ while cutting
    off huge amplifications greater than $\approx\pi\Omega$. This result has the caveat that Eq.~\ref{eqy:f_wave_pm} is used to approximate Eq.~\ref{eqy:f_point_mass}, which is an excellent approximation for $\Omega \gtrsim 1$.}\label{figy:tau_ratio}
\end{figure}

In Sec.~\ref{secy:finite}, we discussed how the infinite magnifications in the point-mass model which appear in geometric optics get regularized by wave optics. One might initially suspect that this decreases the optical depth in a wave optics calculation relative to the geometric-optics calculation, but the reality is that it depends on the magnification scale~\citep{sugiyama2020on}, and the angular motion of the source over the observation period. Diffraction eliminates the possibility of chance alignments creating extremely high magnification events ($|F|^2 \gtrsim
\pi\Omega$)~\citep{bontz1981diffraction,nakamura1999wave,depaolis2002note}, as can be seen by the flattening of the wave optics magnification curves at small $y$ in the bottom panel of Fig.~\ref{figy:finite} (for compact sources, magnifications of up to $|F|^2 \approx \pi\Omega$ can be achieved). However, for more modest amplification factors, the sidelobes of $F(\Omega,y)$ contribute non-negligibly to the cross-section. Let us take a graphical approach to understand this.
Note the rapidly-varying oscillations in the top panel of Fig.~\ref{figy:finite}. In a typical optical depth calculation with a certain sensitivity threshold $|F|^2_\textrm{min}$, we integrate over all $y$ in the source plane for which $|F(y)|^2 > |F|^2_\textrm{min}$ to obtain the optical depth $\sigma(>|F_\textrm{min}|^2)$. In the geometric optics case, $|F_\textrm{geo}|^2(y)$ is a decreasing function. Therefore, the region of the source plane that counts towards the optical depth is a circle of radius $y_*$ such that $|F_\textrm{geo}(y_*)|^2 = |F|^2_\textrm{min}$. Its area, the cross-section to lensing, is $\sigma(>|F_\textrm{min}|^2) = \sigma_{\textrm{geo}} = \pi y_*^2$. In the wave optics case, it is not as straightforward to calculate a threshold value of $y$, because the amplification factor is oscillatory, as seen in the top panel of Fig.~\ref{figy:finite}. We numerically evaluate Eq.~\ref{eqy:f_wave_pm} on a dense grid of points $\{y_i\}$, checking for each $y_i$ whether $|F(y_i)|^2 > |F_\textrm{min}|^2$ and accumulating $2\pi y_i\Delta y_i$ to $\sigma(>|F_\textrm{min}|^2)$ if the magnification is above the threshold.
Repeating this as a function of $|F_\textrm{min}|^2$, for $|F_\textrm{geo}|^2$ (Eq.~\ref{eqy:f_geo_pm}), as well as $|F_\textrm{wave}|^2$ (we use the approximation in Eq.~\ref{eqy:f_wave_pm}) allows us to calculate the optical depth in wave optics and in geometric optics as a function of the sensitivity cutoff $|F_\textrm{min}|^2$; the ratio is plotted in Fig.~\ref{figy:tau_ratio} (see also Appendix B of~\citep{sammons2022effect}, which studies the effect of high magnification lensing events on fast transient
luminosity functions). 

Finally, for $\Omega \lesssim 1$, the lensing is somewhat suppressed. This was considered in the context of kilohertz gravitational waves, which do not ``feel'' mass distributions smaller than $\sim 10^2 M_{\odot}$~\citep{nakamura1998gravitational}. The bad news is that this limits the mass reach of searches for gravitational wave lensing. The good news is that existing search pipelines are reasonably sensitive to diffractively-lensed gravitational waves, and that the subtle distortions due to diffractive lensing will be detectable for louder events~\citep{dai2018detecting}. Moreover, the chromaticity induces a mass scale dependence of the lensing effect. This enables lensing magnifications to probe intervening density fluctuations as a function of mass scale (i.e. the matter power spectrum)~\citep{takahashi2006amplitude,oguri2020probing} and dark matter on dwarf galaxy scales~\citep{han2021small,guo2022probing}. In addition, diffraction suppresses the effect of microlensing on the measured magnification ratios of gravitational waves macro-lensed by galaxies or clusters, unless the macro-magnifications $\gtrsim 15$~\citep{mishra2021gravitational}. This can offer cleaner time-delay cosmography measurements and allow gravitational waves to be localized~\citep{cheung2021stellar}.

Since $F(\Omega,\boldsymbol y)$ is oscillatory as a function of $\Omega$, the cross-section of microlensing is higher at low frequencies. However, as the frequency approaches $\Omega \leq 1$, no microlensing occurs (see next section). According to Fig.~\ref{figy:regimes}, the reach of optical microlensing is also cut off at low masses due to diffractive effects. These two effects lead to overestimating cross-sections calculated in the geometric optics
limit. Due to finite source size effects (e.g.~\citep{witt1994can}) and wave effects, the maximum magnification is less than infinity but drops off more slowly. See e.g.\ Fig. 5 of~\citep{sugiyama2020on} for a detailed study of the effect of finite source size effect on the light curve of a transiting optical source.

\section{Observational prospects and discussion}\label{secy:discussion}
We have presented a summary of the various aspects of interference in gravitational lensing. In some ways, interference in gravitational lensing makes lensing easier to detect. For example, in the Eikonal limit, the presence of interference terms between images makes searching for the interference terms much more sensitive than searches for individual images: one simply searches for a regular sinusoidal pattern in the spectrum whose spacing allows for a direct measurement of the lensing 
time delay. For diffractive optics with a point-mass lens, the sinusoidal pattern becomes distorted at low frequencies as $\Omega$ tends to 1; the unique form of this spectral distortion can break degeneraces in microlensing, allowing for a measurement of the impact parameter and (redshifted) mass of the lens simultaneously. 

In addition, diffraction limits the maximum attainable magnification near caustics, where interference effects become the most pronounced. We find in
diffractive lensing that extremely large magnifications near cusps and folds are no longer
possible, but that ``sidelobes'' in the magnification pattern can boost the cross-section to lensing. We calculate this boost, and find (in agreement with~\citep{sammons2022effect}, which appeared while this manuscript was being prepared) that to a very good approximation, the lensing cross-section is higher by a factor of $3/2$ as compared to geometric optics for large ($1 << |F|^2 < \pi \Omega$) magnifications.

Finally, we characterized the primary challenge to observing interference in gravitational lensing: the paucity of point-like, or angularly-coherent, sources of radiation. A true point source, even one that does not exhibit any significant light curve variations, could be used to measure lensing delays to Nyquist-limited time resolution (nanoseconds at radio frequencies). Such compact steady sources are not yet known, but compact \textit{transient} sources at cosmological distances such as supernovae and fast radio bursts could be used. Gravitational waves from compact binary mergers observed by current and upcoming ground- and space-based detectors are particularly promising candidates for observing diffractive lensing. We comment briefly on the different observational prospects here, but refer the reader to~\citep{oguri2019strong} for a comprehensive discussion of transient lensing.

In the field of lensed transients, supernovae have a head start since a handful of lensed supernovae have already been discovered (e.g.~\citep{kelly2015multiple}). Since $\Omega$ is large for optical and near-infrared supernovae, they can potentially be extremely highly magnified~\citep{diego2019universe}. However, according to Fig.~\ref{figy:finite}, supernovae photospheres are not sufficiently compact, except at cosmological distances.

Gamma-ray bursts (GRBs) are among the most luminous and distant transients known, leading to favorable lensing rates and femtolensing searches over the last few decades~\citep{gould1992femtolensing}, but because of the extremely short wavelengths involved, their angular sizes are not negligible~\citep{stanek1993features,witt1994can,matsunaga2006finite}. While the apparent size of the emission is dependent on the beaming factor~\citep{katz2018femtolensing,khangulyan2022fast}, and can vary from burst to burst, most GRBs are not point-like enough to observe the spectral features of femtolensing; furthermore the range of lens sizes required to observe interference effects in GRBs may not be astrophysical (see Fig.~\ref{figy:regimes}). 

FRBs are extremely compact, but like active galactic nuclei, they may be angularly broadened~\citep{rickett1990radio,lazio2008angular} into a ``scattering disk'' at the low (sub-GHz) radio frequencies at which they are typically observed\citep{katz2020looking}. Nevertheless, angular coherence in the sense of Eq.~\ref{eqy:deltay_geo} is preserved for sub-solar mass lenses~\citep{kader2022high,leung2022constraining}. The bulk of FRBs detected to date have been surveyed by the Canadian Hydrogen Intensity Mapping Experiment at low (sub-gigahertz) frequencies~\citep{chimefrb2021first}, but because of the strong frequency dependence of the size of the scattering disk ($R \propto \lambda^2$), these FRBs are sensitive to angular broadening by inhomogeneous plasma, which potentially washes out interference fringes. This problem can be mitigated with FRBs surveyed at even slightly higher frequencies by upcoming experiments like CHORD and the DSA-2000~\citep{vanderlinde2019chord,hallinan2019dsa}; these may enable observation of coherent interference patterns in FRBs lensed by stars~\citep{connor2022stellar}.

Perhaps the most promising way to detect diffractive effects is via gravitational wave detectors like LIGO~\citep{ligo2015advanced}, VIRGO~\citep{virgo2015advanced}, and KAGRA~\citep{kagra2019advanced}, whose pristine sources are not contaminated by propagation effects, except perhaps by microlenses near the images of a macro-lens~\citep{mishra2021gravitational}. Gravitational-wave sources are extremely compact ($\sim \SI{100}{\kilo\meter}$; see~\citep{oguri2019strong}), and searches have already been
designed and conducted in both the Eikonal and diffractive regimes~\citep{haris2018identifying,hannuksela2019search,dai2020search,abbott2021search}, though no candidates have been conclusively confirmed. The observational challenge of GW lensing is that with the current generation of gravitational wave detectors (LIGO, VIRGO, KAGRA, and GEO600), gravitational waves are difficult to localize to a host galaxy without an electromagnetic counterpart. However, rapid progress, particularly in FRBs and gravitational-wave instruments, will soon lead to detections of interference and diffractive effects in gravitational lensing, and the creative application of lensing to unique measurements for astronomy and fundamental physics.

\begin{acknowledgements}
We thank the International Space Science Institute in Bern (ISSI) for their hospitality and the conveners for organizing the workshop on Strong Gravitational Lensing where this paper was initially written. 
\allacks
We thank Paul Schechter for helpful discussions throughout the preparation of this manuscript.
\end{acknowledgements}

\bibliographystyle{aasjournal}
\bibliography{waves}                % name your BibTeX data base

\begin{thebibliography}{}
\expandafter\ifx\csname natexlab\endcsname\relax\def\natexlab#1{#1}\fi
\providecommand{\url}[1]{\href{#1}{#1}}
\providecommand{\dodoi}[1]{doi:~\href{http://doi.org/#1}{\nolinkurl{#1}}}
\providecommand{\doeprint}[1]{\href{http://ascl.net/#1}{\nolinkurl{http://ascl.net/#1}}}
\providecommand{\doarXiv}[1]{\href{https://arxiv.org/abs/#1}{\nolinkurl{https://arxiv.org/abs/#1}}}

\bibitem[{{Abbott} {et~al.}(2016){Abbott}, {Abbott}, {Abbott}, {Abernathy},
  {Acernese}, {Ackley}, {Adams}, {Adams}, {Addesso}, {Adhikari}, {Adya},
  {Affeldt}, {Agathos}, {Agatsuma}, {Aggarwal}, {Aguiar}, {Aiello}, {Ain},
  {Ajith}, {Allen}, {Allocca}, {Altin}, {Anderson}, {Anderson}, {Arai},
  {Arain}, {Araya}, {Arceneaux}, {Areeda}, {Arnaud}, {Arun}, {Ascenzi},
  {Ashton}, {Ast}, {Aston}, {Astone}, {Aufmuth}, {Aulbert}, {Babak}, {Bacon},
  {Bader}, {Baker}, {Baldaccini}, {Ballardin}, {Ballmer}, {Barayoga},
  {Barclay}, {Barish}, {Barker}, {Barone}, {Barr}, {Barsotti}, {Barsuglia},
  {Barta}, {Bartlett}, {Barton}, {Bartos}, {Bassiri}, {Basti}, {Batch},
  {Baune}, {Bavigadda}, {Bazzan}, {Behnke}, {Bejger}, {Belczynski}, {Bell},
  {Bell}, {Berger}, {Bergman}, {Bergmann}, {Berry}, {Bersanetti}, {Bertolini},
  {Betzwieser}, {Bhagwat}, {Bhandare}, {Bilenko}, {Billingsley}, {Birch},
  {Birney}, {Birnholtz}, {Biscans}, {Bisht}, {Bitossi}, {Biwer}, {Bizouard},
  {Blackburn}, {Blair}, {Blair}, {Blair}, {Bloemen}, {Bock}, {Bodiya}, {Boer},
  {Bogaert}, {Bogan}, {Bohe}, {Bojtos}, {Bond}, {Bondu}, {Bonnand}, {Boom},
  {Bork}, {Boschi}, {Bose}, {Bouffanais}, {Bozzi}, {Bradaschia}, {Brady},
  {Braginsky}, {Branchesi}, {Brau}, {Briant}, {Brillet}, {Brinkmann},
  {Brisson}, {Brockill}, {Brooks}, {Brown}, {Brown}, {Brown}, {Buchanan},
  {Buikema}, {Bulik}, {Bulten}, {Buonanno}, {Buskulic}, {Buy}, {Byer},
  {Cabero}, {Cadonati}, {Cagnoli}, {Cahillane}, {Bustillo}, {Callister},
  {Calloni}, {Camp}, {Cannon}, {Cao}, {Capano}, {Capocasa}, {Carbognani},
  {Caride}, {Casanueva Diaz}, {Casentini}, {Caudill}, {Cavagli{\`a}},
  {Cavalier}, {Cavalieri}, {Cella}, {Cepeda}, {Baiardi}, {Cerretani},
  {Cesarini}, {Chakraborty}, {Chalermsongsak}, {Chamberlin}, {Chan}, {Chao},
  {Charlton}, {Chassande-Mottin}, {Chen}, {Chen}, {Cheng}, {Chincarini},
  {Chiummo}, {Cho}, {Cho}, {Chow}, {Christensen}, {Chu}, {Chua}, {Chung},
  {Ciani}, {Clara}, {Clark}, {Cleva}, {Coccia}, {Cohadon}, {Colla}, {Collette},
  {Cominsky}, {Constancio}, {Conte}, {Conti}, {Cook}, {Corbitt}, {Cornish},
  {Corsi}, {Cortese}, {Costa}, {Coughlin}, {Coughlin}, {Coulon}, {Countryman},
  {Couvares}, {Cowan}, {Coward}, {Cowart}, {Coyne}, {Coyne}, {Craig},
  {Creighton}, {Creighton}, {Cripe}, {Crowder}, {Cruise}, {Cumming},
  {Cunningham}, {Cuoco}, {Dal Canton}, {Danilishin}, {D'Antonio}, {Danzmann},
  {Darman}, {Da Silva Costa}, {Dattilo}, {Dave}, {Daveloza}, {Davier},
  {Davies}, {Daw}, {Day}, {De}, {DeBra}, {Debreczeni}, {Degallaix}, {De
  Laurentis}, {Del{\'e}glise}, {Del Pozzo}, {Denker}, {Dent}, {Dereli},
  {Dergachev}, {DeRosa}, {De Rosa}, {DeSalvo}, {Dhurandhar}, {D{\'\i}az}, {Di
  Fiore}, {Di Giovanni}, {Di Lieto}, {Di Pace}, {Di Palma}, {Di Virgilio},
  {Dojcinoski}, {Dolique}, {Donovan}, {Dooley}, {Doravari}, {Douglas},
  {Downes}, {Drago}, {Drever}, {Driggers}, {Du}, {Ducrot}, {Dwyer}, {Edo},
  {Edwards}, {Effler}, {Eggenstein}, {Ehrens}, {Eichholz}, {Eikenberry},
  {Engels}, {Essick}, {Etzel}, {Evans}, {Evans}, {Everett}, {Factourovich},
  {Fafone}, {Fair}, {Fairhurst}, {Fan}, {Fang}, {Farinon}, {Farr}, {Farr},
  {Favata}, {Fays}, {Fehrmann}, {Fejer}, {Feldbaum}, {Ferrante}, {Ferreira},
  {Ferrini}, {Fidecaro}, {Finn}, {Fiori}, {Fiorucci}, {Fisher}, {Flaminio},
  {Fletcher}, {Fong}, {Fournier}, {Franco}, {Frasca}, {Frasconi}, {Frede},
  {Frei}, {Freise}, {Frey}, {Frey}, {Fricke}, {Fritschel}, {Frolov}, {Fulda},
  {Fyffe}, {Gabbard}, {Gair}, {Gammaitoni}, {Gaonkar}, {Garufi}, {Gatto},
  {Gaur}, {Gehrels}, {Gemme}, {Gendre}, {Genin}, {Gennai}, {George}, {Gergely},
  {Germain}, {Ghosh}, {Ghosh}, {Ghosh}, {Giaime}, {Giardina}, {Giazotto},
  {Gill}, {Glaefke}, {Gleason}, {Goetz}, {Goetz}, {Gondan}, {Gonz{\'a}lez},
  {Castro}, {Gopakumar}, {Gordon}, {Gorodetsky}, {Gossan}, {Gosselin},
  {Gouaty}, {Graef}, {Graff}, {Granata}, {Grant}, {Gras}, {Gray}, {Greco},
  {Green}, {Greenhalgh}, {Groot}, {Grote}, {Grunewald}, {Guidi}, {Guo},
  {Gupta}, {Gupta}, {Gushwa}, {Gustafson}, {Gustafson}, {Hacker}, {Hall},
  {Hall}, {Hammond}, {Haney}, {Hanke}, {Hanks}, {Hanna}, {Hannam}, {Hanson},
  {Hardwick}, {Harms}, {Harry}, {Harry}, {Hart}, {Hartman}, {Haster},
  {Haughian}, {Healy}, {Heefner}, {Heidmann}, {Heintze}, {Heinzel}, {Heitmann},
  {Hello}, {Hemming}, {Hendry}, {Heng}, {Hennig}, {Heptonstall}, {Heurs},
  {Hild}, {Hoak}, {Hodge}, {Hofman}, {Hollitt}, {Holt}, {Holz}, {Hopkins},
  {Hosken}, {Hough}, {Houston}, {Howell}, {Hu}, {Huang}, {Huerta}, {Huet},
  {Hughey}, {Husa}, {Huttner}, {Huynh-Dinh}, {Idrisy}, {Indik}, {Ingram},
  {Inta}, {Isa}, {Isac}, {Isi}, {Islas}, {Isogai}, {Iyer}, {Izumi}, {Jacobson},
  {Jacqmin}, {Jang}, {Jani}, {Jaranowski}, {Jawahar}, {Jim{\'e}nez-Forteza},
  {Johnson}, {Johnson-McDaniel}, {Jones}, {Jones}, {Jonker}, {Ju}, {Haris},
  {Kalaghatgi}, {Kalogera}, {Kandhasamy}, {Kang}, {Kanner}, {Karki},
  {Kasprzack}, {Katsavounidis}, {Katzman}, {Kaufer}, {Kaur}, {Kawabe},
  {Kawazoe}, {K{\'e}f{\'e}lian}, {Kehl}, {Keitel}, {Kelley}, {Kells},
  {Kennedy}, {Keppel}, {Key}, {Khalaidovski}, {Khalili}, {Khan}, {Khan},
  {Khan}, {Khazanov}, {Kijbunchoo}, {Kim}, {Kim}, {Kim}, {Kim}, {Kim}, {Kim},
  {King}, {King}, {Kinzel}, {Kissel}, {Kleybolte}, {Klimenko}, {Koehlenbeck},
  {Kokeyama}, {Koley}, {Kondrashov}, {Kontos}, {Koranda}, {Korobko}, {Korth},
  {Kowalska}, {Kozak}, {Kringel}, {Krishnan}, {Kr{\'o}lak}, {Krueger}, {Kuehn},
  {Kumar}, {Kumar}, {Kuo}, {Kutynia}, {Kwee}, {Lackey}, {Landry}, {Lange},
  {Lantz}, {Lasky}, {Lazzarini}, {Lazzaro}, {Leaci}, {Leavey}, {Lebigot},
  {Lee}, {Lee}, {Lee}, {Lee}, {Lenon}, {Leonardi}, {Leong}, {Leroy},
  {Letendre}, {Levin}, {Levine}, {Li}, {Libson}, {Littenberg}, {Lockerbie},
  {Logue}, {Lombardi}, {London}, {Lord}, {Lorenzini}, {Loriette}, {Lormand},
  {Losurdo}, {Lough}, {Lousto}, {Lovelace}, {L{\"u}ck}, {Lundgren}, {Luo},
  {Lynch}, {Ma}, {MacDonald}, {Machenschalk}, {MacInnis}, {Macleod},
  {Maga{\~n}a-Sandoval}, {Magee}, {Mageswaran}, {Majorana}, {Maksimovic},
  {Malvezzi}, {Man}, {Mandel}, {Mandic}, {Mangano}, {Mansell}, {Manske},
  {Mantovani}, {Marchesoni}, {Marion}, {M{\'a}rka}, {M{\'a}rka}, {Markosyan},
  {Maros}, {Martelli}, {Martellini}, {Martin}, {Martin}, {Martynov}, {Marx},
  {Mason}, {Masserot}, {Massinger}, {Masso-Reid}, {Matichard}, {Matone},
  {Mavalvala}, {Mazumder}, {Mazzolo}, {McCarthy}, {McClelland}, {McCormick},
  {McGuire}, {McIntyre}, {McIver}, {McManus}, {McWilliams}, {Meacher},
  {Meadors}, {Meidam}, {Melatos}, {Mendell}, {Mendoza-Gandara}, {Mercer},
  {Merilh}, {Merzougui}, {Meshkov}, {Messenger}, {Messick}, {Meyers},
  {Mezzani}, {Miao}, {Michel}, {Middleton}, {Mikhailov}, {Milano}, {Miller},
  {Millhouse}, {Minenkov}, {Ming}, {Mirshekari}, {Mishra}, {Mitra},
  {Mitrofanov}, {Mitselmakher}, {Mittleman}, {Moggi}, {Mohan}, {Mohapatra},
  {Montani}, {Moore}, {Moore}, {Moraru}, {Moreno}, {Morriss}, {Mossavi},
  {Mours}, {Mow-Lowry}, {Mueller}, {Mueller}, {Muir}, {Mukherjee}, {Mukherjee},
  {Mukherjee}, {Mukund}, {Mullavey}, {Munch}, {Murphy}, {Murray}, {Mytidis},
  {Nardecchia}, {Naticchioni}, {Nayak}, {Necula}, {Nedkova}, {Nelemans},
  {Neri}, {Neunzert}, {Newton}, {Nguyen}, {Nielsen}, {Nissanke}, {Nitz},
  {Nocera}, {Nolting}, {Normandin}, {Nuttall}, {Oberling}, {Ochsner}, {O'Dell},
  {Oelker}, {Ogin}, {Oh}, {Oh}, {Ohme}, {Oliver}, {Oppermann}, {Oram},
  {O'Reilly}, {O'Shaughnessy}, {Ott}, {Ottaway}, {Ottens}, {Overmier}, {Owen},
  {Pai}, {Pai}, {Palamos}, {Palashov}, {Palomba}, {Pal-Singh}, {Pan}, {Pan},
  {Pankow}, {Pannarale}, {Pant}, {Paoletti}, {Paoli}, {Papa}, {Paris},
  {Parker}, {Pascucci}, {Pasqualetti}, {Passaquieti}, {Passuello},
  {Patricelli}, {Patrick}, {Pearlstone}, {Pedraza}, {Pedurand}, {Pekowsky},
  {Pele}, {Penn}, {Perreca}, {Pfeiffer}, {Phelps}, {Piccinni}, {Pichot},
  {Pickenpack}, {Piergiovanni}, {Pierro}, {Pillant}, {Pinard}, {Pinto},
  {Pitkin}, {Poeld}, {Poggiani}, {Popolizio}, {Post}, {Powell}, {Prasad},
  {Predoi}, {Premachandra}, {Prestegard}, {Price}, {Prijatelj}, {Principe},
  {Privitera}, {Prix}, {Prodi}, {Prokhorov}, {Puncken}, {Punturo}, {Puppo},
  {P{\"u}rrer}, {Qi}, {Qin}, {Quetschke}, {Quintero}, {Quitzow-James}, {Raab},
  {Rabeling}, {Radkins}, {Raffai}, {Raja}, {Rakhmanov}, {Ramet}, {Rapagnani},
  {Raymond}, {Razzano}, {Re}, {Read}, {Reed}, {Regimbau}, {Rei}, {Reid},
  {Reitze}, {Rew}, {Reyes}, {Ricci}, {Riles}, {Robertson}, {Robie}, {Robinet},
  {Rocchi}, {Rolland}, {Rollins}, {Roma}, {Romano}, {Romano}, {Romanov},
  {Romie}, {Rosi{\'n}ska}, {Rowan}, {R{\"u}diger}, {Ruggi}, {Ryan}, {Sachdev},
  {Sadecki}, {Sadeghian}, {Salconi}, {Saleem}, {Salemi}, {Samajdar}, {Sammut},
  {Sampson}, {Sanchez}, {Sandberg}, {Sandeen}, {Sanders}, {Sanders},
  {Sassolas}, {Sathyaprakash}, {Saulson}, {Sauter}, {Savage}, {Sawadsky},
  {Schale}, {Schilling}, {Schmidt}, {Schmidt}, {Schnabel}, {Schofield},
  {Sch{\"o}nbeck}, {Schreiber}, {Schuette}, {Schutz}, {Scott}, {Scott},
  {Sellers}, {Sengupta}, {Sentenac}, {Sequino}, {Sergeev}, {Serna},
  {Setyawati}, {Sevigny}, {Shaddock}, {Shaffer}, {Shah}, {Shahriar}, {Shaltev},
  {Shao}, {Shapiro}, {Shawhan}, {Sheperd}, {Shoemaker}, {Shoemaker}, {Siellez},
  {Siemens}, {Sigg}, {Silva}, {Simakov}, {Singer}, {Singer}, {Singh}, {Singh},
  {Singhal}, {Sintes}, {Slagmolen}, {Smith}, {Smith}, {Smith}, {Smith}, {Son},
  {Sorazu}, {Sorrentino}, {Souradeep}, {Srivastava}, {Staley}, {Steinke},
  {Steinlechner}, {Steinlechner}, {Steinmeyer}, {Stephens}, {Stevenson},
  {Stone}, {Strain}, {Straniero}, {Stratta}, {Strauss}, {Strigin}, {Sturani},
  {Stuver}, {Summerscales}, {Sun}, {Sutton}, {Swinkels}, {Szczepa{\'n}czyk},
  {Tacca}, {Talukder}, {Tanner}, {T{\'a}pai}, {Tarabrin}, {Taracchini},
  {Taylor}, {Theeg}, {Thirugnanasambandam}, {Thomas}, {Thomas}, {Thomas},
  {Thorne}, {Thorne}, {Thrane}, {Tiwari}, {Tiwari}, {Tokmakov}, {Tomlinson},
  {Tonelli}, {Torres}, {Torrie}, {T{\"o}yr{\"a}}, {Travasso}, {Traylor},
  {Trifir{\`o}}, {Tringali}, {Trozzo}, {Tse}, {Turconi}, {Tuyenbayev},
  {Ugolini}, {Unnikrishnan}, {Urban}, {Usman}, {Vahlbruch}, {Vajente},
  {Valdes}, {Vallisneri}, {van Bakel}, {van Beuzekom}, {van den Brand}, {Van
  Den Broeck}, {Vander-Hyde}, {van der Schaaf}, {van Heijningen}, {van Veggel},
  {Vardaro}, {Vass}, {Vas{\'u}th}, {Vaulin}, {Vecchio}, {Vedovato}, {Veitch},
  {Veitch}, {Venkateswara}, {Verkindt}, {Vetrano}, {Vicer{\'e}}, {Vinciguerra},
  {Vine}, {Vinet}, {Vitale}, {Vo}, {Vocca}, {Vorvick}, {Voss}, {Vousden},
  {Vyatchanin}, {Wade}, {Wade}, {Wade}, {Waldman}, {Walker}, {Wallace},
  {Walsh}, {Wang}, {Wang}, {Wang}, {Wang}, {Wang}, {Ward}, {Ward}, {Warner},
  {Was}, {Weaver}, {Wei}, {Weinert}, {Weinstein}, {Weiss}, {Welborn}, {Wen},
  {We{\ss}els}, {Westphal}, {Wette}, {Whelan}, {Whitcomb}, {White}, {Whiting},
  {Wiesner}, {Wilkinson}, {Willems}, {Williams}, {Williams}, {Williamson},
  {Willis}, {Willke}, {Wimmer}, {Winkelmann}, {Winkler}, {Wipf}, {Wiseman},
  {Wittel}, {Woan}, {Worden}, {Wright}, {Wu}, {Yablon}, {Yakushin}, {Yam},
  {Yamamoto}, {Yancey}, {Yap}, {Yu}, {Yvert}, {Zadro{\.Z}ny}, {Zangrando},
  {Zanolin}, {Zendri}, {Zevin}, {Zhang}, {Zhang}, {Zhang}, {Zhang}, {Zhao},
  {Zhou}, {Zhou}, {Zhu}, {Zucker}, {Zuraw}, {Zweizig}, {LIGO Scientific
  Collaboration}, \& {Virgo Collaboration}}]{abbott2016observation}
{Abbott}, B.~P., {Abbott}, R., {Abbott}, T.~D., {et~al.} 2016, \prl, 116,
  061102, \dodoi{10.1103/PhysRevLett.116.061102}

\bibitem[{{Abbott} {et~al.}(2017){Abbott}, {Abbott}, {Abbott}, {Acernese},
  {Ackley}, {Adams}, {Adams}, {Addesso}, {Adhikari}, {Adya}, {Affeldt},
  {Afrough}, {Agarwal}, {Agathos}, {Agatsuma}, {Aggarwal}, {Aguiar}, {Aiello},
  {Ain}, {Ajith}, {Allen}, {Allen}, {Allocca}, {Altin}, {Amato}, {Ananyeva},
  {Anderson}, {Anderson}, {Angelova}, {Antier}, {Appert}, {Arai}, {Araya},
  {Areeda}, {Arnaud}, {Arun}, {Ascenzi}, {Ashton}, {Ast}, {Aston}, {Astone},
  {Atallah}, {Aufmuth}, {Aulbert}, {AultONeal}, {Austin}, {Avila-Alvarez},
  {Babak}, {Bacon}, {Bader}, {Bae}, {Bailes}, {Baker}, {Baldaccini},
  {Ballardin}, {Ballmer}, {Banagiri}, {Barayoga}, {Barclay}, {Barish},
  {Barker}, {Barkett}, {Barone}, {Barr}, {Barsotti}, {Barsuglia}, {Barta},
  {Barthelmy}, {Bartlett}, {Bartos}, {Bassiri}, {Basti}, {Batch}, {Bawaj},
  {Bayley}, {Bazzan}, {B{\'e}csy}, {Beer}, {Bejger}, {Belahcene}, {Bell},
  {Berger}, {Bergmann}, {Bernuzzi}, {Bero}, {Berry}, {Bersanetti}, {Bertolini},
  {Betzwieser}, {Bhagwat}, {Bhandare}, {Bilenko}, {Billingsley}, {Billman},
  {Birch}, {Birney}, {Birnholtz}, {Biscans}, {Biscoveanu}, {Bisht}, {Bitossi},
  {Biwer}, {Bizouard}, {Blackburn}, {Blackman}, {Blair}, {Blair}, {Blair},
  {Bloemen}, {Bock}, {Bode}, {Boer}, {Bogaert}, {Bohe}, {Bondu}, {Bonilla},
  {Bonnand}, {Boom}, {Bork}, {Boschi}, {Bose}, {Bossie}, {Bouffanais}, {Bozzi},
  {Bradaschia}, {Brady}, {Branchesi}, {Brau}, {Briant}, {Brillet}, {Brinkmann},
  {Brisson}, {Brockill}, {Broida}, {Brooks}, {Brown}, {Brown}, {Brunett},
  {Buchanan}, {Buikema}, {Bulik}, {Bulten}, {Buonanno}, {Buskulic}, {Buy},
  {Byer}, {Cabero}, {Cadonati}, {Cagnoli}, {Cahillane}, {Calder{\'o}n
  Bustillo}, {Callister}, {Calloni}, {Camp}, {Canepa}, {Canizares}, {Cannon},
  {Cao}, {Cao}, {Capano}, {Capocasa}, {Carbognani}, {Caride}, {Carney},
  {Carullo}, {Casanueva Diaz}, {Casentini}, {Caudill}, {Cavagli{\`a}},
  {Cavalier}, {Cavalieri}, {Cella}, {Cepeda}, {Cerd{\'a}-Dur{\'a}n},
  {Cerretani}, {Cesarini}, {Chamberlin}, {Chan}, {Chao}, {Charlton}, {Chase},
  {Chassande-Mottin}, {Chatterjee}, {Chatziioannou}, {Cheeseboro}, {Chen},
  {Chen}, {Chen}, {Cheng}, {Chia}, {Chincarini}, {Chiummo}, {Chmiel}, {Cho},
  {Cho}, {Chow}, {Christensen}, {Chu}, {Chua}, {Chua}, {Chung}, {Chung},
  {Ciani}, {Ciolfi}, {Cirelli}, {Cirone}, {Clara}, {Clark}, {Clearwater},
  {Cleva}, {Cocchieri}, {Coccia}, {Cohadon}, {Cohen}, {Colla}, {Collette},
  {Cominsky}, {Constancio}, {Conti}, {Cooper}, {Corban}, {Corbitt},
  {Cordero-Carri{\'o}n}, {Corley}, {Cornish}, {Corsi}, {Cortese}, {Costa},
  {Coughlin}, {Coughlin}, {Coulon}, {Countryman}, {Couvares}, {Covas}, {Cowan},
  {Coward}, {Cowart}, {Coyne}, {Coyne}, {Creighton}, {Creighton}, {Cripe},
  {Crowder}, {Cullen}, {Cumming}, {Cunningham}, {Cuoco}, {Dal Canton},
  {D{\'a}lya}, {Danilishin}, {D'Antonio}, {Danzmann}, {Dasgupta}, {Da Silva
  Costa}, {Dattilo}, {Dave}, {Davier}, {Davis}, {Daw}, {Day}, {De}, {DeBra},
  {Degallaix}, {De Laurentis}, {Del{\'e}glise}, {Del Pozzo}, {Demos}, {Denker},
  {Dent}, {De Pietri}, {Dergachev}, {De Rosa}, {DeRosa}, {De Rossi}, {DeSalvo},
  {de Varona}, {Devenson}, {Dhurandhar}, {D{\'\i}az}, {Dietrich}, {Di Fiore},
  {Di Giovanni}, {Di Girolamo}, {Di Lieto}, {Di Pace}, {Di Palma}, {Di Renzo},
  {Doctor}, {Dolique}, {Donovan}, {Dooley}, {Doravari}, {Dorrington},
  {Douglas}, {Dovale {\'A}lvarez}, {Downes}, {Drago}, {Dreissigacker},
  {Driggers}, {Du}, {Ducrot}, {Dudi}, {Dupej}, {Dwyer}, {Edo}, {Edwards},
  {Effler}, {Eggenstein}, {Ehrens}, {Eichholz}, {Eikenberry}, {Eisenstein},
  {Essick}, {Estevez}, {Etienne}, {Etzel}, {Evans}, {Evans}, {Factourovich},
  {Fafone}, {Fair}, {Fairhurst}, {Fan}, {Farinon}, {Farr}, {Farr},
  {Fauchon-Jones}, {Favata}, {Fays}, {Fee}, {Fehrmann}, {Feicht}, {Fejer},
  {Fernandez-Galiana}, {Ferrante}, {Ferreira}, {Ferrini}, {Fidecaro},
  {Finstad}, {Fiori}, {Fiorucci}, {Fishbach}, {Fisher}, {Fitz-Axen},
  {Flaminio}, {Fletcher}, {Fong}, {Font}, {Forsyth}, {Forsyth}, {Fournier},
  {Frasca}, {Frasconi}, {Frei}, {Freise}, {Frey}, {Frey}, {Fries}, {Fritschel},
  {Frolov}, {Fulda}, {Fyffe}, {Gabbard}, {Gadre}, {Gaebel}, {Gair},
  {Gammaitoni}, {Ganija}, {Gaonkar}, {Garcia-Quiros}, {Garufi}, {Gateley},
  {Gaudio}, {Gaur}, {Gayathri}, {Gehrels}, {Gemme}, {Genin}, {Gennai},
  {George}, {George}, {Gergely}, {Germain}, {Ghonge}, {Ghosh}, {Ghosh},
  {Ghosh}, {Giaime}, {Giardina}, {Giazotto}, {Gill}, {Glover}, {Goetz},
  {Goetz}, {Gomes}, {Goncharov}, {Gonz{\'a}lez}, {Gonzalez Castro},
  {Gopakumar}, {Gorodetsky}, {Gossan}, {Gosselin}, {Gouaty}, {Grado}, {Graef},
  {Granata}, {Grant}, {Gras}, {Gray}, {Greco}, {Green}, {Gretarsson}, {Groot},
  {Grote}, {Grunewald}, {Gruning}, {Guidi}, {Guo}, {Gupta}, {Gupta}, {Gushwa},
  {Gustafson}, {Gustafson}, {Halim}, {Hall}, {Hall}, {Hamilton}, {Hammond},
  {Haney}, {Hanke}, {Hanks}, {Hanna}, {Hannam}, {Hannuksela}, {Hanson},
  {Hardwick}, {Harms}, {Harry}, {Harry}, {Hart}, {Haster}, {Haughian}, {Healy},
  {Heidmann}, {Heintze}, {Heitmann}, {Hello}, {Hemming}, {Hendry}, {Heng},
  {Hennig}, {Heptonstall}, {Heurs}, {Hild}, {Hinderer}, {Ho}, {Hoak}, {Hofman},
  {Holt}, {Holz}, {Hopkins}, {Horst}, {Hough}, {Houston}, {Howell}, {Hreibi},
  {Hu}, {Huerta}, {Huet}, {Hughey}, {Husa}, {Huttner}, {Huynh-Dinh}, {Indik},
  {Inta}, {Intini}, {Isa}, {Isac}, {Isi}, {Iyer}, {Izumi}, {Jacqmin}, {Jani},
  {Jaranowski}, {Jawahar}, {Jim{\'e}nez-Forteza}, {Johnson},
  {Johnson-McDaniel}, {Jones}, {Jones}, {Jonker}, {Ju}, {Junker}, {Kalaghatgi},
  {Kalogera}, {Kamai}, {Kandhasamy}, {Kang}, {Kanner}, {Kapadia}, {Karki},
  {Karvinen}, {Kasprzack}, {Kastaun}, {Katolik}, {Katsavounidis}, {Katzman},
  {Kaufer}, {Kawabe}, {K{\'e}f{\'e}lian}, {Keitel}, {Kemball}, {Kennedy},
  {Kent}, {Key}, {Khalili}, {Khan}, {Khan}, {Khan}, {Khazanov}, {Kijbunchoo},
  {Kim}, {Kim}, {Kim}, {Kim}, {Kim}, {Kim}, {Kimbrell}, {King}, {King},
  {Kinley-Hanlon}, {Kirchhoff}, {Kissel}, {Kleybolte}, {Klimenko}, {Knowles},
  {Koch}, {Koehlenbeck}, {Koley}, {Kondrashov}, {Kontos}, {Korobko}, {Korth},
  {Kowalska}, {Kozak}, {Kr{\"a}mer}, {Kringel}, {Krishnan}, {Kr{\'o}lak},
  {Kuehn}, {Kumar}, {Kumar}, {Kumar}, {Kuo}, {Kutynia}, {Kwang}, {Lackey},
  {Lai}, {Landry}, {Lang}, {Lange}, {Lantz}, {Lanza}, {Larson},
  {Lartaux-Vollard}, {Lasky}, {Laxen}, {Lazzarini}, {Lazzaro}, {Leaci},
  {Leavey}, {Lee}, {Lee}, {Lee}, {Lee}, {Lee}, {Lehmann}, {Lenon}, {Leon},
  {Leonardi}, {Leroy}, {Letendre}, {Levin}, {Li}, {Linker}, {Littenberg},
  {Liu}, {Liu}, {Lo}, {Lockerbie}, {London}, {Lord}, {Lorenzini}, {Loriette},
  {Lormand}, {Losurdo}, {Lough}, {Lousto}, {Lovelace}, {L{\"u}ck}, {Lumaca},
  {Lundgren}, {Lynch}, {Ma}, {Macas}, {Macfoy}, {Machenschalk}, {MacInnis},
  {Macleod}, {Maga{\~n}a Hernandez}, {Maga{\~n}a-Sandoval}, {Maga{\~n}a
  Zertuche}, {Magee}, {Majorana}, {Maksimovic}, {Man}, {Mandic}, {Mangano},
  {Mansell}, {Manske}, {Mantovani}, {Marchesoni}, {Marion}, {M{\'a}rka},
  {M{\'a}rka}, {Markakis}, {Markosyan}, {Markowitz}, {Maros}, {Marquina},
  {Marsh}, {Martelli}, {Martellini}, {Martin}, {Martin}, {Martynov}, {Marx},
  {Mason}, {Massera}, {Masserot}, {Massinger}, {Masso-Reid}, {Mastrogiovanni},
  {Matas}, {Matichard}, {Matone}, {Mavalvala}, {Mazumder}, {McCarthy},
  {McClelland}, {McCormick}, {McCuller}, {McGuire}, {McIntyre}, {McIver},
  {McManus}, {McNeill}, {McRae}, {McWilliams}, {Meacher}, {Meadors}, {Mehmet},
  {Meidam}, {Mejuto-Villa}, {Melatos}, {Mendell}, {Mercer}, {Merilh},
  {Merzougui}, {Meshkov}, {Messenger}, {Messick}, {Metzdorff}, {Meyers},
  {Miao}, {Michel}, {Middleton}, {Mikhailov}, {Milano}, {Miller}, {Miller},
  {Miller}, {Millhouse}, {Milovich-Goff}, {Minazzoli}, {Minenkov}, {Ming},
  {Mishra}, {Mitra}, {Mitrofanov}, {Mitselmakher}, {Mittleman}, {Moffa},
  {Moggi}, {Mogushi}, {Mohan}, {Mohapatra}, {Molina}, {Montani}, {Moore},
  {Moraru}, {Moreno}, {Morisaki}, {Morriss}, {Mours}, {Mow-Lowry}, {Mueller},
  {Muir}, {Mukherjee}, {Mukherjee}, {Mukherjee}, {Mukund}, {Mullavey}, {Munch},
  {Mu{\~n}iz}, {Muratore}, {Murray}, {Nagar}, {Napier}, {Nardecchia},
  {Naticchioni}, {Nayak}, {Neilson}, {Nelemans}, {Nelson}, {Nery}, {Neunzert},
  {Nevin}, {Newport}, {Newton}, {Ng}, {Nguyen}, {Nguyen}, {Nichols}, {Nielsen},
  {Nissanke}, {Nitz}, {Noack}, {Nocera}, {Nolting}, {North}, {Nuttall},
  {Oberling}, {O'Dea}, {Ogin}, {Oh}, {Oh}, {Ohme}, {Okada}, {Oliver},
  {Oppermann}, {Oram}, {O'Reilly}, {Ormiston}, {Ortega}, {O'Shaughnessy},
  {Ossokine}, {Ottaway}, {Overmier}, {Owen}, {Pace}, {Page}, {Page}, {Pai},
  {Pai}, {Palamos}, {Palashov}, {Palomba}, {Pal-Singh}, {Pan}, {Pan}, {Pang},
  {Pang}, {Pankow}, {Pannarale}, {Pant}, {Paoletti}, {Paoli}, {Papa}, {Parida},
  {Parker}, {Pascucci}, {Pasqualetti}, {Passaquieti}, {Passuello}, {Patil},
  {Patricelli}, {Pearlstone}, {Pedraza}, {Pedurand}, {Pekowsky}, {Pele},
  {Penn}, {Perez}, {Perreca}, {Perri}, {Pfeiffer}, {Phelps}, {Piccinni},
  {Pichot}, {Piergiovanni}, {Pierro}, {Pillant}, {Pinard}, {Pinto}, {Pirello},
  {Pitkin}, {Poe}, {Poggiani}, {Popolizio}, {Porter}, {Post}, {Powell},
  {Prasad}, {Pratt}, {Pratten}, {Predoi}, {Prestegard}, {Prijatelj},
  {Principe}, {Privitera}, {Prix}, {Prodi}, {Prokhorov}, {Puncken}, {Punturo},
  {Puppo}, {P{\"u}rrer}, {Qi}, {Quetschke}, {Quintero}, {Quitzow-James},
  {Raab}, {Rabeling}, {Radkins}, {Raffai}, {Raja}, {Rajan}, {Rajbhandari},
  {Rakhmanov}, {Ramirez}, {Ramos-Buades}, {Rapagnani}, {Raymond}, {Razzano},
  {Read}, {Regimbau}, {Rei}, {Reid}, {Reitze}, {Ren}, {Reyes}, {Ricci},
  {Ricker}, {Rieger}, {Riles}, {Rizzo}, {Robertson}, {Robie}, {Robinet},
  {Rocchi}, {Rolland}, {Rollins}, {Roma}, {Romano}, {Romano}, {Romel}, {Romie},
  {Rosi{\'n}ska}, {Ross}, {Rowan}, {R{\"u}diger}, {Ruggi}, {Rutins}, {Ryan},
  {Sachdev}, {Sadecki}, {Sadeghian}, {Sakellariadou}, {Salconi}, {Saleem},
  {Salemi}, {Samajdar}, {Sammut}, {Sampson}, {Sanchez}, {Sanchez},
  {Sanchis-Gual}, {Sandberg}, {Sanders}, {Sassolas}, {Sathyaprakash},
  {Saulson}, {Sauter}, {Savage}, {Sawadsky}, {Schale}, {Scheel}, {Scheuer},
  {Schmidt}, {Schmidt}, {Schnabel}, {Schofield}, {Sch{\"o}nbeck}, {Schreiber},
  {Schuette}, {Schulte}, {Schutz}, {Schwalbe}, {Scott}, {Scott}, {Seidel},
  {Sellers}, {Sengupta}, {Sentenac}, {Sequino}, {Sergeev}, {Shaddock},
  {Shaffer}, {Shah}, {Shahriar}, {Shaner}, {Shao}, {Shapiro}, {Shawhan},
  {Sheperd}, {Shoemaker}, {Shoemaker}, {Siellez}, {Siemens}, {Sieniawska},
  {Sigg}, {Silva}, {Singer}, {Singh}, {Singhal}, {Sintes}, {Slagmolen},
  {Smith}, {Smith}, {Smith}, {Somala}, {Son}, {Sonnenberg}, {Sorazu},
  {Sorrentino}, {Souradeep}, {Spencer}, {Srivastava}, {Staats}, {Staley},
  {Steinke}, {Steinlechner}, {Steinlechner}, {Steinmeyer}, {Stevenson},
  {Stone}, {Stops}, {Strain}, {Stratta}, {Strigin}, {Strunk}, {Sturani},
  {Stuver}, {Summerscales}, {Sun}, {Sunil}, {Suresh}, {Sutton}, {Swinkels},
  {Szczepa{\'n}czyk}, {Tacca}, {Tait}, {Talbot}, {Talukder}, {Tanner},
  {T{\'a}pai}, {Taracchini}, {Tasson}, {Taylor}, {Taylor}, {Tewari}, {Theeg},
  {Thies}, {Thomas}, {Thomas}, {Thomas}, {Thorne}, {Thorne}, {Thrane},
  {Tiwari}, {Tiwari}, {Tokmakov}, {Toland}, {Tonelli}, {Tornasi},
  {Torres-Forn{\'e}}, {Torrie}, {T{\"o}yr{\"a}}, {Travasso}, {Traylor},
  {Trinastic}, {Tringali}, {Trozzo}, {Tsang}, {Tse}, {Tso}, {Tsukada}, {Tsuna},
  {Tuyenbayev}, {Ueno}, {Ugolini}, {Unnikrishnan}, {Urban}, {Usman},
  {Vahlbruch}, {Vajente}, {Valdes}, {Vallisneri}, {van Bakel}, {van Beuzekom},
  {van den Brand}, {Van Den Broeck}, {Vander-Hyde}, {van der Schaaf}, {van
  Heijningen}, {van Veggel}, {Vardaro}, {Varma}, {Vass}, {Vas{\'u}th},
  {Vecchio}, {Vedovato}, {Veitch}, {Veitch}, {Venkateswara}, {Venugopalan},
  {Verkindt}, {Vetrano}, {Vicer{\'e}}, {Viets}, {Vinciguerra}, {Vine}, {Vinet},
  {Vitale}, {Vo}, {Vocca}, {Vorvick}, {Vyatchanin}, {Wade}, {Wade}, {Wade},
  {Walet}, {Walker}, {Wallace}, {Walsh}, {Wang}, {Wang}, {Wang}, {Wang},
  {Wang}, {Ward}, {Warner}, {Was}, {Watchi}, {Weaver}, {Wei}, {Weinert},
  {Weinstein}, {Weiss}, {Wen}, {Wessel}, {We{\ss}els}, {Westerweck},
  {Westphal}, {Wette}, {Whelan}, {Whitcomb}, {Whiting}, {Whittle}, {Wilken},
  {Williams}, {Williams}, {Williamson}, {Willis}, {Willke}, {Wimmer},
  {Winkler}, {Wipf}, {Wittel}, {Woan}, {Woehler}, {Wofford}, {Wong}, {Worden},
  {Wright}, {Wu}, {Wysocki}, {Xiao}, {Yamamoto}, {Yancey}, {Yang}, {Yap},
  {Yazback}, {Yu}, {Yu}, {Yvert}, {Zadro{\.Z}ny}, {Zanolin}, {Zelenova},
  {Zendri}, {Zevin}, {Zhang}, {Zhang}, {Zhang}, {Zhang}, {Zhao}, {Zhou},
  {Zhou}, {Zhu}, {Zhu}, {Zimmerman}, {Zucker}, {Zweizig}, {LIGO Scientific
  Collaboration}, \& {Virgo Collaboration}}]{abbott2017observation}
---. 2017, \prl, 119, 161101, \dodoi{10.1103/PhysRevLett.119.161101}

\bibitem[{{Abbott} {et~al.}(2021{\natexlab{a}}){Abbott}, {Abbott}, {Abraham},
  {Acernese}, {Ackley}, {Adams}, {Adams}, {Adhikari}, {Adya}, {Affeldt},
  {Agarwal}, {Agathos}, {Agatsuma}, {Aggarwal}, {Aguiar}, {Aiello}, {Ain},
  {Ajith}, {Akutsu}, {Aleman}, {Allen}, {Allocca}, {Altin}, {Amato}, {Anand},
  {Ananyeva}, {Anderson}, {Anderson}, {Ando}, {Angelova}, {Ansoldi}, {Antelis},
  {Antier}, {Appert}, {Arai}, {Arai}, {Arai}, {Araki}, {Araya}, {Araya},
  {Areeda}, {Ar{\`e}ne}, {Aritomi}, {Arnaud}, {Aronson}, {Arun}, {Asada},
  {Asali}, {Ashton}, {Aso}, {Aston}, {Astone}, {Aubin}, {Aufmuth}, {Aultoneal},
  {Austin}, {Babak}, {Badaracco}, {Bader}, {Bae}, {Bae}, {Baer}, {Bagnasco},
  {Bai}, {Baiotti}, {Baird}, {Bajpai}, {Ball}, {Ballardin}, {Ballmer}, {Bals},
  {Balsamo}, {Baltus}, {Banagiri}, {Bankar}, {Bankar}, {Barayoga}, {Barbieri},
  {Barish}, {Barker}, {Barneo}, {Barone}, {Barr}, {Barsotti}, {Barsuglia},
  {Barta}, {Bartlett}, {Barton}, {Bartos}, {Bassiri}, {Basti}, {Bawaj},
  {Bayley}, {Baylor}, {Bazzan}, {B{\'e}csy}, {Bedakihale}, {Bejger},
  {Belahcene}, {Benedetto}, {Beniwal}, {Benjamin}, {Benkel}, {Bennett},
  {Bentley}, {Benyaala}, {Bergamin}, {Berger}, {Bernuzzi}, {Berry},
  {Bersanetti}, {Bertolini}, {Betzwieser}, {Bhandare}, {Bhandari},
  {Bhattacharjee}, {Bhaumik}, {Bidler}, {Bilenko}, {Billingsley}, {Birney},
  {Birnholtz}, {Biscans}, {Bischi}, {Biscoveanu}, {Bisht}, {Biswas}, {Bitossi},
  {Bizouard}, {Blackburn}, {Blackman}, {Blair}, {Blair}, {Blair}, {Bobba},
  {Bode}, {Boer}, {Bogaert}, {Boldrini}, {Bondu}, {Bonilla}, {Bonnand},
  {Booker}, {Boom}, {Bork}, {Boschi}, {Bose}, {Bose}, {Bossilkov}, {Boudart},
  {Bouffanais}, {Bozzi}, {Bradaschia}, {Brady}, {Bramley}, {Branch},
  {Branchesi}, {Brau}, {Breschi}, {Briant}, {Briggs}, {Brillet}, {Brinkmann},
  {Brockill}, {Brooks}, {Brooks}, {Brown}, {Brunett}, {Bruno}, {Bruntz},
  {Bryant}, {Buikema}, {Bulik}, {Bulten}, {Buonanno}, {Buscicchio}, {Buskulic},
  {Byer}, {Cadonati}, {Caesar}, {Cagnoli}, {Cahillane}, {Cain}, {Calder{\'o}n
  Bustillo}, {Callaghan}, {Callister}, {Calloni}, {Camp}, {Canepa},
  {Cannavacciuolo}, {Cannon}, {Cao}, {Cao}, {Cao}, {Capocasa}, {Capote},
  {Carapella}, {Carbognani}, {Carlin}, {Carney}, {Carpinelli}, {Carullo},
  {Carver}, {Casanueva Diaz}, {Casentini}, {Castaldi}, {Caudill},
  {Cavagli{\`a}}, {Cavalier}, {Cavalieri}, {Cella}, {Cerd{\'a}-Dur{\'a}n},
  {Cesarini}, {Chaibi}, {Chakravarti}, {Champion}, {Chan}, {Chan}, {Chan},
  {Chan}, {Chandra}, {Chanial}, {Chao}, {Charlton}, {Chase},
  {Chassande-Mottin}, {Chatterjee}, {Chaturvedi}, {Chatziioannou}, {Chen},
  {Chen}, {Chen}, {Chen}, {Chen}, {Chen}, {Chen}, {Chen}, {Chen}, {Cheng},
  {Cheong}, {Cheung}, {Chia}, {Chiadini}, {Chiang}, {Chierici}, {Chincarini},
  {Chiofalo}, {Chiummo}, {Cho}, {Cho}, {Choate}, {Choudhary}, {Choudhary},
  {Christensen}, {Chu}, {Chu}, {Chu}, {Chua}, {Chung}, {Ciani}, {Ciecielag},
  {Cie{\'s}lar}, {Cifaldi}, {Ciobanu}, {Ciolfi}, {Cipriano}, {Cirone}, {Clara},
  {Clark}, {Clark}, {Clarke}, {Clearwater}, {Clesse}, {Cleva}, {Coccia},
  {Cohadon}, {Cohen}, {Cohen}, {Colleoni}, {Collette}, {Colpi}, {Compton},
  {Constancio}, {Conti}, {Cooper}, {Corban}, {Corbitt}, {Cordero-Carri{\'o}n},
  {Corezzi}, {Corley}, {Cornish}, {Corre}, {Corsi}, {Cortese}, {Costa},
  {Cotesta}, {Coughlin}, {Coughlin}, {Coulon}, {Countryman}, {Cousins},
  {Couvares}, {Covas}, {Coward}, {Cowart}, {Coyne}, {Coyne}, {Creighton},
  {Creighton}, {Criswell}, {Croquette}, {Crowder}, {Cudell}, {Cullen},
  {Cumming}, {Cummings}, {Cuoco}, {Cury{\l}o}, {Dal Canton}, {D{\'a}lya},
  {Dana}, {Daneshgaranbajastani}, {D'Angelo}, {Danilishin}, {D'Antonio},
  {Danzmann}, {Darsow-Fromm}, {Dasgupta}, {Datrier}, {Dattilo}, {Dave},
  {Davier}, {Davies}, {Davis}, {Daw}, {Dean}, {Debra}, {Deenadayalan},
  {Degallaix}, {de Laurentis}, {Del{\'e}glise}, {Del Favero}, {de Lillo}, {de
  Lillo}, {Del Pozzo}, {Demarchi}, {de Matteis}, {D'Emilio}, {Demos}, {Dent},
  {Depasse}, {de Pietri}, {De Rosa}, {de Rossi}, {Desalvo}, {de Simone},
  {Dhurandhar}, {D{\'\i}az}, {Diaz-Ortiz}, {Didio}, {Dietrich}, {di Fiore}, {di
  Fronzo}, {di Giorgio}, {di Giovanni}, {di Girolamo}, {di Lieto}, {Ding}, {di
  Pace}, {di Palma}, {di Renzo}, {Divakarla}, {Dmitriev}, {Doctor},
  {D'Onofrio}, {Donovan}, {Dooley}, {Doravari}, {Dorrington}, {Drago},
  {Driggers}, {Drori}, {Du}, {Ducoin}, {Dupej}, {Durante}, {D'Urso}, {Duverne},
  {Dwyer}, {Easter}, {Ebersold}, {Eddolls}, {Edelman}, {Edo}, {Edy}, {Effler},
  {Eguchi}, {Eichholz}, {Eikenberry}, {Eisenmann}, {Eisenstein}, {Ejlli},
  {Enomoto}, {Errico}, {Essick}, {Estell{\'e}s}, {Estevez}, {Etienne}, {Etzel},
  {Evans}, {Evans}, {Ewing}, {Fafone}, {Fair}, {Fairhurst}, {Fan}, {Farah},
  {Farinon}, {Farr}, {Farr}, {Farrow}, {Fauchon-Jones}, {Favata}, {Fays},
  {Fazio}, {Feicht}, {Fejer}, {Feng}, {Fenyvesi}, {Ferguson},
  {Fernandez-Galiana}, {Ferrante}, {Ferreira}, {Fidecaro}, {Figura}, {Fiori},
  {Fishbach}, {Fisher}, {Fittipaldi}, {Fiumara}, {Flaminio}, {Floden}, {Flynn},
  {Fong}, {Font}, {Fornal}, {Forsyth}, {Franke}, {Frasca}, {Frasconi},
  {Frederick}, {Frei}, {Freise}, {Frey}, {Fritschel}, {Frolov}, {Fronz{\'e}},
  {Fujii}, {Fujikawa}, {Fukunaga}, {Fukushima}, {Fulda}, {Fyffe}, {Gabbard},
  {Gadre}, {Gaebel}, {Gair}, {Gais}, {Galaudage}, {Gamba}, {Ganapathy},
  {Ganguly}, {Gao}, {Gaonkar}, {Garaventa}, {Garc{\'\i}a-N{\'u}{\~n}ez},
  {Garc{\'\i}a-Quir{\'o}s}, {Garufi}, {Gateley}, {Gaudio}, {Gayathri}, {Ge},
  {Gemme}, {Gennai}, {George}, {Gergely}, {Gewecke}, {Ghonge}, {Ghosh},
  {Ghosh}, {Ghosh}, {Ghosh}, {Ghosh}, {Giacomazzo}, {Giacoppo}, {Giaime},
  {Giardina}, {Gibson}, {Gier}, {Giesler}, {Giri}, {Gissi}, {Glanzer},
  {Gleckl}, {Godwin}, {Goetz}, {Goetz}, {Gohlke}, {Goncharov}, {Gonz{\'a}lez},
  {Gopakumar}, {Gosselin}, {Gouaty}, {Grace}, {Grado}, {Granata}, {Granata},
  {Grant}, {Gras}, {Grassia}, {Gray}, {Gray}, {Greco}, {Green}, {Green},
  {Gretarsson}, {Gretarsson}, {Griffith}, {Griffiths}, {Griggs}, {Grignani},
  {Grimaldi}, {Grimes}, {Grimm}, {Grote}, {Grunewald}, {Gruning}, {Guerrero},
  {Guidi}, {Guimaraes}, {Guix{\'e}}, {Gulati}, {Guo}, {Guo}, {Gupta}, {Gupta},
  {Gupta}, {Gustafson}, {Gustafson}, {Guzman}, {Ha}, {Haegel}, {Hagiwara},
  {Haino}, {Halim}, {Hall}, {Hamilton}, {Hammond}, {Han}, {Haney}, {Hanks},
  {Hanna}, {Hannam}, {Hannuksela}, {Hansen}, {Hansen}, {Hanson}, {Harder},
  {Hardwick}, {Haris}, {Harms}, {Harry}, {Harry}, {Hartwig}, {Hasegawa},
  {Haskell}, {Hasskew}, {Haster}, {Hattori}, {Haughian}, {Hayakawa}, {Hayama},
  {Hayes}, {Healy}, {Heidmann}, {Heintze}, {Heinze}, {Heinzel}, {Heitmann},
  {Hellman}, {Hello}, {Helmling-Cornell}, {Hemming}, {Hendry}, {Heng},
  {Hennes}, {Hennig}, {Hennig}, {Hernandez Vivanco}, {Heurs}, {Hild}, {Hill},
  {Himemoto}, {Hinderer}, {Hines}, {Hiranuma}, {Hirata}, {Hirose}, {Ho},
  {Hochheim}, {Hofman}, {Hohmann}, {Holgado}, {Holland}, {Hollows}, {Holmes},
  {Holt}, {Holz}, {Hong}, {Hopkins}, {Hough}, {Howell}, {Hoy}, {Hoyland},
  {Hreibi}, {Hsieh}, {Hsu}, {Huang}, {Huang}, {Huang}, {Huang}, {Huang},
  {Huang}, {H{\"u}bner}, {Huddart}, {Huerta}, {Hughey}, {Hui}, {Hui}, {Husa},
  {Huttner}, {Huxford}, {Huynh-Dinh}, {Ide}, {Idzkowski}, {Iess}, {Ikenoue},
  {Imam}, {Inayoshi}, {Inchauspe}, {Ingram}, {Inoue}, {Intini}, {Ioka}, {Isi},
  {Isleif}, {Ito}, {Itoh}, {Iyer}, {Izumi}, {Jaberianhamedan}, {Jacqmin},
  {Jadhav}, {Jadhav}, {James}, {Jan}, {Jani}, {Janssens}, {Janthalur},
  {Jaranowski}, {Jariwala}, {Jaume}, {Jenkins}, {Jeon}, {Jeunon}, {Jia},
  {Jiang}, {Jin}, {Johns}, {Jones}, {Jones}, {Jones}, {Jones}, {Jones},
  {Jonker}, {Ju}, {Jung}, {Jung}, {Junker}, {Kaihotsu}, {Kajita}, {Kakizaki},
  {Kalaghatgi}, {Kalogera}, {Kamai}, {Kamiizumi}, {Kanda}, {Kandhasamy},
  {Kang}, {Kanner}, {Kao}, {Kapadia}, {Kapasi}, {Karat}, {Karathanasis},
  {Karki}, {Kashyap}, {Kasprzack}, {Kastaun}, {Katsanevas}, {Katsavounidis},
  {Katzman}, {Kaur}, {Kawabe}, {Kawaguchi}, {Kawai}, {Kawasaki},
  {K{\'e}f{\'e}lian}, {Keitel}, {Key}, {Khadka}, {Khalili}, {Khan}, {Khan},
  {Khazanov}, {Khetan}, {Khursheed}, {Kijbunchoo}, {Kim}, {Kim}, {Kim}, {Kim},
  {Kim}, {Kim}, {Kimball}, {Kimura}, {King}, {Kinley-Hanlon}, {Kirchhoff},
  {Kissel}, {Kita}, {Kitazawa}, {Kleybolte}, {Klimenko}, {Knee}, {Knowles},
  {Knyazev}, {Koch}, {Koekoek}, {Kojima}, {Kokeyama}, {Koley}, {Kolitsidou},
  {Kolstein}, {Komori}, {Kondrashov}, {Kong}, {Kontos}, {Koper}, {Korobko},
  {Kotake}, {Kovalam}, {Kozak}, {Kozakai}, {Kozu}, {Kringel}, {Krishnendu},
  {Kr{\'o}lak}, {Kuehn}, {Kuei}, {Kumar}, {Kumar}, {Kumar}, {Kumar}, {Kume},
  {Kuns}, {Kuo}, {Kuo}, {Kuromiya}, {Kuroyanagi}, {Kusayanagi}, {Kwak},
  {Kwang}, {Laghi}, {Lalande}, {Lam}, {Lamberts}, {Landry}, {Landry}, {Lane},
  {Lang}, {Lange}, {Lantz}, {La Rosa}, {Lartaux-Vollard}, {Lasky}, {Laxen},
  {Lazzarini}, {Lazzaro}, {Leaci}, {Leavey}, {Lecoeuche}, {Lee}, {Lee}, {Lee},
  {Lee}, {Lee}, {Lee}, {Lehmann}, {Lema{\^\i}tre}, {Leon}, {Leonardi}, {Leroy},
  {Letendre}, {Levin}, {Leviton}, {Li}, {Li}, {Li}, {Li}, {Li}, {Li}, {Lin},
  {Lin}, {Lin}, {Lin}, {Lin}, {Linde}, {Linker}, {Linley}, {Littenberg}, {Liu},
  {Liu}, {Liu}, {Liu}, {Llorens-Monteagudo}, {Lo}, {Lockwood}, {Lollie},
  {London}, {Longo}, {Lopez}, {Lorenzini}, {Loriette}, {Lormand}, {Losurdo},
  {Lough}, {Lousto}, {Lovelace}, {L{\"u}ck}, {Lumaca}, {Lundgren}, {Luo},
  {Macas}, {Macinnis}, {MacLeod}, {MacMillan}, {Macquet}, {Maga{\~n}a
  Hernandez}, {Maga{\~n}a-Sandoval}, {Magazz{\`u}}, {Magee}, {Maggiore},
  {Majorana}, {Makarem}, {Maksimovic}, {Maliakal}, {Malik}, {Man}, {Mandic},
  {Mangano}, {Mango}, {Mansell}, {Manske}, {Mantovani}, {Mapelli},
  {Marchesoni}, {Marchio}, {Marion}, {Mark}, {M{\'a}rka}, {M{\'a}rka},
  {Markakis}, {Markosyan}, {Markowitz}, {Maros}, {Marquina}, {Marsat},
  {Martelli}, {Martin}, {Martin}, {Martinez}, {Martinez}, {Martinovic},
  {Martynov}, {Marx}, {Masalehdan}, {Mason}, {Massera}, {Masserot},
  {Massinger}, {Masso-Reid}, {Mastrogiovanni}, {Matas}, {Mateu-Lucena},
  {Matichard}, {Matiushechkina}, {Mavalvala}, {McCann}, {McCarthy},
  {McClelland}, {McClincy}, {McCormick}, {McCuller}, {McGhee}, {McGuire},
  {McIsaac}, {McIver}, {McManus}, {McRae}, {McWilliams}, {Meacher}, {Mehmet},
  {Mehta}, {Melatos}, {Melchor}, {Mendell}, {Menendez-Vazquez}, {Menoni},
  {Mercer}, {Mereni}, {Merfeld}, {Merilh}, {Merritt}, {Merzougui}, {Meshkov},
  {Messenger}, {Messick}, {Meyers}, {Meylahn}, {Mhaske}, {Miani}, {Miao},
  {Michaloliakos}, {Michel}, {Michimura}, {Middleton}, {Milano}, {Miller},
  {Millhouse}, {Mills}, {Milotti}, {Milovich-Goff}, {Minazzoli}, {Minenkov},
  {Mio}, {Mir}, {Mishkin}, {Mishra}, {Mishra}, {Mistry}, {Mitra}, {Mitrofanov},
  {Mitselmakher}, {Mittleman}, {Miyakawa}, {Miyamoto}, {Miyazaki}, {Miyo},
  {Miyoki}, {Mo}, {Mogushi}, {Mohapatra}, {Mohite}, {Molina}, {Molina-Ruiz},
  {Mondin}, {Montani}, {Moore}, {Moraru}, {Morawski}, {More}, {Moreno},
  {Moreno}, {Mori}, {Morisaki}, {Moriwaki}, {Mours}, {Mow-Lowry}, {Mozzon},
  {Muciaccia}, {Mukherjee}, {Mukherjee}, {Mukherjee}, {Mukherjee}, {Mukund},
  {Mullavey}, {Munch}, {Mu{\~n}iz}, {Murray}, {Musenich}, {Nadji}, {Nagano},
  {Nagano}, {Nagar}, {Nakamura}, {Nakano}, {Nakano}, {Nakashima}, {Nakayama},
  {Nardecchia}, {Narikawa}, {Naticchioni}, {Nayak}, {Nayak}, {Negishi}, {Neil},
  {Neilson}, {Nelemans}, {Nelson}, {Nery}, {Neunzert}, {Ng}, {Ng}, {Nguyen},
  {Nguyen}, {Nguyen}, {Nguyen Quynh}, {Ni}, {Nichols}, {Nishizawa}, {Nissanke},
  {Nocera}, {Noh}, {Norman}, {North}, {Nozaki}, {Nuttall}, {Oberling},
  {O'Brien}, {Obuchi}, {O'Dell}, {Ogaki}, {Oganesyan}, {Oh}, {Oh}, {Oh},
  {Ohashi}, {Ohishi}, {Ohkawa}, {Ohme}, {Ohta}, {Okada}, {Okutani}, {Okutomi},
  {Olivetto}, {Oohara}, {Ooi}, {Oram}, {O'Reilly}, {Ormiston}, {Ormsby},
  {Ortega}, {O'Shaughnessy}, {O'Shea}, {Oshino}, {Ossokine}, {Osthelder},
  {Otabe}, {Ottaway}, {Overmier}, {Pace}, {Pagano}, {Page}, {Pagliaroli},
  {Pai}, {Pai}, {Palamos}, {Palashov}, {Palomba}, {Pan}, {Panda}, {Pang},
  {Pang}, {Pankow}, {Pannarale}, {Pant}, {Paoletti}, {Paoli}, {Paolone},
  {Parisi}, {Park}, {Parker}, {Pascucci}, {Pasqualetti}, {Passaquieti},
  {Passuello}, {Patel}, {Patricelli}, {Payne}, {Pechsiri}, {Pedraza},
  {Pegoraro}, {Pele}, {Pe{\~n}a Arellano}, {Penn}, {Perego}, {Pereira},
  {Pereira}, {Perez}, {P{\'e}rigois}, {Perreca}, {Perri{\`e}s}, {Petermann},
  {Petterson}, {Pfeiffer}, {Pham}, {Phukon}, {Piccinni}, {Pichot},
  {Piendibene}, {Piergiovanni}, {Pierini}, {Pierro}, {Pillant}, {Pilo},
  {Pinard}, {Pinto}, {Piotrzkowski}, {Piotrzkowski}, {Pirello}, {Pitkin},
  {Placidi}, {Plastino}, {Pluchar}, {Poggiani}, {Polini}, {Pong}, {Ponrathnam},
  {Popolizio}, {Porter}, {Powell}, {Pracchia}, {Pradier}, {Prajapati},
  {Prasai}, {Prasanna}, {Pratten}, {Prestegard}, {Principe}, {Prodi},
  {Prokhorov}, {Prosposito}, {Prudenzi}, {Puecher}, {Punturo}, {Puosi},
  {Puppo}, {P{\"u}rrer}, {Qi}, {Quetschke}, {Quinonez}, {Quitzow-James},
  {Raab}, {Raaijmakers}, {Radkins}, {Radulesco}, {Raffai}, {Rail}, {Raja},
  {Rajan}, {Ramirez}, {Ramirez}, {Ramos-Buades}, {Rana}, {Rapagnani}, {Rapol},
  {Ratto}, {Ray}, {Raymond}, {Raza}, {Razzano}, {Read}, {Rees}, {Regimbau},
  {Rei}, {Reid}, {Reitze}, {Relton}, {Rettegno}, {Ricci}, {Richardson},
  {Richardson}, {Richardson}, {Ricker}, {Riemenschneider}, {Riles}, {Rizzo},
  {Robertson}, {Robie}, {Robinet}, {Rocchi}, {Rocha}, {Rodriguez},
  {Rodriguez-Soto}, {Rolland}, {Rollins}, {Roma}, {Romanelli}, {Romano},
  {Romel}, {Romero}, {Romero-Shaw}, {Romie}, {Rose}, {Rosi{\'n}ska},
  {Rosofsky}, {Ross}, {Rowan}, {Rowlinson}, {Roy}, {Roy}, {Rozza}, {Ruggi},
  {Ryan}, {Sachdev}, {Sadecki}, {Sadiq}, {Sago}, {Saito}, {Saito}, {Sakai},
  {Sakai}, {Sakellariadou}, {Sakuno}, {Salafia}, {Salconi}, {Saleem}, {Salemi},
  {Samajdar}, {Sanchez}, {Sanchez}, {Sanchez}, {Sanchis-Gual}, {Sanders},
  {Sanuy}, {Saravanan}, {Sarin}, {Sassolas}, {Satari}, {Sathyaprakash}, {Sato},
  {Sato}, {Sauter}, {Savage}, {Savant}, {Sawada}, {Sawant}, {Sawant}, {Sayah},
  {Schaetzl}, {Scheel}, {Scheuer}, {Schindler-Tyka}, {Schmidt}, {Schnabel},
  {Schneewind}, {Schofield}, {Sch{\"o}nbeck}, {Schulte}, {Schutz}, {Schwartz},
  {Scott}, {Scott}, {Seglar-Arroyo}, {Seidel}, {Sekiguchi}, {Sekiguchi},
  {Sellers}, {Sengupta}, {Sennett}, {Sentenac}, {Seo}, {Sequino}, {Sergeev},
  {Setyawati}, {Shaffer}, {Shahriar}, {Shams}, {Shao}, {Sharifi}, {Sharma},
  {Sharma}, {Shawhan}, {Shcheblanov}, {Shen}, {Shibagaki}, {Shikauchi},
  {Shimizu}, {Shimoda}, {Shimode}, {Shink}, {Shinkai}, {Shishido}, {Shoda},
  {Shoemaker}, {Shoemaker}, {Shukla}, {Shyamsundar}, {Sieniawska}, {Sigg},
  {Singer}, {Singh}, {Singh}, {Singha}, {Sintes}, {Sipala}, {Skliris},
  {Slagmolen}, {Slaven-Blair}, {Smetana}, {Smith}, {Smith}, {Somala}, {Somiya},
  {Son}, {Soni}, {Soni}, {Sorazu}, {Sordini}, {Sorrentino}, {Sorrentino},
  {Sotani}, {Soulard}, {Souradeep}, {Sowell}, {Spagnuolo}, {Spencer}, {Spera},
  {Srivastava}, {Srivastava}, {Staats}, {Stachie}, {Steer}, {Steinlechner},
  {Steinlechner}, {Stops}, {Stevenson}, {Stover}, {Strain}, {Strang},
  {Stratta}, {Strunk}, {Sturani}, {Stuver}, {S{\"u}dbeck}, {Sudhagar},
  {Sudhir}, {Sugimoto}, {Suh}, {Summerscales}, {Sun}, {Sun}, {Sunil}, {Sur},
  {Suresh}, {Sutton}, {Suzuki}, {Suzuki}, {Swinkels}, {Szczepa{\'n}czyk},
  {Szewczyk}, {Tacca}, {Tagoshi}, {Tait}, {Takahashi}, {Takahashi}, {Takamori},
  {Takano}, {Takeda}, {Takeda}, {Talbot}, {Tanaka}, {Tanaka}, {Tanaka},
  {Tanaka}, {Tanaka}, {Tanasijczuk}, {Tanioka}, {Tanner}, {Tao}, {Tapia},
  {Tapia San Martin}, {Tasson}, {Telada}, {Tenorio}, {Terkowski}, {Test},
  {Thirugnanasambandam}, {Thomas}, {Thomas}, {Thompson}, {Thondapu}, {Thorne},
  {Thrane}, {Tiwari}, {Tiwari}, {Tiwari}, {Toland}, {Tolley}, {Tomaru},
  {Tomigami}, {Tomura}, {Tonelli}, {Torres-Forn{\'e}}, {Torrie}, {Tosta E
  Melo}, {T{\"o}yr{\"a}}, {Trapananti}, {Travasso}, {Traylor}, {Tringali},
  {Tripathee}, {Troiano}, {Trovato}, {Trozzo}, {Trudeau}, {Tsai}, {Tsai},
  {Tsang}, {Tsang}, {Tsao}, {Tse}, {Tso}, {Tsubono}, {Tsuchida}, {Tsukada},
  {Tsuna}, {Tsutsui}, {Tsuzuki}, {Turconi}, {Tuyenbayev}, {Ubhi}, {Uchikata},
  {Uchiyama}, {Udall}, {Ueda}, {Uehara}, {Ueno}, {Ueshima}, {Ugolini},
  {Unnikrishnan}, {Uraguchi}, {Urban}, {Ushiba}, {Usman}, {Utina}, {Vahlbruch},
  {Vajente}, {Vajpeyi}, {Valdes}, {Valentini}, {Valsan}, {van Bakel}, {van
  Beuzekom}, {van den Brand}, {van den Broeck}, {Vander-Hyde}, {van der
  Schaaf}, {van Heijningen}, {Vanosky}, {van Putten}, {Vardaro}, {Vargas},
  {Varma}, {Vas{\'u}th}, {Vecchio}, {Vedovato}, {Veitch}, {Veitch},
  {Venkateswara}, {Venneberg}, {Venugopalan}, {Verkindt}, {Verma}, {Veske},
  {Vetrano}, {Vicer{\'e}}, {Viets}, {Villa-Ortega}, {Vinet}, {Vitale}, {Vo},
  {Vocca}, {von Reis}, {von Wrangel}, {Vorvick}, {Vyatchanin}, {Wade}, {Wade},
  {Wagner}, {Walet}, {Walker}, {Wallace}, {Wallace}, {Walsh}, {Wang}, {Wang},
  {Wang}, {Ward}, {Warner}, {Was}, {Washimi}, {Washington}, {Watchi}, {Weaver},
  {Wei}, {Weinert}, {Weinstein}, {Weiss}, {Weller}, {Wellmann}, {Wen},
  {We{\ss}els}, {Westhouse}, {Wette}, {Whelan}, {White}, {Whiting}, {Whittle},
  {Wilken}, {Williams}, {Williams}, {Williamson}, {Willis}, {Willke}, {Wilson},
  {Winkler}, {Wipf}, {Wlodarczyk}, {Woan}, {Woehler}, {Wofford}, {Wong}, {Wu},
  {Wu}, {Wu}, {Wu}, {Wysocki}, {Xiao}, {Xu}, {Yamada}, {Yamamoto}, {Yamamoto},
  {Yamamoto}, {Yamamoto}, {Yamashita}, {Yamazaki}, {Yang}, {Yang}, {Yang},
  {Yang}, {Yang}, {Yap}, {Yeeles}, {Yelikar}, {Ying}, {Yokogawa}, {Yokoyama},
  {Yokozawa}, {Yoon}, {Yoshioka}, {Yu}, {Yu}, {Yuzurihara}, {Zadro{\.z}ny},
  {Zanolin}, {Zappa}, {Zeidler}, {Zelenova}, {Zendri}, {Zevin}, {Zhan},
  {Zhang}, {Zhang}, {Zhang}, {Zhang}, {Zhang}, {Zhao}, {Zhao}, {Zhao}, {Zhao},
  {Zhou}, {Zhu}, {Zhu}, {Zimmerman}, {Zlochower}, {Zucker}, {Zweizig}, {Ligo
  Scientific Collaboration}, {VIRGO Collaboration}, \& {KAGRA
  Collaboration}}]{abbott2021observation}
{Abbott}, R., {Abbott}, T.~D., {Abraham}, S., {et~al.} 2021{\natexlab{a}},
  \apjl, 915, L5, \dodoi{10.3847/2041-8213/ac082e}

\bibitem[{{Abbott} {et~al.}(2021{\natexlab{b}}){Abbott}, {Abbott}, {Abraham},
  {Acernese}, {Ackley}, {Adams}, {Adams}, {Adhikari}, {Adya}, {Affeldt},
  {Agarwal}, {Agathos}, {Agatsuma}, {Aggarwal}, {Aguiar}, {Aiello}, {Ain},
  {Ajith}, {Aleman}, {Allen}, {Allocca}, {Altin}, {Amato}, {Anand}, {Ananyeva},
  {Anderson}, {Anderson}, {Angelova}, {Ansoldi}, {Antelis}, {Antier}, {Appert},
  {Arai}, {Araya}, {Areeda}, {Ar{\`e}ne}, {Arnaud}, {Aronson}, {Arun}, {Asali},
  {Ashton}, {Aston}, {Astone}, {Aubin}, {Aufmuth}, {AultONeal}, {Austin},
  {Babak}, {Badaracco}, {Bader}, {Bae}, {Baer}, {Bagnasco}, {Bai}, {Baird},
  {Ball}, {Ballardin}, {Ballmer}, {Bals}, {Balsamo}, {Baltus}, {Banagiri},
  {Bankar}, {Bankar}, {Barayoga}, {Barbieri}, {Barish}, {Barker}, {Barneo},
  {Barone}, {Barr}, {Barsotti}, {Barsuglia}, {Barta}, {Bartlett}, {Barton},
  {Bartos}, {Bassiri}, {Basti}, {Bawaj}, {Bayley}, {Baylor}, {Bazzan},
  {B{\'e}csy}, {Bedakihale}, {Bejger}, {Belahcene}, {Benedetto}, {Beniwal},
  {Benjamin}, {Bennett}, {Bentley}, {BenYaala}, {Bergamin}, {Berger},
  {Bernuzzi}, {Berry}, {Bersanetti}, {Bertolini}, {Betzwieser}, {Bhandare},
  {Bhandari}, {Bhattacharjee}, {Bhaumik}, {Bidler}, {Bilenko}, {Billingsley},
  {Birney}, {Birnholtz}, {Biscans}, {Bischi}, {Biscoveanu}, {Bisht}, {Biswas},
  {Bitossi}, {Bizouard}, {Blackburn}, {Blackman}, {Blair}, {Blair}, {Blair},
  {Bobba}, {Bode}, {Boer}, {Bogaert}, {Boldrini}, {Bondu}, {Bonilla},
  {Bonnand}, {Booker}, {Boom}, {Bork}, {Boschi}, {Bose}, {Bose}, {Bossilkov},
  {Boudart}, {Bouffanais}, {Bozzi}, {Bradaschia}, {Brady}, {Bramley}, {Branch},
  {Branchesi}, {Brau}, {Breschi}, {Briant}, {Briggs}, {Brillet}, {Brinkmann},
  {Brockill}, {Brooks}, {Brooks}, {Brown}, {Brunett}, {Bruno}, {Bruntz},
  {Bryant}, {Buikema}, {Bulik}, {Bulten}, {Buonanno}, {Buscicchio}, {Buskulic},
  {Byer}, {Cadonati}, {Caesar}, {Cagnoli}, {Cahillane}, {}, {Bustillo},
  {Callaghan}, {Callister}, {Calloni}, {Camp}, {Canepa}, {Cannavacciuolo},
  {Cannon}, {Cao}, {Cao}, {Capote}, {Carapella}, {Carbognani}, {Carlin},
  {Carney}, {Carpinelli}, {Carullo}, {Carver}, {Diaz}, {Casentini}, {Castaldi},
  {Caudill}, {Cavagli{\`a}}, {Cavalier}, {Cavalieri}, {Cella},
  {Cerd{\'a}-Dur{\'a}n}, {Cesarini}, {Chaibi}, {Chakravarti}, {Champion},
  {Chan}, {Chan}, {Chan}, {Chandra}, {Chanial}, {Chao}, {Charlton}, {Chase},
  {Chassande-Mottin}, {Chatterjee}, {Chaturvedi}, {Chen}, {Chen}, {Chen},
  {Chen}, {Chen}, {Chen}, {Cheng}, {Cheong}, {Cheung}, {Chia}, {Chiadini},
  {Chierici}, {Chincarini}, {Chiofalo}, {Chiummo}, {Cho}, {Cho}, {Choate},
  {Choudhary}, {Choudhary}, {Christensen}, {Chu}, {Chua}, {Chung}, {Ciani},
  {Ciecielag}, {Cie{\'s}lar}, {Cifaldi}, {Ciobanu}, {Ciolfi}, {Cipriano},
  {Cirone}, {Clara}, {Clark}, {Clark}, {Clarke}, {Clearwater}, {Clesse},
  {Cleva}, {Coccia}, {Cohadon}, {Cohen}, {Cohen}, {Colleoni}, {Collette},
  {Colpi}, {Compton}, {Constancio}, {Conti}, {Cooper}, {Corban}, {Corbitt},
  {Cordero-Carri{\'o}n}, {Corezzi}, {Corley}, {Cornish}, {Corre}, {Corsi},
  {Cortese}, {Costa}, {Cotesta}, {Coughlin}, {Coughlin}, {Coulon},
  {Countryman}, {Cousins}, {Couvares}, {Covas}, {Coward}, {Cowart}, {Coyne},
  {Coyne}, {Creighton}, {Creighton}, {Criswell}, {Croquette}, {Crowder},
  {Cudell}, {Cullen}, {Cumming}, {Cummings}, {Cuoco}, {Cury{\l}o}, {Canton},
  {D{\'a}lya}, {Dana}, {DaneshgaranBajastani}, {D'Angelo}, {Danilishin},
  {D'Antonio}, {Danzmann}, {Darsow-Fromm}, {Dasgupta}, {Datrier}, {Dattilo},
  {Dave}, {Davier}, {Davies}, {Davis}, {Daw}, {Dean}, {DeBra}, {Deenadayalan},
  {Degallaix}, {Laurentis}, {Del{\'e}glise}, {Favero}, {Lillo}, {Lillo},
  {Pozzo}, {DeMarchi}, {Matteis}, {D'Emilio}, {Demos}, {Dent}, {Depasse},
  {Pietri}, {Rosa}, {Rossi}, {DeSalvo}, {Simone}, {Dhurandhar}, {D{\'\i}az},
  {Diaz-Ortiz}, {Didio}, {Dietrich}, {Di Fiore}, {Di Fronzo}, {Di Giorgio}, {Di
  Giovanni}, {Di Girolamo}, {Di Lieto}, {Ding}, {Di Pace}, {Di Palma}, {Di
  Renzo}, {Divakarla}, {Dmitriev}, {Doctor}, {D'Onofrio}, {Donovan}, {Dooley},
  {Doravari}, {Dorrington}, {Drago}, {Driggers}, {Drori}, {Du}, {Ducoin},
  {Dupej}, {Durante}, {D'Urso}, {Duverne}, {Dwyer}, {Easter}, {Ebersold},
  {Eddolls}, {Edelman}, {Edo}, {Edy}, {Effler}, {Eichholz}, {Eikenberry},
  {Eisenmann}, {Eisenstein}, {Ejlli}, {Errico}, {Essick}, {Estell{\'e}s},
  {Estevez}, {Etienne}, {Etzel}, {Evans}, {Evans}, {Ewing}, {Ezquiaga},
  {Fafone}, {Fair}, {Fairhurst}, {Fan}, {Farah}, {Farinon}, {Farr}, {Farr},
  {Farrow}, {Fauchon-Jones}, {Favata}, {Fays}, {Fazio}, {Feicht}, {Fejer},
  {Feng}, {Fenyvesi}, {Ferguson}, {Fernandez-Galiana}, {Ferrante}, {Ferreira},
  {Fidecaro}, {Figura}, {Fiori}, {Fishbach}, {Fisher}, {Fittipaldi}, {Fiumara},
  {Flaminio}, {Floden}, {Flynn}, {Fong}, {Font}, {Fornal}, {Forsyth}, {Franke},
  {Frasca}, {Frasconi}, {Frederick}, {Frei}, {Freise}, {Frey}, {Fritschel},
  {Frolov}, {Fronz{\'e}}, {Fulda}, {Fyffe}, {Gabbard}, {Gadre}, {Gaebel},
  {Gair}, {Gais}, {Galaudage}, {Gamba}, {Ganapathy}, {Ganguly}, {Gaonkar},
  {Garaventa}, {Garc{\'\i}a-N{\'u}{\~n}ez}, {Garc{\'\i}a-Quir{\'o}s}, {Garufi},
  {Gateley}, {Gaudio}, {Gayathri}, {Gemme}, {Gennai}, {George}, {Gergely},
  {Gewecke}, {Ghonge}, {Ghosh}, {Ghosh}, {Ghosh}, {Ghosh}, {Ghosh},
  {Giacomazzo}, {Giacoppo}, {Giaime}, {Giardina}, {Gibson}, {Gier}, {Giesler},
  {Giri}, {Gissi}, {Glanzer}, {Gleckl}, {Godwin}, {Goetz}, {Goetz}, {Gohlke},
  {Goncharov}, {Gonz{\'a}lez}, {Gopakumar}, {Gosselin}, {Gouaty}, {Goyal},
  {Grace}, {Grado}, {Granata}, {Granata}, {Grant}, {Gras}, {Grassia}, {Gray},
  {Gray}, {Greco}, {Green}, {Green}, {Gretarsson}, {Gretarsson}, {Griffith},
  {Griffiths}, {Griggs}, {Grignani}, {Grimaldi}, {Grimes}, {Grimm}, {Grote},
  {Grunewald}, {Gruning}, {Guerrero}, {Guidi}, {Guimaraes}, {Guix{\'e}},
  {Gulati}, {Guo}, {Guo}, {Gupta}, {Gupta}, {Gupta}, {Gustafson}, {Gustafson},
  {Guzman}, {Haegel}, {Halim}, {Hall}, {Hamilton}, {Hammond}, {Haney}, {Hanks},
  {Hanna}, {Hannam}, {Hannuksela}, {Hansen}, {Hansen}, {Hanson}, {Harder},
  {Hardwick}, {Haris}, {Harms}, {Harry}, {Harry}, {Hartwig}, {Haskell},
  {Hasskew}, {Haster}, {Haughian}, {Hayes}, {Healy}, {Heidmann}, {Heintze},
  {Heinze}, {Heinzel}, {Heitmann}, {Hellman}, {Hello}, {Helmling-Cornell},
  {Hemming}, {Hendry}, {Heng}, {Hennes}, {Hennig}, {Hennig}, {Vivanco},
  {Heurs}, {Hild}, {Hill}, {Hines}, {Hochheim}, {Hofman}, {Hohmann}, {Holgado},
  {Holland}, {Hollows}, {Holmes}, {Holt}, {Holz}, {Hopkins}, {Hough}, {Howell},
  {Hoy}, {Hoyland}, {Hreibi}, {Hsu}, {Huang}, {H{\"u}bner}, {Huddart},
  {Huerta}, {Hughey}, {Hui}, {Husa}, {Huttner}, {Huxford}, {Huynh-Dinh},
  {Idzkowski}, {Iess}, {Inchauspe}, {Ingram}, {Intini}, {Isi}, {Isleif},
  {Iyer}, {JaberianHamedan}, {Jacqmin}, {Jadhav}, {Jadhav}, {James}, {Jan},
  {Jani}, {Janquart}, {Janssens}, {Janthalur}, {Jaranowski}, {Jariwala},
  {Jaume}, {Jenkins}, {Jeunon}, {Jia}, {Jiang}, {Johns}, {Jones}, {Jones},
  {Jones}, {Jones}, {Jones}, {Jonker}, {Ju}, {Junker}, {Kalaghatgi},
  {Kalogera}, {Kamai}, {Kandhasamy}, {Kang}, {Kanner}, {Kao}, {Kapadia},
  {Kapasi}, {Karat}, {Karathanasis}, {Karki}, {Kashyap}, {Kasprzack},
  {Kastaun}, {Katsanevas}, {Katsavounidis}, {Katzman}, {Kaur}, {Kawabe},
  {K{\'e}f{\'e}lian}, {Keitel}, {Key}, {Khadka}, {Khalili}, {Khan}, {Khan},
  {Khazanov}, {Khetan}, {Khursheed}, {Kijbunchoo}, {Kim}, {Kim}, {Kim}, {Kim},
  {Kim}, {Kimball}, {King}, {Kinley-Hanlon}, {Kirchhoff}, {Kissel},
  {Kleybolte}, {Klimenko}, {Knee}, {Knowles}, {Knyazev}, {Koch}, {Koekoek},
  {Koley}, {Kolitsidou}, {Kolstein}, {Komori}, {Kondrashov}, {Kontos}, {Koper},
  {Korobko}, {Kovalam}, {Kozak}, {Kringel}, {Krishnendu}, {Kr{\'o}lak},
  {Kuehn}, {Kuei}, {Kumar}, {Kumar}, {Kumar}, {Kumar}, {Kuns}, {Kwang},
  {Laghi}, {Lalande}, {Lam}, {Lamberts}, {Landry}, {Lane}, {Lang}, {Lange},
  {Lantz}, {Rosa}, {Lartaux-Vollard}, {Lasky}, {Laxen}, {Lazzarini}, {Lazzaro},
  {Leaci}, {Leavey}, {Lecoeuche}, {Lee}, {Lee}, {Lee}, {Lee}, {Lehmann},
  {Lema{\^\i}tre}, {Leon}, {Leroy}, {Letendre}, {Levin}, {Leviton}, {Li}, {Li},
  {Li}, {Li}, {Li}, {Linde}, {Linker}, {Linley}, {Littenberg}, {Liu}, {Liu},
  {Liu}, {Llorens-Monteagudo}, {Lo}, {Lockwood}, {Lollie}, {London}, {Longo},
  {Lopez}, {Lorenzini}, {Loriette}, {Lormand}, {Losurdo}, {Lough}, {Lousto},
  {Lovelace}, {L{\"u}ck}, {Lumaca}, {Lundgren}, {Macas}, {MacInnis}, {Macleod},
  {MacMillan}, {Macquet}, {Hernandez}, {Maga{\~n}a-Sandoval}, {Magazz{\`u}},
  {Magee}, {Maggiore}, {Majorana}, {Makarem}, {Maksimovic}, {Maliakal},
  {Malik}, {Man}, {Mandic}, {Mangano}, {Mango}, {Mansell}, {Manske},
  {Mantovani}, {Mapelli}, {Marchesoni}, {Marion}, {Mark}, {M{\'a}rka},
  {M{\'a}rka}, {Markakis}, {Markosyan}, {Markowitz}, {Maros}, {Marquina},
  {Marsat}, {Martelli}, {Martin}, {Martin}, {Martinez}, {Martinez},
  {Martinovic}, {Martynov}, {Marx}, {Masalehdan}, {Mason}, {Massera},
  {Masserot}, {Massinger}, {Masso-Reid}, {Mastrogiovanni}, {Matas},
  {Mateu-Lucena}, {Matichard}, {Matiushechkina}, {Mavalvala}, {McCann},
  {McCarthy}, {McClelland}, {McClincy}, {McCormick}, {McCuller}, {McGhee},
  {McGuire}, {McIsaac}, {McIver}, {McManus}, {McRae}, {McWilliams}, {Meacher},
  {Mehmet}, {Mehta}, {Melatos}, {Melchor}, {Mendell}, {Menendez-Vazquez},
  {Menoni}, {Mercer}, {Mereni}, {Merfeld}, {Merilh}, {Merritt}, {Merzougui},
  {Meshkov}, {Messenger}, {Messick}, {Meyers}, {Meylahn}, {Mhaske}, {Miani},
  {Miao}, {Michaloliakos}, {Michel}, {Middleton}, {Milano}, {Miller},
  {Millhouse}, {Mills}, {Milotti}, {Milovich-Goff}, {Minazzoli}, {Minenkov},
  {Mir}, {Mishkin}, {Mishra}, {Mishra}, {Mistry}, {Mitra}, {Mitrofanov},
  {Mitselmakher}, {Mittleman}, {Mo}, {Mogushi}, {Mohapatra}, {Mohite},
  {Molina}, {Molina-Ruiz}, {Mondin}, {Montani}, {Moore}, {Moraru}, {Morawski},
  {More}, {Moreno}, {Moreno}, {Morisaki}, {Mours}, {Mow-Lowry}, {Mozzon},
  {Muciaccia}, {Mukherjee}, {Mukherjee}, {Mukherjee}, {Mukherjee}, {Mukund},
  {Mullavey}, {Munch}, {Mu{\~n}iz}, {Murray}, {Musenich}, {Nadji}, {Nagar},
  {Nardecchia}, {Naticchioni}, {Nayak}, {Nayak}, {Neil}, {Neilson}, {Nelemans},
  {Nelson}, {Nery}, {Neunzert}, {Ng}, {Ng}, {Nguyen}, {Nguyen}, {Nguyen},
  {Nichols}, {Nissanke}, {Nocera}, {Noh}, {Norman}, {North}, {Nuttall},
  {Oberling}, {O'Brien}, {O'Dell}, {Oganesyan}, {Oh}, {Oh}, {Ohme}, {Ohta},
  {Okada}, {Olivetto}, {Oram}, {O'Reilly}, {Ormiston}, {Ormsby}, {Ortega},
  {O'Shaughnessy}, {O'Shea}, {Ossokine}, {Osthelder}, {Ottaway}, {Overmier},
  {Pace}, {Pagano}, {Page}, {Pagliaroli}, {Pai}, {Pai}, {Palamos}, {Palashov},
  {Palomba}, {Panda}, {Pang}, {Pankow}, {Pannarale}, {Pant}, {Paoletti},
  {Paoli}, {Paolone}, {Parker}, {Pascucci}, {Pasqualetti}, {Passaquieti},
  {Passuello}, {Patel}, {Patricelli}, {Payne}, {Pechsiri}, {Pedraza},
  {Pegoraro}, {Pele}, {Penn}, {Perego}, {Pereira}, {Pereira}, {Perez},
  {P{\'e}rigois}, {Perreca}, {Perri{\`e}s}, {Petermann}, {Petterson},
  {Pfeiffer}, {Pham}, {Phukon}, {Piccinni}, {Pichot}, {Piendibene},
  {Piergiovanni}, {Pierini}, {Pierro}, {Pillant}, {Pilo}, {Pinard}, {Pinto},
  {Piotrzkowski}, {Piotrzkowski}, {Pirello}, {Pitkin}, {Placidi}, {Plastino},
  {Pluchar}, {Poggiani}, {Polini}, {Pong}, {Ponrathnam}, {Popolizio}, {Porter},
  {Powell}, {Pracchia}, {Pradier}, {Prajapati}, {Prasai}, {Prasanna},
  {Pratten}, {Prestegard}, {Principe}, {Prodi}, {Prokhorov}, {Prosposito},
  {Prudenzi}, {Puecher}, {Punturo}, {Puosi}, {Puppo}, {P{\"u}rrer}, {Qi},
  {Quetschke}, {Quinonez}, {Quitzow-James}, {Raab}, {Raaijmakers}, {Radkins},
  {Radulesco}, {Raffai}, {Rail}, {Raja}, {Rajan}, {Ramirez}, {Ramirez},
  {Ramos-Buades}, {Rana}, {Rapagnani}, {Rapol}, {Ratto}, {Raymond}, {Raza},
  {Razzano}, {Read}, {Rees}, {Regimbau}, {Rei}, {Reid}, {Reitze}, {Relton},
  {Rettegno}, {Ricci}, {Richardson}, {Richardson}, {Richardson}, {Ricker},
  {Riemenschneider}, {Riles}, {Rizzo}, {Robertson}, {Robie}, {Robinet},
  {Rocchi}, {Rocha}, {Rodriguez}, {Rodriguez-Soto}, {Rolland}, {Rollins},
  {Roma}, {Romanelli}, {Romano}, {Romel}, {Romero}, {Romero-Shaw}, {Romie},
  {Rose}, {Rosi{\'n}ska}, {Rosofsky}, {Ross}, {Rowan}, {Rowlinson}, {Roy},
  {Roy}, {Rozza}, {Ruggi}, {Ryan}, {Sachdev}, {Sadecki}, {Sadiq},
  {Sakellariadou}, {Salafia}, {Salconi}, {Saleem}, {Salemi}, {Samajdar},
  {Sanchez}, {Sanchez}, {Sanchez}, {Sanchis-Gual}, {Sanders}, {Sanuy},
  {Saravanan}, {Sarin}, {Sassolas}, {Satari}, {Sathyaprakash}, {Sauter},
  {Savage}, {Savant}, {Sawant}, {Sawant}, {Sayah}, {Schaetzl}, {Scheel},
  {Scheuer}, {Schindler-Tyka}, {Schmidt}, {Schnabel}, {Schneewind},
  {Schofield}, {Sch{\"o}nbeck}, {Schulte}, {Schutz}, {Schwartz}, {Scott},
  {Scott}, {Seglar-Arroyo}, {Seidel}, {Sellers}, {Sengupta}, {Sennett},
  {Sentenac}, {Seo}, {Sequino}, {Sergeev}, {Setyawati}, {Shaffer}, {Shahriar},
  {Shams}, {Sharifi}, {Sharma}, {Sharma}, {Shawhan}, {Shcheblanov}, {Shen},
  {Shikauchi}, {Shink}, {Shoemaker}, {Shoemaker}, {Shukla}, {ShyamSundar},
  {Sieniawska}, {Sigg}, {Singer}, {Singh}, {Singh}, {Singha}, {Sintes},
  {Sipala}, {Skliris}, {Slagmolen}, {Slaven-Blair}, {Smetana}, {Smith},
  {Smith}, {Somala}, {Son}, {Soni}, {Soni}, {Sorazu}, {Sordini}, {Sorrentino},
  {Sorrentino}, {Soulard}, {Souradeep}, {Sowell}, {Spagnuolo}, {Spencer},
  {Spera}, {Srivastava}, {Srivastava}, {Staats}, {Stachie}, {Steer},
  {Steinlechner}, {Steinlechner}, {Stops}, {Stover}, {Strain}, {Strang},
  {Stratta}, {Strunk}, {Sturani}, {Stuver}, {S{\"u}dbeck}, {Sudhagar},
  {Sudhir}, {Suh}, {Summerscales}, {Sun}, {Sun}, {Sunil}, {Sur}, {Suresh},
  {Sutton}, {Swinkels}, {Szczepa{\'n}czyk}, {Szewczyk}, {Tacca}, {Tait},
  {Talbot}, {Tanasijczuk}, {Tanner}, {Tao}, {Tapia}, {San Martin}, {Tasson},
  {Tenorio}, {Terkowski}, {Test}, {Thirugnanasambandam}, {Thomas}, {Thomas},
  {Thompson}, {Thondapu}, {Thorne}, {Thrane}, {Tiwari}, {Tiwari}, {Tiwari},
  {Toland}, {Tolley}, {Tonelli}, {Torres-Forn{\'e}}, {Torrie}, {e Melo},
  {T{\"o}yr{\"a}}, {Trapananti}, {Travasso}, {Traylor}, {Tringali},
  {Tripathee}, {Troiano}, {Trovato}, {Trudeau}, {Tsai}, {Tsai}, {Tsang}, {Tse},
  {Tso}, {Tsukada}, {Tsuna}, {Tsutsui}, {Turconi}, {Ubhi}, {Udall}, {Ueno},
  {Ugolini}, {Unnikrishnan}, {Urban}, {Usman}, {Utina}, {Vahlbruch}, {Vajente},
  {Vajpeyi}, {Valdes}, {Valentini}, {Valsan}, {van Bakel}, {van Beuzekom}, {van
  den Brand}, {Van Den Broeck}, {Vander-Hyde}, {van der Schaaf}, {van
  Heijningen}, {Vanosky}, {Vardaro}, {Vargas}, {Varma}, {Vas{\'u}th},
  {Vecchio}, {Vedovato}, {Veitch}, {Veitch}, {Venkateswara}, {Venneberg},
  {Venugopalan}, {Verkindt}, {Verma}, {Veske}, {Vetrano}, {Vicer{\'e}},
  {Viets}, {Villa-Ortega}, {Vinet}, {Vitale}, {Vo}, {Vocca}, {von Reis}, {von
  Wrangel}, {Vorvick}, {Vyatchanin}, {Wade}, {Wade}, {Wagner}, {Walet},
  {Walker}, {Wallace}, {Wallace}, {Walsh}, {Wang}, {Wang}, {Ward}, {Warner},
  {Was}, {Washington}, {Watchi}, {Weaver}, {Wei}, {Weinert}, {Weinstein},
  {Weiss}, {Weller}, {Wellmann}, {Wen}, {We{\ss}els}, {Westhouse}, {Wette},
  {Whelan}, {White}, {Whiting}, {Whittle}, {Wilken}, {Williams}, {Williams},
  {Williamson}, {Willis}, {Willke}, {Wilson}, {Winkler}, {Wipf}, {Wlodarczyk},
  {Woan}, {Woehler}, {Wofford}, {Wong}, {Wright}, {Wu}, {Wysocki}, {Xiao},
  {Yamamoto}, {Yang}, {Yang}, {Yang}, {Yang}, {Yap}, {Yeeles}, {Yelikar},
  {Yeung}, {Ying}, {Yoon}, {Yu}, {Yu}, {Zadro{\.z}ny}, {Zanolin}, {Zelenova},
  {Zendri}, {Zevin}, {Zhang}, {Zhang}, {Zhang}, {Zhang}, {Zhao}, {Zhao},
  {Zhao}, {Zhou}, {Zhu}, {Zimmerman}, {Zucker}, {Zweizig}, {LIGO Scientific
  Collaboration}, \& {Virgo Collaboration}}]{abbott2021search}
---. 2021{\natexlab{b}}, \apj, 923, 14, \dodoi{10.3847/1538-4357/ac23db}

\bibitem[{{Acernese} {et~al.}(2015){Acernese}, {Agathos}, {Agatsuma}, {Aisa},
  {Allemandou}, {Allocca}, {Amarni}, {Astone}, {Balestri}, {Ballardin},
  {Barone}, {Baronick}, {Barsuglia}, {Basti}, {Basti}, {Bauer}, {Bavigadda},
  {Bejger}, {Beker}, {Belczynski}, {Bersanetti}, {Bertolini}, {Bitossi},
  {Bizouard}, {Bloemen}, {Blom}, {Boer}, {Bogaert}, {Bondi}, {Bondu},
  {Bonelli}, {Bonnand}, {Boschi}, {Bosi}, {Bouedo}, {Bradaschia}, {Branchesi},
  {Briant}, {Brillet}, {Brisson}, {Bulik}, {Bulten}, {Buskulic}, {Buy},
  {Cagnoli}, {Calloni}, {Campeggi}, {Canuel}, {Carbognani}, {Cavalier},
  {Cavalieri}, {Cella}, {Cesarini}, {Mottin}, {Chincarini}, {Chiummo}, {Chua},
  {Cleva}, {Coccia}, {Cohadon}, {Colla}, {Colombini}, {Conte}, {Coulon},
  {Cuoco}, {Dalmaz}, {D'Antonio}, {Dattilo}, {Davier}, {Day}, {Debreczeni},
  {Degallaix}, {Del{\'e}glise}, {Pozzo}, {Dereli}, {Rosa}, {Fiore}, {Lieto},
  {Virgilio}, {Doets}, {Dolique}, {Drago}, {Ducrot}, {Endr{\H{o}}czi},
  {Fafone}, {Farinon}, {Ferrante}, {Ferrini}, {Fidecaro}, {Fiori}, {Flaminio},
  {Fournier}, {Franco}, {Frasca}, {Frasconi}, {Gammaitoni}, {Garufi},
  {Gaspard}, {Gatto}, {Gemme}, {Gendre}, {Genin}, {Gennai}, {Ghosh},
  {Giacobone}, {Giazotto}, {Gouaty}, {Granata}, {Greco}, {Groot}, {Guidi},
  {Harms}, {Heidmann}, {Heitmann}, {Hello}, {Hemming}, {Hennes}, {Hofman},
  {Jaranowski}, {Jonker}, {Kasprzack}, {K{\'e}f{\'e}lian}, {Kowalska}, {Kraan},
  {Kr{\'o}lak}, {Kutynia}, {Lazzaro}, {Leonardi}, {Leroy}, {Letendre}, {Li},
  {Lieunard}, {Lorenzini}, {Loriette}, {Losurdo}, {Magazz{\`u}}, {Majorana},
  {Maksimovic}, {Malvezzi}, {Man}, {Mangano}, {Mantovani}, {Marchesoni},
  {Marion}, {Marque}, {Martelli}, {Martellini}, {Masserot}, {Meacher},
  {Meidam}, {Mezzani}, {Michel}, {Milano}, {Minenkov}, {Moggi}, {Mohan},
  {Montani}, {Morgado}, {Mours}, {Mul}, {Nagy}, {Nardecchia}, {Naticchioni},
  {Nelemans}, {Neri}, {Neri}, {Nocera}, {Pacaud}, {Palomba}, {Paoletti},
  {Paoli}, {Pasqualetti}, {Passaquieti}, {Passuello}, {Perciballi}, {Petit},
  {Pichot}, {Piergiovanni}, {Pillant}, {Piluso}, {Pinard}, {Poggiani},
  {Prijatelj}, {Prodi}, {Punturo}, {Puppo}, {Rabeling}, {R{\'a}cz},
  {Rapagnani}, {Razzano}, {Re}, {Regimbau}, {Ricci}, {Robinet}, {Rocchi},
  {Rolland}, {Romano}, {Rosi{\'n}ska}, {Ruggi}, {Saracco}, {Sassolas},
  {Schimmel}, {Sentenac}, {Sequino}, {Shah}, {Siellez}, {Straniero},
  {Swinkels}, {Tacca}, {Tonelli}, {Travasso}, {Turconi}, {Vajente}, {van
  Bakel}, {van Beuzekom}, {van den Brand}, {Van Den Broeck}, {van der Sluys},
  {van Heijningen}, {Vas{\'u}th}, {Vedovato}, {Veitch}, {Verkindt}, {Vetrano},
  {Vicer{\'e}}, {Vinet}, {Visser}, {Vocca}, {Ward}, {Was}, {Wei}, {Yvert},
  {{\.z}ny}, \& {Zendri}}]{virgo2015advanced}
{Acernese}, F., {Agathos}, M., {Agatsuma}, K., {et~al.} 2015, Classical and
  Quantum Gravity, 32, 024001, \dodoi{10.1088/0264-9381/32/2/024001}

\bibitem[{{Asada} \& {Kasai}(2000)}]{Asada2000}
{Asada}, H., \& {Kasai}, M. 2000, Progress of Theoretical Physics, 104, 95,
  \dodoi{10.1143/PTP.104.95}

\bibitem[{{Baraldo} {et~al.}(1999){Baraldo}, {Hosoya}, \&
  {Nakamura}}]{baraldo1999gravitationally}
{Baraldo}, C., {Hosoya}, A., \& {Nakamura}, T.~T. 1999, \prd, 59, 083001,
  \dodoi{10.1103/PhysRevD.59.083001}

\bibitem[{{Beach} \& {Lovelace}(1997)}]{beach1997diffraction}
{Beach}, T.~L., \& {Lovelace}, R. V.~E. 1997, Radio Science, 32, 913,
  \dodoi{10.1029/97RS00063}

\bibitem[{{Benson} \& {Cooke}(1979)}]{benson1979high}
{Benson}, J.~R., \& {Cooke}, J.~H. 1979, \apj, 227, 360, \dodoi{10.1086/156739}

\bibitem[{{Berry}(2021)}]{berry2021scalings}
{Berry}, M.~V. 2021, Journal of Optics, 23, 065604,
  \dodoi{10.1088/2040-8986/abfee1}

\bibitem[{Berry \& Upstill(1980)}]{berryupstill1980}
Berry, M.~V., \& Upstill, C. 1980, Progress in Optics, 18, 257

\bibitem[{{Blandford} \& {Narayan}(1986)}]{blandford1986fermats}
{Blandford}, R., \& {Narayan}, R. 1986, \apj, 310, 568, \dodoi{10.1086/164709}

\bibitem[{{Bontz} \& {Haugan}(1981)}]{bontz1981diffraction}
{Bontz}, R.~J., \& {Haugan}, M.~P. 1981, \apss, 78, 199,
  \dodoi{10.1007/BF00654034}

\bibitem[{{Boyle} \& {Pen}(2012)}]{boyle2012pulsar}
{Boyle}, L., \& {Pen}, U.-L. 2012, \prd, 86, 124028,
  \dodoi{10.1103/PhysRevD.86.124028}

\bibitem[{{Chang} \& {Refsdal}(1979)}]{chang1979flux}
{Chang}, K., \& {Refsdal}, S. 1979, \nat, 282, 561, \dodoi{10.1038/282561a0}

\bibitem[{{Cheung} {et~al.}(2021){Cheung}, {Gais}, {Hannuksela}, \&
  {Li}}]{cheung2021stellar}
{Cheung}, M. H.~Y., {Gais}, J., {Hannuksela}, O.~A., \& {Li}, T. G.~F. 2021,
  \mnras, 503, 3326, \dodoi{10.1093/mnras/stab579}

\bibitem[{{CHIME/FRB Collaboration} {et~al.}(2021){CHIME/FRB Collaboration},
  {Amiri}, {Andersen}, {Bandura}, {Berger}, {Bhardwaj}, {Boyce}, {Boyle},
  {Brar}, {Breitman}, {Cassanelli}, {Chawla}, {Chen}, {Cliche}, {Cook},
  {Cubranic}, {Curtin}, {Deng}, {Dobbs}, {Dong}, {Eadie}, {Fandino}, {Fonseca},
  {Gaensler}, {Giri}, {Good}, {Halpern}, {Hill}, {Hinshaw}, {Josephy},
  {Kaczmarek}, {Kader}, {Kania}, {Kaspi}, {Landecker}, {Lang}, {Leung}, {Li},
  {Lin}, {Masui}, {McKinven}, {Mena-Parra}, {Merryfield}, {Meyers}, {Michilli},
  {Milutinovic}, {Mirhosseini}, {M{\"u}nchmeyer}, {Naidu}, {Newburgh}, {Ng},
  {Patel}, {Pen}, {Petroff}, {Pinsonneault-Marotte}, {Pleunis},
  {Rafiei-Ravandi}, {Rahman}, {Ransom}, {Renard}, {Sanghavi}, {Scholz}, {Shaw},
  {Shin}, {Siegel}, {Sikora}, {Singh}, {Smith}, {Stairs}, {Tan}, {Tendulkar},
  {Vanderlinde}, {Wang}, {Wulf}, \& {Zwaniga}}]{chimefrb2021first}
{CHIME/FRB Collaboration}, {Amiri}, M., {Andersen}, B.~C., {et~al.} 2021,
  \apjs, 257, 59, \dodoi{10.3847/1538-4365/ac33ab}

\bibitem[{{Chwolson}(1924)}]{chwolson1924uber}
{Chwolson}, O. 1924, Astronomische Nachrichten, 221, 329,
  \dodoi{10.1002/asna.19242212003}

\bibitem[{{Congedo} {et~al.}(2006){Congedo}, {de Paolis}, {Longo}, {Nucita},
  {Vetrugno}, \& {Qadir}}]{congedo2006gravitational}
{Congedo}, G., {de Paolis}, F., {Longo}, P., {et~al.} 2006, International
  Journal of Modern Physics D, 15, 1937, \dodoi{10.1142/S0218271806009248}

\bibitem[{{Connor} \& {Ravi}(2022)}]{connor2022stellar}
{Connor}, L., \& {Ravi}, V. 2022, arXiv e-prints, arXiv:2206.14310.
\newblock \doarXiv{2206.14310}

\bibitem[{{Cremonese} {et~al.}(2021){Cremonese}, {Ezquiaga}, \&
  {Salzano}}]{cremonese2021breaking}
{Cremonese}, P., {Ezquiaga}, J.~M., \& {Salzano}, V. 2021, \prd, 104, 023503,
  \dodoi{10.1103/PhysRevD.104.023503}

\bibitem[{{Dai}(2014)}]{Dai2014cmb}
{Dai}, L. 2014, \prl, 112, 041303, \dodoi{10.1103/PhysRevLett.112.041303}

\bibitem[{{Dai} {et~al.}(2018){Dai}, {Li}, {Zackay}, {Mao}, \&
  {Lu}}]{dai2018detecting}
{Dai}, L., {Li}, S.-S., {Zackay}, B., {Mao}, S., \& {Lu}, Y. 2018, \prd, 98,
  104029, \dodoi{10.1103/PhysRevD.98.104029}

\bibitem[{{Dai} \& {Miralda-Escud{\'e}}(2020)}]{dai2020gravitational}
{Dai}, L., \& {Miralda-Escud{\'e}}, J. 2020, \aj, 159, 49,
  \dodoi{10.3847/1538-3881/ab5e83}

\bibitem[{{Dai} \& {Venumadhav}(2017)}]{dai2017on}
{Dai}, L., \& {Venumadhav}, T. 2017, arXiv e-prints, arXiv:1702.04724.
\newblock \doarXiv{1702.04724}

\bibitem[{{Dai} {et~al.}(2020){Dai}, {Zackay}, {Venumadhav}, {Roulet}, \&
  {Zaldarriaga}}]{dai2020search}
{Dai}, L., {Zackay}, B., {Venumadhav}, T., {Roulet}, J., \& {Zaldarriaga}, M.
  2020, arXiv e-prints, arXiv:2007.12709.
\newblock \doarXiv{2007.12709}

\bibitem[{Davydov(2013)}]{davydov2013quantum}
Davydov, A.~S. 2013, Quantum mechanics: international series in natural
  philosophy, Vol.~1 (Elsevier)

\bibitem[{{De Paolis} {et~al.}(2002){De Paolis}, {Ingrosso}, {Nucita}, \&
  {Qadir}}]{depaolis2002note}
{De Paolis}, F., {Ingrosso}, G., {Nucita}, A.~A., \& {Qadir}, A. 2002, \aap,
  394, 749, \dodoi{10.1051/0004-6361:20021258}

\bibitem[{{Deguchi} \& {Watson}(1986{\natexlab{a}})}]{deguchi1986wave}
{Deguchi}, S., \& {Watson}, W.~D. 1986{\natexlab{a}}, \prd, 34, 1708,
  \dodoi{10.1103/PhysRevD.34.1708}

\bibitem[{{Deguchi} \& {Watson}(1986{\natexlab{b}})}]{deguchi1986diffraction}
---. 1986{\natexlab{b}}, \apj, 307, 30, \dodoi{10.1086/164389}

\bibitem[{{Diego}(2019)}]{diego2019universe}
{Diego}, J.~M. 2019, \aap, 625, A84, \dodoi{10.1051/0004-6361/201833670}

\bibitem[{{Doyle} \& {Carico}(2009)}]{doyle2009quantum}
{Doyle}, L.~R., \& {Carico}, D.~P. 2009, The Open Astronomy Journal, 2, 63,
  \dodoi{10.2174/1874381100902010063}

\bibitem[{{Engeli} \& {Saha}(2022)}]{engeli2022optical}
{Engeli}, S., \& {Saha}, P. 2022, \mnras, 516, 4679,
  \dodoi{10.1093/mnras/stac2522}

\bibitem[{{Ezquiaga} {et~al.}(2021){Ezquiaga}, {Holz}, {Hu}, {Lagos}, \&
  {Wald}}]{ezquiaga2021phase}
{Ezquiaga}, J.~M., {Holz}, D.~E., {Hu}, W., {Lagos}, M., \& {Wald}, R.~M. 2021,
  \prd, 103, 064047, \dodoi{10.1103/PhysRevD.103.064047}

\bibitem[{{Ezquiaga} {et~al.}(2020){Ezquiaga}, {Hu}, \&
  {Lagos}}]{ezquiaga2020apparent}
{Ezquiaga}, J.~M., {Hu}, W., \& {Lagos}, M. 2020, \prd, 102, 023531,
  \dodoi{10.1103/PhysRevD.102.023531}

\bibitem[{{Falco} {et~al.}(1985){Falco}, {Gorenstein}, \&
  {Shapiro}}]{falco1985model}
{Falco}, E.~E., {Gorenstein}, M.~V., \& {Shapiro}, I.~I. 1985, \apjl, 289, L1,
  \dodoi{10.1086/184422}

\bibitem[{{Feldbrugge} {et~al.}(2019){Feldbrugge}, {Pen}, \&
  {Turok}}]{feldbrugge2019oscillatory}
{Feldbrugge}, J., {Pen}, U.-L., \& {Turok}, N. 2019, arXiv e-prints,
  arXiv:1909.04632.
\newblock \doarXiv{1909.04632}

\bibitem[{{Feldbrugge} \& {Turok}(2020)}]{feldbrugge2020gravitational}
{Feldbrugge}, J., \& {Turok}, N. 2020, arXiv e-prints, arXiv:2008.01154.
\newblock \doarXiv{2008.01154}

\bibitem[{{Gil Choi} {et~al.}(2021){Gil Choi}, {Park}, \&
  {Jung}}]{han2021small}
{Gil Choi}, H., {Park}, C., \& {Jung}, S. 2021, arXiv e-prints,
  arXiv:2103.08618.
\newblock \doarXiv{2103.08618}

\bibitem[{{Goldreich} \& {Keeley}(1972)}]{goldreich1972astrophysical}
{Goldreich}, P., \& {Keeley}, D.~A. 1972, \apj, 174, 517,
  \dodoi{10.1086/151514}

\bibitem[{{Goodman} {et~al.}(1987){Goodman}, {Romani}, {Blandford}, \&
  {Narayan}}]{goodman1987effects}
{Goodman}, J.~J., {Romani}, R.~W., {Blandford}, R.~D., \& {Narayan}, R. 1987,
  \mnras, 229, 73, \dodoi{10.1093/mnras/229.1.73}

\bibitem[{{Gosselin} {et~al.}(2007){Gosselin}, {B{\'e}rard}, \&
  {Mohrbach}}]{gosselin2007spin}
{Gosselin}, P., {B{\'e}rard}, A., \& {Mohrbach}, H. 2007, \prd, 75, 084035,
  \dodoi{10.1103/PhysRevD.75.084035}

\bibitem[{{Gould}(1992)}]{gould1992femtolensing}
{Gould}, A. 1992, \apjl, 386, L5, \dodoi{10.1086/186279}

\bibitem[{{Grillo} \& {Cordes}(2018)}]{grillo2018wave}
{Grillo}, G., \& {Cordes}, J. 2018, arXiv e-prints, arXiv:1810.09058.
\newblock \doarXiv{1810.09058}

\bibitem[{{Guo} \& {Lu}(2022)}]{guo2022probing}
{Guo}, X., \& {Lu}, Y. 2022, \prd, 106, 023018,
  \dodoi{10.1103/PhysRevD.106.023018}

\bibitem[{{Hallinan} {et~al.}(2019){Hallinan}, {Ravi}, {Weinreb}, {Kocz},
  {Huang}, {Woody}, {Lamb}, {D'Addario}, {Catha}, {Law}, {Kulkarni}, {Phinney},
  {Eastwood}, {Bouman}, {McLaughlin}, {Ransom}, {Siemens}, {Cordes}, {Lynch},
  {Kaplan}, {Brazier}, {Bhatnagar}, {Myers}, {Walter}, \&
  {Gaensler}}]{hallinan2019dsa}
{Hallinan}, G., {Ravi}, V., {Weinreb}, S., {et~al.} 2019, in Bulletin of the
  American Astronomical Society, Vol.~51, 255,
  \dodoi{10.48550/arXiv.1907.07648}

\bibitem[{{Hannuksela} {et~al.}(2019){Hannuksela}, {Haris}, {Ng}, {Kumar},
  {Mehta}, {Keitel}, {Li}, \& {Ajith}}]{hannuksela2019search}
{Hannuksela}, O.~A., {Haris}, K., {Ng}, K.~K.~Y., {et~al.} 2019, \apjl, 874,
  L2, \dodoi{10.3847/2041-8213/ab0c0f}

\bibitem[{{Haris} {et~al.}(2018){Haris}, {Mehta}, {Kumar}, {Venumadhav}, \&
  {Ajith}}]{haris2018identifying}
{Haris}, K., {Mehta}, A.~K., {Kumar}, S., {Venumadhav}, T., \& {Ajith}, P.
  2018, arXiv e-prints, arXiv:1807.07062.
\newblock \doarXiv{1807.07062}

\bibitem[{{Herlt} \& {Stephani}(1976)}]{herlt1976wave}
{Herlt}, E., \& {Stephani}, H. 1976, International Journal of Theoretical
  Physics, 15, 45, \dodoi{10.1007/BF01807086}

\bibitem[{{Hewitt} {et~al.}(1988){Hewitt}, {Turner}, {Schneider}, {Burke}, \&
  {Langston}}]{hewitt1988unusual}
{Hewitt}, J.~N., {Turner}, E.~L., {Schneider}, D.~P., {Burke}, B.~F., \&
  {Langston}, G.~I. 1988, \nat, 333, 537, \dodoi{10.1038/333537a0}

\bibitem[{{Inamori} \& {Suyama}(2021)}]{inamori2021universal}
{Inamori}, M., \& {Suyama}, T. 2021, \apjl, 918, L30,
  \dodoi{10.3847/2041-8213/ac2142}

\bibitem[{{Ishihara} {et~al.}(1988){Ishihara}, {Takahashi}, \&
  {Tomimatsu}}]{ishihara1988gravitational}
{Ishihara}, H., {Takahashi}, M., \& {Tomimatsu}, A. 1988, \prd, 38, 472,
  \dodoi{10.1103/PhysRevD.38.472}

\bibitem[{{Itoh} {et~al.}(2009){Itoh}, {Futamase}, \&
  {Hattori}}]{itoh2009method}
{Itoh}, Y., {Futamase}, T., \& {Hattori}, M. 2009, \prd, 80, 044009,
  \dodoi{10.1103/PhysRevD.80.044009}

\bibitem[{{Jacques} {et~al.}(2007){Jacques}, {Wu}, {Grosshans}, {Treussart},
  {Grangier}, {Aspect}, \& {Roch}}]{jacques2007experimental}
{Jacques}, V., {Wu}, E., {Grosshans}, F., {et~al.} 2007, Science, 315, 966,
  \dodoi{10.1126/science.1136303}

\bibitem[{{Janquart} {et~al.}(2021){Janquart}, {Seo}, {Hannuksela}, {Li}, \&
  {Van Den Broeck}}]{janquart2021on}
{Janquart}, J., {Seo}, E., {Hannuksela}, O.~A., {Li}, T. G.~F., \& {Van Den
  Broeck}, C. 2021, \apjl, 923, L1, \dodoi{10.3847/2041-8213/ac3bcf}

\bibitem[{{Jaroszynski} \& {Paczynski}(1995)}]{jaroszynski1995diffraction}
{Jaroszynski}, M., \& {Paczynski}, B. 1995, \apj, 455, 443,
  \dodoi{10.1086/176593}

\bibitem[{{Johnston} {et~al.}(1997){Johnston}, {Gaume}, {Wilson}, {Nguyen}, \&
  {Nedoluha}}]{johnston1997apparent}
{Johnston}, K.~J., {Gaume}, R.~A., {Wilson}, T.~L., {Nguyen}, H.~A., \&
  {Nedoluha}, G.~E. 1997, \apj, 490, 758, \dodoi{10.1086/304891}

\bibitem[{{Jow} {et~al.}(2020){Jow}, {Foreman}, {Pen}, \& {Zhu}}]{jow2020wave}
{Jow}, D.~L., {Foreman}, S., {Pen}, U.-L., \& {Zhu}, W. 2020, \mnras, 497,
  4956, \dodoi{10.1093/mnras/staa2230}

\bibitem[{{Jow} {et~al.}(2021){Jow}, {Lin}, {Tyhurst}, \&
  {Pen}}]{jow2021imaginary}
{Jow}, D.~L., {Lin}, F.~X., {Tyhurst}, E., \& {Pen}, U.-L. 2021, \mnras, 507,
  5390, \dodoi{10.1093/mnras/stab2337}

\bibitem[{{Jow} {et~al.}(2022){Jow}, {Pen}, \& {Feldbrugge}}]{jow2022regimes}
{Jow}, D.~L., {Pen}, U.-L., \& {Feldbrugge}, J. 2022, arXiv e-prints,
  arXiv:2204.12004.
\newblock \doarXiv{2204.12004}

\bibitem[{{Kader} {et~al.}(2022){Kader}, {Leung}, {Dobbs}, {Masui}, {Michilli},
  {Mena-Parra}, {Mckinven}, {Ng}, {Bandura}, {Bhardwaj}, {Brar}, {Cassanelli},
  {Chawla}, {Dong}, {Good}, {Kaspi}, {Lanman}, {Lin}, {Meyers}, {Pearlman},
  {Pen}, {Petroff}, {Pleunis}, {Rafiei-Ravandi}, {Rahman}, {Sanghavi},
  {Scholz}, {Shin}, {Siegel}, {Smith}, {Stairs}, {Tendulkar}, {Vanderlinde}, \&
  {Wulf}}]{kader2022high}
{Kader}, Z., {Leung}, C., {Dobbs}, M., {et~al.} 2022, arXiv e-prints,
  arXiv:2204.06014.
\newblock \doarXiv{2204.06014}

\bibitem[{{Kagra Collaboration} {et~al.}(2019){Kagra Collaboration}, {Akutsu},
  {Ando}, {Arai}, {Arai}, {Araki}, {Araya}, {Aritomi}, {Asada}, {Aso},
  {Atsuta}, {Awai}, {Bae}, {Baiotti}, {Barton}, {Cannon}, {Capocasa}, {Chen},
  {Chiu}, {Cho}, {Chu}, {Craig}, {Creus}, {Doi}, {Eda}, {Enomoto}, {Flaminio},
  {Fujii}, {Fujimoto}, {Fukunaga}, {Fukushima}, {Furuhata}, {Haino},
  {Hasegawa}, {Hashino}, {Hayama}, {Hirobayashi}, {Hirose}, {Hsieh}, {Huang},
  {Ikenoue}, {Inoue}, {Ioka}, {Itoh}, {Izumi}, {Kaji}, {Kajita}, {Kakizaki},
  {Kamiizumi}, {Kanbara}, {Kanda}, {Kanemura}, {Kaneyama}, {Kang}, {Kasuya},
  {Kataoka}, {Kawai}, {Kawamura}, {Kawasaki}, {Kim}, {Kim}, {Kim}, {Kim},
  {Kim}, {Kimura}, {Kinugawa}, {Kirii}, {Kitaoka}, {Kitazawa}, {Kojima},
  {Kokeyama}, {Komori}, {Kong}, {Kotake}, {Kozu}, {Kumar}, {Kuo}, {Kuroyanagi},
  {Lee}, {Lee}, {Lee}, {Leonardi}, {Lin}, {Lin}, {Liu}, {Liu}, {Majorana},
  {Mano}, {Marchio}, {Matsui}, {Matsushima}, {Michimura}, {Mio}, {Miyakawa},
  {Miyamoto}, {Miyamoto}, {Miyo}, {Miyoki}, {Morii}, {Morisaki}, {Moriwaki},
  {Morozumi}, {Musha}, {Nagano}, {Nagano}, {Nakamura}, {Nakamura}, {Nakano},
  {Nakano}, {Nakao}, {Narikawa}, {Naticchioni}, {Nguyen Quynh}, {Ni},
  {Nishizawa}, {Obuchi}, {Ochi}, {Oh}, {Oh}, {Ohashi}, {Ohishi}, {Ohkawa},
  {Okutomi}, {Ono}, {Oohara}, {Ooi}, {Pan}, {Park}, {Pe{\~n}a Arellano},
  {Pinto}, {Sago}, {Saijo}, {Saitou}, {Saito}, {Sakai}, {Sakai}, {Sakai},
  {Sasai}, {Sasaki}, {Sasaki}, {Sato}, {Sato}, {Sato}, {Sekiguchi}, {Seto},
  {Shibata}, {Shimoda}, {Shinkai}, {Shishido}, {Shoda}, {Somiya}, {Son},
  {Suemasa}, {Suzuki}, {Suzuki}, {Tagoshi}, {Tahara}, {Takahashi}, {Takahashi},
  {Takamori}, {Takeda}, {Tanaka}, {Tanaka}, {Tanaka}, {Tanioka}, {Tapia San
  Martin}, {Tatsumi}, {Tomaru}, {Tomura}, {Travasso}, {Tsubono}, {Tsuchida},
  {Uchikata}, {Uchiyama}, {Uehara}, {Ueki}, {Ueno}, {Uraguchi}, {Ushiba}, {van
  Putten}, {Vocca}, {Wada}, {Wakamatsu}, {Watanabe}, {Xu}, {Yamada},
  {Yamamoto}, {Yamamoto}, {Yamamoto}, {Yamamoto}, {Yamamoto}, {Yokogawa},
  {Yokoyama}, {Yokozawa}, {Yoon}, {Yoshioka}, {Yuzurihara}, {Zeidler}, \&
  {Zhu}}]{kagra2019advanced}
{Kagra Collaboration}, {Akutsu}, T., {Ando}, M., {et~al.} 2019, Nature
  Astronomy, 3, 35, \dodoi{10.1038/s41550-018-0658-y}

\bibitem[{{Katz} {et~al.}(2018){Katz}, {Kopp}, {Sibiryakov}, \&
  {Xue}}]{katz2018femtolensing}
{Katz}, A., {Kopp}, J., {Sibiryakov}, S., \& {Xue}, W. 2018, \jcap, 2018, 005,
  \dodoi{10.1088/1475-7516/2018/12/005}

\bibitem[{{Katz} {et~al.}(2020){Katz}, {Kopp}, {Sibiryakov}, \&
  {Xue}}]{katz2020looking}
---. 2020, \mnras, 496, 564, \dodoi{10.1093/mnras/staa1497}

\bibitem[{{Kelly} {et~al.}(2015){Kelly}, {Rodney}, {Treu}, {Foley}, {Brammer},
  {Schmidt}, {Zitrin}, {Sonnenfeld}, {Strolger}, {Graur}, {Filippenko}, {Jha},
  {Riess}, {Bradac}, {Weiner}, {Scolnic}, {Malkan}, {von der Linden}, {Trenti},
  {Hjorth}, {Gavazzi}, {Fontana}, {Merten}, {McCully}, {Jones}, {Postman},
  {Dressler}, {Patel}, {Cenko}, {Graham}, \& {Tucker}}]{kelly2015multiple}
{Kelly}, P.~L., {Rodney}, S.~A., {Treu}, T., {et~al.} 2015, Science, 347, 1123,
  \dodoi{10.1126/science.aaa3350}

\bibitem[{{Khangulyan} {et~al.}(2022){Khangulyan}, {Barkov}, \&
  {Popov}}]{khangulyan2022fast}
{Khangulyan}, D., {Barkov}, M.~V., \& {Popov}, S.~B. 2022, \apj, 927, 2,
  \dodoi{10.3847/1538-4357/ac4bdf}

\bibitem[{{Lazio} {et~al.}(2008){Lazio}, {Ojha}, {Fey}, {Kedziora-Chudczer},
  {Cordes}, {Jauncey}, \& {Lovell}}]{lazio2008angular}
{Lazio}, T. J.~W., {Ojha}, R., {Fey}, A.~L., {et~al.} 2008, \apj, 672, 115,
  \dodoi{10.1086/520572}

\bibitem[{{Leung} {et~al.}(2018){Leung}, {Brown}, {Nguyen}, {Friedman},
  {Kaiser}, \& {Gallicchio}}]{leung2018astronomical}
{Leung}, C., {Brown}, A., {Nguyen}, H., {et~al.} 2018, \pra, 97, 042120,
  \dodoi{10.1103/PhysRevA.97.042120}

\bibitem[{{Leung} {et~al.}(2022){Leung}, {Kader}, {Masui}, {Dobbs}, {Michilli},
  {Mena-Parra}, {Mckinven}, {Ng}, {Bandura}, {Bhardwaj}, {Brar}, {Cassanelli},
  {Chawla}, {Dong}, {Good}, {Kaspi}, {Lanman}, {Lin}, {Meyers}, {Pearlman},
  {Pen}, {Petroff}, {Pleunis}, {Rafiei-Ravandi}, {Rahman}, {Sanghavi},
  {Scholz}, {Shin}, {Siegel}, {Smith}, {Stairs}, {Tendulkar}, \&
  {Vanderlinde}}]{leung2022constraining}
{Leung}, C., {Kader}, Z., {Masui}, K.~W., {et~al.} 2022, arXiv e-prints,
  arXiv:2204.06001.
\newblock \doarXiv{2204.06001}

\bibitem[{{Li} {et~al.}(2019){Li}, {Lin}, {Main}, {Pen}, {van Kerkwijk}, \&
  {Yang}}]{Li2019bwbiref}
{Li}, D., {Lin}, F.~X., {Main}, R., {et~al.} 2019, \mnras, 484, 5723,
  \dodoi{10.1093/mnras/stz374}

\bibitem[{{Li} {et~al.}(2022){Li}, {Qiao}, {Zhao}, \& {Er}}]{Li2022gfr}
{Li}, Z., {Qiao}, J., {Zhao}, W., \& {Er}, X. 2022, \jcap, 2022, 095,
  \dodoi{10.1088/1475-7516/2022/10/095}

\bibitem[{{Lightman} {et~al.}(1975){Lightman}, {Press}, {Price}, \&
  {Teukolsky}}]{lightman1975problem}
{Lightman}, A.~P., {Press}, W.~H., {Price}, R.~H., \& {Teukolsky}, S.~A. 1975,
  {Problem Book in Relativity and Gravitation}

\bibitem[{{LIGO Scientific Collaboration} {et~al.}(2015){LIGO Scientific
  Collaboration}, {Aasi}, {Abbott}, {Abbott}, {Abbott}, {Abernathy}, {Ackley},
  {Adams}, {Adams}, {Addesso}, {Adhikari}, {Adya}, {Affeldt}, {Aggarwal},
  {Aguiar}, {Ain}, {Ajith}, {Alemic}, {Allen}, {Amariutei}, {Anderson},
  {Anderson}, {Arai}, {Araya}, {Arceneaux}, {Areeda}, {Ashton}, {Ast}, {Aston},
  {Aufmuth}, {Aulbert}, {Aylott}, {Babak}, {Baker}, {Ballmer}, {Barayoga},
  {Barbet}, {Barclay}, {Barish}, {Barker}, {Barr}, {Barsotti}, {Bartlett},
  {Barton}, {Bartos}, {Bassiri}, {Batch}, {Baune}, {Behnke}, {Bell}, {Bell},
  {Benacquista}, {Bergman}, {Bergmann}, {Berry}, {Betzwieser}, {Bhagwat},
  {Bhandare}, {Bilenko}, {Billingsley}, {Birch}, {Biscans}, {Biwer},
  {Blackburn}, {Blackburn}, {Blair}, {Blair}, {Bock}, {Bodiya}, {Bojtos},
  {Bond}, {Bork}, {Born}, {Bose}, {Brady}, {Braginsky}, {Brau}, {Bridges},
  {Brinkmann}, {Brooks}, {Brown}, {Brown}, {Brown}, {Buchman}, {Buikema},
  {Buonanno}, {Cadonati}, {Calder{\'o}n Bustillo}, {Camp}, {Cannon}, {Cao},
  {Capano}, {Caride}, {Caudill}, {Cavagli{\`a}}, {Cepeda}, {Chakraborty},
  {Chalermsongsak}, {Chamberlin}, {Chao}, {Charlton}, {Chen}, {Cho}, {Cho},
  {Chow}, {Christensen}, {Chu}, {Chung}, {Ciani}, {Clara}, {Clark}, {Collette},
  {Cominsky}, {Constancio}, {Cook}, {Corbitt}, {Cornish}, {Corsi}, {Costa},
  {Coughlin}, {Countryman}, {Couvares}, {Coward}, {Cowart}, {Coyne}, {Coyne},
  {Craig}, {Creighton}, {Creighton}, {Cripe}, {Crowder}, {Cumming},
  {Cunningham}, {Cutler}, {Dahl}, {Dal Canton}, {Damjanic}, {Danilishin},
  {Danzmann}, {Dartez}, {Dave}, {Daveloza}, {Davies}, {Daw}, {DeBra}, {Del
  Pozzo}, {Denker}, {Dent}, {Dergachev}, {DeRosa}, {DeSalvo}, {Dhurandhar},
  {D{\textasciiacute}{\i}az}, {Di Palma}, {Dojcinoski}, {Dominguez}, {Donovan},
  {Dooley}, {Doravari}, {Douglas}, {Downes}, {Driggers}, {Du}, {Dwyer},
  {Eberle}, {Edo}, {Edwards}, {Edwards}, {Effler}, {Eggenstein}, {Ehrens},
  {Eichholz}, {Eikenberry}, {Essick}, {Etzel}, {Evans}, {Evans},
  {Factourovich}, {Fairhurst}, {Fan}, {Fang}, {Farr}, {Farr}, {Favata}, {Fays},
  {Fehrmann}, {Fejer}, {Feldbaum}, {Ferreira}, {Fisher}, {Frei}, {Freise},
  {Frey}, {Fricke}, {Fritschel}, {Frolov}, {Fuentes-Tapia}, {Fulda}, {Fyffe},
  {Gair}, {Gaonkar}, {Gehrels}, {Gergely}, {Giaime}, {Giardina}, {Gleason},
  {Goetz}, {Goetz}, {Gondan}, {Gonz{\'a}lez}, {Gordon}, {Gorodetsky}, {Gossan},
  {Go{\ss}ler}, {Gr{\"a}f}, {Graff}, {Grant}, {Gras}, {Gray}, {Greenhalgh},
  {Gretarsson}, {Grote}, {Grunewald}, {Guido}, {Guo}, {Gushwa}, {Gustafson},
  {Gustafson}, {Hacker}, {Hall}, {Hammond}, {Hanke}, {Hanks}, {Hanna},
  {Hannam}, {Hanson}, {Hardwick}, {Harry}, {Harry}, {Hart}, {Hartman},
  {Haster}, {Haughian}, {Hee}, {Heintze}, {Heinzel}, {Hendry}, {Heng},
  {Heptonstall}, {Heurs}, {Hewitson}, {Hild}, {Hoak}, {Hodge}, {Hollitt},
  {Holt}, {Hopkins}, {Hosken}, {Hough}, {Houston}, {Howell}, {Hu}, {Huerta},
  {Hughey}, {Husa}, {Huttner}, {Huynh}, {Huynh-Dinh}, {Idrisy}, {Indik},
  {Ingram}, {Inta}, {Islas}, {Isler}, {Isogai}, {Iyer}, {Izumi}, {Jacobson},
  {Jang}, {Jawahar}, {Ji}, {Jim{\'e}nez-Forteza}, {Johnson}, {Jones}, {Jones},
  {Ju}, {Haris}, {Kalogera}, {Kandhasamy}, {Kang}, {Kanner}, {Katsavounidis},
  {Katzman}, {Kaufer}, {Kaufer}, {Kaur}, {Kawabe}, {Kawazoe}, {Keiser},
  {Keitel}, {Kelley}, {Kells}, {Keppel}, {Key}, {Khalaidovski}, {Khalili},
  {Khazanov}, {Kim}, {Kim}, {Kim}, {Kim}, {Kim}, {King}, {King}, {Kinzel},
  {Kissel}, {Klimenko}, {Kline}, {Koehlenbeck}, {Kokeyama}, {Kondrashov},
  {Korobko}, {Korth}, {Kozak}, {Kringel}, {Krishnan}, {Krueger}, {Kuehn},
  {Kumar}, {Kumar}, {Kuo}, {Landry}, {Lantz}, {Larson}, {Lasky}, {Lazzarini},
  {Lazzaro}, {Le}, {Leaci}, {Leavey}, {Lebigot}, {Lee}, {Lee}, {Lee}, {Leong},
  {Levin}, {Levine}, {Lewis}, {Li}, {Libbrecht}, {Libson}, {Lin}, {Littenberg},
  {Lockerbie}, {Lockett}, {Logue}, {Lombardi}, {Lormand}, {Lough}, {Lubinski},
  {L{\"u}ck}, {Lundgren}, {Lynch}, {Ma}, {Macarthur}, {MacDonald},
  {Machenschalk}, {MacInnis}, {Macleod}, {Maga{\~n}a-Sandoval}, {Magee},
  {Mageswaran}, {Maglione}, {Mailand}, {Mandel}, {Mandic}, {Mangano},
  {Mansell}, {M{\'a}rka}, {M{\'a}rka}, {Markosyan}, {Maros}, {Martin},
  {Martin}, {Martynov}, {Marx}, {Mason}, {Massinger}, {Matichard}, {Matone},
  {Mavalvala}, {Mazumder}, {Mazzolo}, {McCarthy}, {McClelland}, {McCormick},
  {McGuire}, {McIntyre}, {McIver}, {McLin}, {McWilliams}, {Meadors},
  {Meinders}, {Melatos}, {Mendell}, {Mercer}, {Meshkov}, {Messenger}, {Meyers},
  {Miao}, {Middleton}, {Mikhailov}, {Miller}, {Miller}, {Millhouse}, {Ming},
  {Mirshekari}, {Mishra}, {Mitra}, {Mitrofanov}, {Mitselmakher}, {Mittleman},
  {Moe}, {Mohanty}, {Mohapatra}, {Moore}, {Moraru}, {Moreno}, {Morriss},
  {Mossavi}, {Mow-Lowry}, {Mueller}, {Mueller}, {Mukherjee}, {Mullavey},
  {Munch}, {Murphy}, {Murray}, {Mytidis}, {Nash}, {Nayak}, {Necula}, {Nedkova},
  {Newton}, {Nguyen}, {Nielsen}, {Nissanke}, {Nitz}, {Nolting}, {Normandin},
  {Nuttall}, {Ochsner}, {O'Dell}, {Oelker}, {Ogin}, {Oh}, {Oh}, {Ohme},
  {Oppermann}, {Oram}, {O'Reilly}, {Ortega}, {O'Shaughnessy}, {Osthelder},
  {Ott}, {Ottaway}, {Ottens}, {Overmier}, {Owen}, {Padilla}, {Pai}, {Pai},
  {Palashov}, {Pal-Singh}, {Pan}, {Pankow}, {Pannarale}, {Pant}, {Papa},
  {Paris}, {Patrick}, {Pedraza}, {Pekowsky}, {Pele}, {Penn}, {Perreca},
  {Phelps}, {Pierro}, {Pinto}, {Pitkin}, {Poeld}, {Post}, {Poteomkin},
  {Powell}, {Prasad}, {Predoi}, {Premachandra}, {Prestegard}, {Price},
  {Principe}, {Privitera}, {Prix}, {Prokhorov}, {Puncken}, {P{\"u}rrer}, {Qin},
  {Quetschke}, {Quintero}, {Quiroga}, {Quitzow-James}, {Raab}, {Rabeling},
  {Radkins}, {Raffai}, {Raja}, {Rajalakshmi}, {Rakhmanov}, {Ramirez},
  {Raymond}, {Reed}, {Reid}, {Reitze}, {Reula}, {Riles}, {Robertson}, {Robie},
  {Rollins}, {Roma}, {Romano}, {Romanov}, {Romie}, {Rowan}, {R{\"u}diger},
  {Ryan}, {Sachdev}, {Sadecki}, {Sadeghian}, {Saleem}, {Salemi}, {Sammut},
  {Sandberg}, {Sanders}, {Sannibale}, {Santiago-Prieto}, {Sathyaprakash},
  {Saulson}, {Savage}, {Sawadsky}, {Scheuer}, {Schilling}, {Schmidt},
  {Schnabel}, {Schofield}, {Schreiber}, {Schuette}, {Schutz}, {Scott}, {Scott},
  {Sellers}, {Sengupta}, {Sergeev}, {Serna}, {Sevigny}, {Shaddock}, {Shahriar},
  {Shaltev}, {Shao}, {Shapiro}, {Shawhan}, {Shoemaker}, {Sidery}, {Siemens},
  {Sigg}, {Silva}, {Simakov}, {Singer}, {Singer}, {Singh}, {Sintes},
  {Slagmolen}, {Smith}, {Smith}, {Smith}, {Smith-Lefebvre}, {Son}, {Sorazu},
  {Souradeep}, {Staley}, {Stebbins}, {Steinke}, {Steinlechner}, {Steinlechner},
  {Steinmeyer}, {Stephens}, {Steplewski}, {Stevenson}, {Stone}, {Strain},
  {Strigin}, {Sturani}, {Stuver}, {Summerscales}, {Sutton}, {Szczepanczyk},
  {Szeifert}, {Talukder}, {Tanner}, {T{\'a}pai}, {Tarabrin}, {Taracchini},
  {Taylor}, {Tellez}, {Theeg}, {Thirugnanasambandam}, {Thomas}, {Thomas},
  {Thorne}, {Thorne}, {Thrane}, {Tiwari}, {Tomlinson}, {Torres}, {Torrie},
  {Traylor}, {Tse}, {Tshilumba}, {Ugolini}, {Unnikrishnan}, {Urban}, {Usman},
  {Vahlbruch}, {Vajente}, {Valdes}, {Vallisneri}, {van Veggel}, {Vass},
  {Vaulin}, {Vecchio}, {Veitch}, {Veitch}, {Venkateswara}, {Vincent-Finley},
  {Vitale}, {Vo}, {Vorvick}, {Vousden}, {Vyatchanin}, {Wade}, {Wade}, {Wade},
  {Walker}, {Wallace}, {Walsh}, {Wang}, {Wang}, {Wang}, {Ward}, {Warner},
  {Was}, {Weaver}, {Weinert}, {Weinstein}, {Weiss}, {Welborn}, {Wen},
  {Wessels}, {Westphal}, {Wette}, {Whelan}, {Whitcomb}, {White}, {Whiting},
  {Wilkinson}, {Williams}, {Williams}, {Williamson}, {Willis}, {Willke},
  {Wimmer}, {Winkler}, {Wipf}, {Wittel}, {Woan}, {Worden}, {Xie}, {Yablon},
  {Yakushin}, {Yam}, {Yamamoto}, {Yancey}, {Yang}, {Zanolin}, {Zhang}, {Zhang},
  {Zhang}, {Zhang}, {Zhao}, {Zhou}, {Zhu}, {Zucker}, {Zuraw}, \&
  {Zweizig}}]{ligo2015advanced}
{LIGO Scientific Collaboration}, {Aasi}, J., {Abbott}, B.~P., {et~al.} 2015,
  Classical and Quantum Gravity, 32, 074001,
  \dodoi{10.1088/0264-9381/32/7/074001}

\bibitem[{{Lin} {et~al.}(2022){Lin}, {van Kerkwijk}, {Main}, {Mahajan}, {Pen},
  \& {Kirsten}}]{Lin2022crab}
{Lin}, R., {van Kerkwijk}, M.~H., {Main}, R., {et~al.} 2022, arXiv e-prints,
  arXiv:2211.05209, \dodoi{10.48550/arXiv.2211.05209}

\bibitem[{{Maccone}(2010)}]{maccone2010focal}
{Maccone}, C. 2010, Acta Astronautica, 67, 521,
  \dodoi{10.1016/j.actaastro.2010.03.013}

\bibitem[{{Macquart}(2004)}]{macquart2004scattering}
{Macquart}, J.~P. 2004, \aap, 422, 761, \dodoi{10.1051/0004-6361:20034512}

\bibitem[{{Matsunaga} \& {Yamamoto}(2006)}]{matsunaga2006finite}
{Matsunaga}, N., \& {Yamamoto}, K. 2006, \jcap, 2006, 023,
  \dodoi{10.1088/1475-7516/2006/01/023}

\bibitem[{{Miller} \& {Wheeler}(1997)}]{miller1997delayed}
{Miller}, W.~A., \& {Wheeler}, J.~A. 1997, Foundations Of Quantum Mechanics In
  The Light Of New Technology. Series: Advanced Series in Applied Physics, 4,
  72, \dodoi{10.1142/9789812819895_0008}

\bibitem[{{Mishra} {et~al.}(2021){Mishra}, {Meena}, {More}, {Bose}, \&
  {Bagla}}]{mishra2021gravitational}
{Mishra}, A., {Meena}, A.~K., {More}, A., {Bose}, S., \& {Bagla}, J.~S. 2021,
  \mnras, 508, 4869, \dodoi{10.1093/mnras/stab2875}

\bibitem[{{Nakamura}(1998)}]{nakamura1998gravitational}
{Nakamura}, T.~T. 1998, \prl, 80, 1138, \dodoi{10.1103/PhysRevLett.80.1138}

\bibitem[{{Nakamura} \& {Deguchi}(1999)}]{nakamura1999wave}
{Nakamura}, T.~T., \& {Deguchi}, S. 1999, Progress of Theoretical Physics
  Supplement, 133, 137, \dodoi{10.1143/PTPS.133.137}

\bibitem[{{Narayan}(1992)}]{narayan1992physics}
{Narayan}, R. 1992, Philosophical Transactions of the Royal Society of London
  Series A, 341, 151, \dodoi{10.1098/rsta.1992.0090}

\bibitem[{{Nemiroff} {et~al.}(1993){Nemiroff}, {Norris}, {Wickramasinghe},
  {Horack}, {Kouveliotou}, {Fishman}, {Meegan}, {Wilson}, \&
  {Paciesas}}]{nemiroff1993searching}
{Nemiroff}, R.~J., {Norris}, J.~P., {Wickramasinghe}, W.~A.~D.~T., {et~al.}
  1993, \apj, 414, 36, \dodoi{10.1086/173054}

\bibitem[{{Nouri-Zonoz}(1999)}]{Nouri-Zonoz1999}
{Nouri-Zonoz}, M. 1999, \prd, 60, 024013, \dodoi{10.1103/PhysRevD.60.024013}

\bibitem[{{Nye}(1999)}]{Nye1999}
{Nye}, J.~F. 1999, {Natural focusing and fine structure of light: caustics and
  wave dislocations}

\bibitem[{{Oancea} {et~al.}(2020){Oancea}, {Joudioux}, {Dodin}, {Ruiz},
  {Paganini}, \& {Andersson}}]{oancea2020gravitational}
{Oancea}, M.~A., {Joudioux}, J., {Dodin}, I.~Y., {et~al.} 2020, \prd, 102,
  024075, \dodoi{10.1103/PhysRevD.102.024075}

\bibitem[{{Oguri}(2019)}]{oguri2019strong}
{Oguri}, M. 2019, Reports on Progress in Physics, 82, 126901,
  \dodoi{10.1088/1361-6633/ab4fc5}

\bibitem[{{Oguri} \& {Takahashi}(2020)}]{oguri2020probing}
{Oguri}, M., \& {Takahashi}, R. 2020, \apj, 901, 58,
  \dodoi{10.3847/1538-4357/abafab}

\bibitem[{{Oguri} \& {Takahashi}(2022)}]{oguri2022amplitude}
---. 2022, \prd, 106, 043532, \dodoi{10.1103/PhysRevD.106.043532}

\bibitem[{{Ohanian}(1973)}]{ohanian1973focusing}
{Ohanian}, H.~C. 1973, \prd, 8, 2734, \dodoi{10.1103/PhysRevD.8.2734}

\bibitem[{{Ohanian}(1974)}]{ohanian1974on}
---. 1974, International Journal of Theoretical Physics, 9, 425,
  \dodoi{10.1007/BF01810927}

\bibitem[{{Ohanian}(1983)}]{ohanian1983caustics}
---. 1983, \apj, 271, 551, \dodoi{10.1086/161221}

\bibitem[{{Pearson} {et~al.}(2014){Pearson}, {Olver}, \&
  {Porter}}]{pearson2014numerical}
{Pearson}, J.~W., {Olver}, S., \& {Porter}, M.~A. 2014, arXiv e-prints,
  arXiv:1407.7786, \dodoi{10.48550/arXiv.1407.7786}

\bibitem[{{Peters}(1974)}]{peters1974index}
{Peters}, P.~C. 1974, \prd, 9, 2207, \dodoi{10.1103/PhysRevD.9.2207}

\bibitem[{{Rahvar}(2018)}]{rahvar2018gravitational}
{Rahvar}, S. 2018, \mnras, 479, 406, \dodoi{10.1093/mnras/sty1369}

\bibitem[{{Rickett}(1990)}]{rickett1990radio}
{Rickett}, B.~J. 1990, \araa, 28, 561,
  \dodoi{10.1146/annurev.aa.28.090190.003021}

\bibitem[{{Saha}(2000)}]{saha2000lensing}
{Saha}, P. 2000, \aj, 120, 1654, \dodoi{10.1086/301581}

\bibitem[{{Sammons} {et~al.}(2022){Sammons}, {James}, {Trott}, \&
  {Walker}}]{sammons2022effect}
{Sammons}, M.~W., {James}, C.~W., {Trott}, C.~M., \& {Walker}, M. 2022, \mnras,
  \dodoi{10.1093/mnras/stac3013}

\bibitem[{{Schneider}(1983)}]{schneider1983mutual}
{Schneider}, P. 1983, in Liege International Astrophysical Colloquia, Vol.~24,
  Liege International Astrophysical Colloquia, ed. J.-P. {Swings}, 131--133

\bibitem[{{Shi} \& {Xu}(2021)}]{shi2021plasma}
{Shi}, X., \& {Xu}, Z. 2021, \mnras, 506, 6039, \dodoi{10.1093/mnras/stab2108}

\bibitem[{{Stanek} {et~al.}(1993){Stanek}, {Paczynski}, \&
  {Goodman}}]{stanek1993features}
{Stanek}, K.~Z., {Paczynski}, B., \& {Goodman}, J. 1993, \apjl, 413, L7,
  \dodoi{10.1086/186946}

\bibitem[{{Stewart}(1989)}]{stewart1989solutions}
{Stewart}, J.~M. 1989, Proceedings of the Royal Society of London Series A,
  424, 239, \dodoi{10.1098/rspa.1989.0078}

\bibitem[{{Sugiyama} {et~al.}(2020){Sugiyama}, {Kurita}, \&
  {Takada}}]{sugiyama2020on}
{Sugiyama}, S., {Kurita}, T., \& {Takada}, M. 2020, \mnras, 493, 3632,
  \dodoi{10.1093/mnras/staa407}

\bibitem[{{Suyama}(2020)}]{suyama2020on}
{Suyama}, T. 2020, \apj, 896, 46, \dodoi{10.3847/1538-4357/ab8d3f}

\bibitem[{{Suyama} {et~al.}(2006){Suyama}, {Tanaka}, \&
  {Takahashi}}]{suyama2006exact}
{Suyama}, T., {Tanaka}, T., \& {Takahashi}, R. 2006, \prd, 73, 024026,
  \dodoi{10.1103/PhysRevD.73.024026}

\bibitem[{{Takahashi}(2006)}]{takahashi2006amplitude}
{Takahashi}, R. 2006, \apj, 644, 80, \dodoi{10.1086/503323}

\bibitem[{{Takahashi}(2017)}]{takahashi2017arrival}
---. 2017, \apj, 835, 103, \dodoi{10.3847/1538-4357/835/1/103}

\bibitem[{{Takahashi} \& {Nakamura}(2003)}]{takahashi2003wave}
{Takahashi}, R., \& {Nakamura}, T. 2003, \apj, 595, 1039,
  \dodoi{10.1086/377430}

\bibitem[{{Takahashi} {et~al.}(2005){Takahashi}, {Suyama}, \&
  {Michikoshi}}]{takahashi2005scattering}
{Takahashi}, R., {Suyama}, T., \& {Michikoshi}, S. 2005, \aap, 438, L5,
  \dodoi{10.1051/0004-6361:200500140}

\bibitem[{{Tanaka} \& {Suyama}(2023)}]{tanaka2023kramers}
{Tanaka}, S., \& {Suyama}, T. 2023, arXiv e-prints, arXiv:2303.05650.
\newblock \doarXiv{2303.05650}

\bibitem[{Thom(1967)}]{thom1967}
Thom, R. 1967, Structural Stability And Morphogenesis (CRC Press).
\newblock \url{https://books.google.ca/books?id=YVNPDwAAQBAJ}

\bibitem[{{Turyshev}(2017)}]{turyshev2017wave}
{Turyshev}, S.~G. 2017, \prd, 95, 084041, \dodoi{10.1103/PhysRevD.95.084041}

\bibitem[{{Ulmer} \& {Goodman}(1995)}]{ulmer1995femtolensing}
{Ulmer}, A., \& {Goodman}, J. 1995, \apj, 442, 67, \dodoi{10.1086/175422}

\bibitem[{{Vanderlinde} {et~al.}(2019){Vanderlinde}, {Liu}, {Gaensler}, {Bond},
  {Hinshaw}, {Ng}, {Chiang}, {Stairs}, {Brown}, {Sievers}, {Mena}, {Smith},
  {Bandura}, {Masui}, {Spekkens}, {Belostotski}, {Dobbs}, {Turok}, {Boyle},
  {Rupen}, {Landecker}, {Pen}, \& {Kaspi}}]{vanderlinde2019chord}
{Vanderlinde}, K., {Liu}, A., {Gaensler}, B., {et~al.} 2019, in Canadian Long
  Range Plan for Astronomy and Astrophysics White Papers, Vol. 2020, 28,
  \dodoi{10.5281/zenodo.3765414}

\bibitem[{{Venumadhav} {et~al.}(2017){Venumadhav}, {Dai}, \&
  {Miralda-Escud{\'e}}}]{venumadhav2017microlensing}
{Venumadhav}, T., {Dai}, L., \& {Miralda-Escud{\'e}}, J. 2017, \apj, 850, 49,
  \dodoi{10.3847/1538-4357/aa9575}

\bibitem[{{Vijaykumar} {et~al.}(2022){Vijaykumar}, {Mehta}, \&
  {Ganguly}}]{vijaykumar2022detection}
{Vijaykumar}, A., {Mehta}, A.~K., \& {Ganguly}, A. 2022, arXiv e-prints,
  arXiv:2202.06334.
\newblock \doarXiv{2202.06334}

\bibitem[{{Walker} {et~al.}(2004){Walker}, {Melrose}, {Stinebring}, \&
  {Zhang}}]{walker2004interpretation}
{Walker}, M.~A., {Melrose}, D.~B., {Stinebring}, D.~R., \& {Zhang}, C.~M. 2004,
  \mnras, 354, 43, \dodoi{10.1111/j.1365-2966.2004.08159.x}

\bibitem[{{Witt} \& {Mao}(1994)}]{witt1994can}
{Witt}, H.~J., \& {Mao}, S. 1994, \apj, 430, 505, \dodoi{10.1086/174426}

\bibitem[{{Wucknitz} {et~al.}(2021){Wucknitz}, {Spitler}, \&
  {Pen}}]{wucknitz2021cosmology}
{Wucknitz}, O., {Spitler}, L.~G., \& {Pen}, U.~L. 2021, \aap, 645, A44,
  \dodoi{10.1051/0004-6361/202038248}

\bibitem[{{Xiao} {et~al.}(2022){Xiao}, {Dai}, \& {McQuinn}}]{xiao2022detecting}
{Xiao}, H., {Dai}, L., \& {McQuinn}, M. 2022, arXiv e-prints, arXiv:2206.13534.
\newblock \doarXiv{2206.13534}

\bibitem[{{Yamamoto}(2003)}]{yamamoto2003path}
{Yamamoto}, K. 2003, arXiv e-prints, astro.
\newblock \doarXiv{astro-ph/0309696}

\bibitem[{{Yang} {et~al.}(2014){Yang}, {Zhang}, {Zimmerman}, \&
  {Chen}}]{yang2014scalar}
{Yang}, H., {Zhang}, F., {Zimmerman}, A., \& {Chen}, Y. 2014, \prd, 89, 064014,
  \dodoi{10.1103/PhysRevD.89.064014}

\bibitem[{{Zengino{\v{g}}lu} \& {Galley}(2012)}]{zenginoglu2012caustic}
{Zengino{\v{g}}lu}, A., \& {Galley}, C.~R. 2012, \prd, 86, 064030,
  \dodoi{10.1103/PhysRevD.86.064030}

\end{thebibliography}

\end{document}